%% file: U1DWFVFESPTGW-arXiv-version2.tex
\documentclass[a4paper,11pt]{article}
\pdfoutput=1
\usepackage{jheppub}
\usepackage[T1]{fontenc}
\usepackage{cancel}

\title{\boldmath A two-component vector WIMP -- fermion FIMP dark matter model with an extended seesaw mechanism}

\author[a]{Francesco Costa,}
\author[a,*]{Sarif Khan,}
\author[b,c,*]{and Jinsu Kim\note[*]{Corresponding author.}}

\affiliation[a]{
Institute for Theoretical Physics, 
Georg-August University G\"ottingen,\\
Friedrich-Hund-Platz 1, G\"ottingen D-37077, Germany}
\affiliation[b]{
School of Physics Science and Engineering, 
Tongji University, \\
Shanghai 200092, China}
\affiliation[c]{
Theoretical Physics Department, CERN, \\
1211 Geneva 23, Switzerland}

\emailAdd{francesco.costa@theorie.physik.uni-goettingen.de}
\emailAdd{sarif.khan@uni-goettingen.de}
\emailAdd{jinsu.kim@cern.ch}

\preprint{CERN-TH-2022-155}

\abstract{
We consider an extension of the Standard Model that explains the neutrino masses and has a rich dark matter phenomenology. The model has two dark matter candidates, a vector WIMP and a fermion FIMP, and the sum of their relic densities matches the total dark matter abundance. We extensively study the dark matter production mechanisms and its connection with the neutrino sector, together with various bounds from present and future experiments. The extra scalar field in the model may induce a first-order phase transition in the early Universe. We study the production of stochastic gravitational waves associated with the first-order phase transition. We show that the phase transition can be strong, and thus the model may satisfy one of the necessary conditions for a successful electroweak baryogenesis. Detectability of the phase transition-associated gravitational waves is also discussed.
}

\begin{document} 
\maketitle
\flushbottom

\section{Introduction}
\label{sec:intro}
The Standard Model (SM) of particle physics has proved extremely successful in the past decades with the experiments matching its predictions and the Higgs boson discovery being the final piece to complete it. Nonetheless, astrophysical and cosmological evidence have posed questions that are not explained by the SM and are still open problems to this date. 

It is well established by neutrino oscillation data (NOD) that the neutrinos have a non-zero mass while in the SM they are massless. A mechanism is therefore needed to generate the masses~\cite{Super-Kamiokande:1998kpq,Gonzalez-Garcia:2002bkq}. The neutrinos are not only massive, but their masses are also much lighter than the other matter particles. The mass splitting between the first and the second eigenstates is $|\Delta m_{21}^{2}|= 7.42_{-0.20}^{+0.21} \times 10^{-5} \, {\rm eV}^{2}$, and the mass gap between the second and the third is $\Delta m_{32}^{2}=2.517_{-0.028}^{+0.026} \times 10^{-3} \, {\rm eV}^{2}$~\cite{Esteban:2020cvm}. Also, from cosmological data, the sum of the neutrino masses is bounded by $\sum_i m_{\nu_i} < 0.23 \, {\rm eV}$~\cite{Planck:2015fie,Planck:2018vyg}. These observations are calling for a new mechanism. Arguably the easiest and first proposed mechanism is the so-called type-I seesaw~\cite{Minkowski:1977sc,Gell-Mann:1979vob}, where heavy singlet leptons are introduced: The mixing between the heavy singlet leptons and the light neutrinos can generate a small mass since the light neutrino masses are suppressed by the heavy mass scale, resulting in $m_{\nu}\sim y v/M$ where $M$ is of the order of the heavy lepton mass, $v$ is the SM Higgs vacuum expectation value (VEV), and $y$ is the light neutrino Yukawa coupling.

An extended version of the type-I seesaw mechanism, dubbed extended double seesaw~\cite{Kang:2006sn,Mitra:2011qr}, where a second set of singlet neutrinos is added, was proposed to achieve a low-scale leptogenesis without a fine tuning of the heavy neutrino masses; see also Ref.~\cite{Majee:2008mn} for an ultraviolet (UV) completion. The attractiveness of the low-scale leptogenesis is its detection possibilities from future collider experiments. In addition, the supersymmetric version of the extended double seesaw mechanism avoids the gravitino problem~\cite{Kang:2006sn,Kawasaki:2004qu}; see also Ref.~\cite{Hook:2018sai}.
Notably, the double seesaw mechanism allows us to consider Yukawa couplings for the extra neutrinos up to $O(1)$ with masses at the TeV scale, having the possibility to be probed by future collider experiments. 

The dark matter (DM) is another missing piece of the SM. We have cosmological evidences that indicate that our Universe is composed of 23\% of DM. These observations point towards cold, dark, and particle-like explanations~\cite{Ostriker:1973uit,Planck:2018vyg,Corbelli:1999af}. The standard solution to the problem is a Weekly Interacting Massive Particle (WIMP) which is initially in thermal contact with the SM thermal bath in the early Universe. At some later time, it freezes out, producing the relic density observed today that is inversely proportional to the thermal cross section~\cite{Gunn:1978gr,Hut:1977zn,Lee:1977ua,Bertone:2004pz}.

Alternative mechanisms have been explored with increasing interest since the effort to detect a WIMP-like particle has been unsuccessful up to now~\cite{XENON:2018voc,CMS:2016lcl,MAGIC:2016xys, Arcadi:2017kky,PandaX-II:2016vec,LUX:2016ggv}. In particular, the freeze-in mechanism has gained tremendous attention in the past two decades~\cite{McDonald:2001vt,Choi:2005vq,Kusenko:2006rh,Hall:2009bx,Cheung:2011nn,Elahi:2014fsa,Arcadi:2015ffa,Bernal:2017kxu,Benakli:2017whb,Bernal:2018qlk,Bernal:2019mhf,Barman:2019lvm,Covi:2020pch,Khan:2020pso,Garcia:2020hyo,Bernal:2020qyu,Barman:2020plp,Barman:2020ifq,Barman:2021yaz,Barman:2021lot,Belanger:2021slj,Barman:2022njh}. The DM particle in this case is called a Feebly Interacting Massive Particle (FIMP) because its interaction is in general much smaller than the electroweak scale, with couplings $\lesssim \mathcal{O}(10^{-8})$. The tiny coupling is due to the requirement that the FIMP remains out-of-equilibrium during the history of the Universe.\footnote{
Such a tiny coupling may naturally be realised in a clockwork framework~\cite{Choi:2015fiu,Kaplan:2015fuy,Giudice:2016yja}; see, {\it e.g.}, Refs.~\cite{Kim:2017mtc,Kim:2018xsp,Goudelis:2018xqi}.
} The DM abundance is then produced by the out-of-equilibrium scattering or decay processes.
When the particles are produced via operators of dimension higher than four, the production mechanism may be of the so-called UV freeze-in type where the relic density is mostly produced at the reheating temperature $T_R$. This is in stark contrast with the standard infrared (IR) freeze-in where the main production occurs at $T\sim m_{{\rm FIMP}}$ where $m_{{\rm FIMP}}$ is the FIMP mass scale~\cite{Elahi:2014fsa}. As we shall discuss later, in the model we study in this work, operators of dimension five give UV contributions to the relic density before the spontaneous symmetry breaking. After the symmetry breaking, the dimension-5 operators also give IR contributions, dominated by the Higgs decay. The dominant contribution will depend on values of the Higgs mass, the reheating temperature $T_R$, and the scale of new physics $\Lambda$.

The visible sector described by the SM is composed of a complex arrangement of particles and gauge groups. Likewise, we could expect the similar complexity to arise in the dark sector. There is no experimental indication that the DM sector is composed of a single field. Since both the freeze-in and freeze-out mechanisms are viable production mechanisms, both the WIMP and FIMP could have been active in the early Universe, producing parts of the total DM relic density $\Omega_{\rm Tot} h^2 = 0.120 \pm 0.001$ as observed by the Planck experiment~\cite{Planck:2018vyg}. Although the simplest setup would be the case where there are two DM candidates both of which contribute to the total DM relic abundance, one may consider a more general multi-component DM scenarios. Recent studies on the multi-component DM scenarios include Refs.~\cite{Zurek:2008qg,Profumo:2009tb,Feldman:2010wy,Ko:2010at,Drozd:2011aa,Aoki:2012ub,Bhattacharya:2013hva,Baek:2013dwa,Esch:2014jpa,Ko:2014bka,Bian:2014cja,Karam:2015jta,Arcadi:2016kmk,DuttaBanik:2016jzv,Karam:2016rsz,Bhattacharya:2016ysw,Ko:2016fcd,Aoki:2016glu,Ahmed:2017dbb,Aoki:2018gjf,Chakraborti:2018lso,Poulin:2018kap,YaserAyazi:2018lrv,Chakraborti:2018aae,Bhattacharya:2019fgs,Chen:2019pnt,Yaguna:2019cvp,Bhattacharya:2019tqq,Betancur:2020fdl,Belanger:2020hyh,Belanger:2021lwd,Bhattacharya:2021rwh,Das:2021zea,Betancur:2021ect,Chakrabarty:2021kmr,Mohamadnejad:2021tke,DiazSaez:2021pfw,Choi:2021yps,Belanger:2022qxt,Das:2022oyx,Ho:2022erb,Costa:2022oaa}. 

In this paper, we consider a beyond the SM (BSM) scenario that addresses the aforementioned problems, exploring its viability and the possible experimental signatures. We introduce two sets of three-generation extra neutrinos $N_L^i$ and $S_L^i$ where the first two generations of neutrinos are used in the extended seesaw mechanism to explain the light neutrino masses while the third generation will be part of the dark sector. In the mass basis, through mixing, $N^3_m$ and $S^3_m$ become FIMP-type particles, and considering $S^3_m$ to be the lighter one, it may become a viable DM candidate. We shall explore possible connections between the neutrino parameters and the DM relic density. The second DM candidate is the vector gauge boson $W_D$ associated with an extra dark $U(1)_D$ gauge symmetry. We study constraints from the lepton flavour violation (LFV) data to the mixing angles and to the DM production via the neutrinos sector.

The dark Higgs field $\phi_D$ associated with the extra dark $U(1)_D$ modifies the scalar sector with respect to the SM. The evolution of the vacuum state may thus change, opening possibilities of having a first-order phase transition (FOPT). We show that the FOPT can be strong and discuss the detectability of the associated stochastic gravitational waves (GWs)~\cite{Kamionkowski:1993fg} by future space-based observatories such as LISA~\cite{Baker:2019nia}, DECIGO~\cite{Seto:2001qf,Kawamura:2006up,Sato:2017dkf,Isoyama:2018rjb,Kawamura:2020pcg}, and BBO~\cite{Corbin:2005ny,Crowder:2005nr,Harry:2006fi}. Recent work on the subject includes {\it e.g.} Refs.~\cite{Grojean:2006bp,Huber:2008hg,Espinosa:2008kw,Caprini:2015zlo,Artymowski:2016tme,Baldes:2017rcu,Beniwal:2018hyi,Hashino:2018zsi,Caprini:2018mtu,Bian:2018mkl,Bian:2018bxr,Bian:2019szo,Bian:2019kmg,Caprini:2019egz,Di:2020ivg,Zhou:2021cfu,Mohamadnejad:2021tke,Bian:2021dmp,Costa:2022oaa}. This is an exciting possibility that opens an experimental window on the cosmological implication of the BSM model we present and may complement the study of (in-)direct detections and collider searches that may probe the nature of DM.

The rest of the paper is organised as follows. In Section~\ref{sec:model}, we present the model under consideration in detail and explain the generation of the neutrino masses as well as the DM candidates. We then scrutinise the LFV bounds on the neutrino sectors in Section~\ref{sec:LFVnu}. In Section~\ref{sec:DMpheno}, we study the two-component DM scenarios and present the result together with various detection bounds. In Section~\ref{sec:FOPTsGWs}, possibilities of having a FOPT are explored. We also discuss the detectability of the stochastic GW signals associated with the FOPTs. In doing so, we present three benchmark points (BPs) that realise the neutrino masses, the correct DM relic density, a strong FOPT, and a detectable GW signal at the same time. Finally, we discuss potential collider searches in Section~\ref{sec:collider} before we conclude in Section~\ref{sec:conc}.

\section{Model}
\label{sec:model}
We consider the following Lagrangian 
\begin{align}\label{eqn:lag}
\mathcal{L}=
\mathcal{L}_{\rm SM} + \mathcal{L}_{N} + \mathcal{L}_{\rm DM}
+ (D_{\mu}\phi_{D})^{\dagger} (D^{\mu}\phi_{D})
-\frac{1}{4} F_D^{\alpha \beta} {F_D}_{\alpha \beta}
-\frac{\zeta}{2}F_D^{\alpha\beta}B_{\alpha\beta}
-V(\phi_{h},\phi_{D})
\,, 
\end{align}
where $\mathcal{L}_{\rm SM}$ is the SM Lagrangian with the SM Higgs field $\phi_h$, $\mathcal{L}_N$ is the Lagrangian associated with the additional singlet neutrinos which take part in the neutrino masses, and $\mathcal{L}_{\rm DM}$ corresponds to the DM Lagrangian. The fourth term describes the kinetic term for the extra $U(1)_D$ Higgs $\phi_D$, and the covariant derivative is given by $D_\mu = \partial_\mu - ig_DW_{D\mu}$ with $g_D$ being the $U(1)_D$ gauge coupling and $W_D$ the vector boson associated with the extra $U(1)_D$ gauge symmetry. $F^{\alpha \beta}_{D}$ is the field strength of the vector boson $W_D$, $B^{\alpha\beta}$ is the field strength tensor associated with the hypercharge $U(1)_Y$ gauge group, and the gauge kinetic mixing between these two field strength tensors is parametrised by $\zeta$. Finally, the last term represents the scalar potential which is given by
\begin{align}\label{eqn:int}
V(\phi_h, \phi_D) = \mu_{D}^{2} \phi_{D}^{\dagger} \phi_{D} 
+ \lambda_{D} (\phi_{D}^{\dagger} \phi_{D})^{2}
+ \lambda_{hD}(\phi_{h}^{\dagger} \phi_{h}) (\phi_{D}^{\dagger} \phi_{D}) 
\,.
\end{align}
The neutrino sector is described by
\begin{align}
\mathcal{L}_{N}&=
\sum_{i=1,2}\frac{i}{2}\bar{N^i_L}\gamma^{\mu}\partial_{\mu} N^{i}_L
+ \sum_{i=1,2}\frac{i}{2}\bar{S^i_L}\gamma^{\mu}\partial_{\mu} S^{i}_L 
- \sum_{i,j = 1,2 }\mu_{ij} S^{i}_{L} S^{j}_{L}
- \sum_{i,j = 1,2 } M_{S}^{ij} S^{i}_{L} N^{j}_{L}  \nonumber \\ 
&\quad
- \sum_{i,j = 1,2 } M_{R}^{ij} N^{i}_{L} N^{j}_{L}
-\sum_{i=e,\,\mu,\,\tau, j=1,2} y_{ij} \bar{L_{i}}
\tilde {\phi_{h}} N_{j} 
+ {\rm h.c.}
\label{eqn:lagN}
\end{align}
where $\tilde{\phi}_{h}=i\sigma_2\phi^*_h$.
We have considered the Yukawa term for $S_i$ to be negligible compared to the one for $N_i$, following the standard extended double seesaw model~\cite{Kang:2006sn,Mitra:2011qr}. The Yukawa terms with the dark Higgs $\phi_D$ are forbidden by symmetries.
The parameters $M_{S}^{ij}$, $M_{R}^{ij}$, and $\mu_{ij}$ are constants with mass-dimension one, while $y_{ij}$ are dimensionless coupling constants that compose the Dirac mass matrix $M_D$ that we will use later.
We have considered that $S^3_{L}$ and $N^3_{L}$ are decoupled from the rest of the particle spectra by assuming that they are $Z_2$-odd while the rest of the particles are $Z_2$-even. Such a discrimination ensures that the lightest particle may be treated as a FIMP-type DM candidate. Productions of $S^3_{L}$ and $N^3_{L}$ are through dimension-5 operators which get naturally suppressed when the scale of new physics $\Lambda$ is large, ensuring feeble interactions with the rest of the particle spectra; in the present work, we consider $\Lambda \geq 10^{14}$ GeV.
\begin{table}[t!]
\centering
\renewcommand{\arraystretch}{1.2}
\tabcolsep=0.098cm
\begin{tabular}{||c|c|c|c||}
\hline
\hline
\begin{tabular}{c}
Groups \&\\
Symmetry \\ 
\hline
$SU(2)_{L}$\\ 
\hline
$U(1)_{Y}$\\ 
\hline
$U(1)_{D}$\\ 
\hline
$Z_2$\\
\end{tabular}
&
\begin{tabular}{c|c|c}
\multicolumn{3}{c}{Baryons}\\ 
\hline
$Q_{L}^{i}=(u_{L}^{i},d_{L}^{i})^{T}$
&$u_{R}^{i}$
&$d_{R}^{i}$\\ 
\hline
$2$&$1$&$1$\\ 
\hline
$1/6$&$2/3$&$-1/3$\\ 
\hline
$0$&$0$&$0$\\
\hline 
$1$&$1$&$1$\\ 
\end{tabular}
&
\begin{tabular}{c|c|c|c|c|c}
\multicolumn{6}{c}{Leptons}\\
\hline
$L_{L}^{i}=(\nu_{L}^{i},e_{L}^{i})^{T}$ 
& $e_{R}^{i}$ 
& $N_{L}^{j}$
& $S_{L}^{j}$
& $N_L^3$
& $S_L^3$\\
\hline
$2$&$1$&$1$&$1$&$1$&$1$\\
\hline
$-1/2$&$-1$&$0$&$0$&$0$&$0$\\
\hline
$0$&$0$&$0$&$0$&$0$&$0$\\
\hline
$1$&$1$&$1$&$1$&$-1$&$-1$\\
\end{tabular}
&
\begin{tabular}{c|c}
\multicolumn{2}{c}{Scalars}\\
\hline
$\phi_{h}$&$\phi_{D}$\\
\hline
$2$&$1$\\
\hline
$1/2$&$0$\\
\hline
$0$&$1$\\
\hline
$1$&$1$\\
\end{tabular}\\
\hline
\hline
\end{tabular}
\caption{Particle contents and their corresponding charges under different gauge groups and discrete symmetry. The index $i$ is for three flavours, running from 1 to 3 whereas the index $j$ runs from 1 to 2.}
\label{tab:contents}
\end{table}
In the neutrino sector, the effect of such dimension-5 operators is negligible.
The Lagrangian associated with $S^3_{L}$ and $N^3_{L}$ is thus given by
\begin{align}
\mathcal{L}_{\rm DM} &= \frac{i}{2}\bar{N^3_L}\gamma^{\mu}\partial_{\mu}
N^{3}_L
+ \frac{i}{2}\bar{S^3_L}\gamma^{\mu}\partial_{\mu} S^{3}_L 
- \mu_{33} S^{3}_{L} S^{3}_{L}
- M_{S}^{33} S^{3}_{L} N^{3}_{L}
- M_{R}^{33} N^{3}_{L} N^{3}_{L} 
\nonumber \\ 
&\quad
+ \frac{\kappa}{\Lambda} S^{3}_{L} S^{3}_{L} (\phi^{\dagger}_h \phi_h)
+ \frac{\kappa^{\prime}}{\Lambda} S^{3}_{L} S^{3}_{L} (\phi^{\dagger}_D \phi_D)
+ \frac{\xi}{\Lambda} N^{3}_{L} N^{3}_{L} (\phi^{\dagger}_h \phi_h)
+ \frac{\xi^{\prime}}{\Lambda} N^{3}_{L} N^{3}_{L} (\phi^{\dagger}_D \phi_D)
\nonumber \\ 
&\quad 
+ \frac{\alpha}{\Lambda} N^{3}_{L} S^{3}_{L} (\phi^{\dagger}_h \phi_h)
+ \frac{\alpha^{\prime}}{\Lambda} N^{3}_{L} S^{3}_{L} (\phi^{\dagger}_D \phi_D)
+ {\rm h.c.}
\label{eqn:ldm}
\end{align}
Finally, the term proportional to the coupling $\zeta$ denotes the gauge kinetic mixing term between the SM $U(1)_Y$ gauge boson and the $U(1)_D$ gauge boson. It has been shown that small values of the parameter $\zeta$ are favoured from the viewpoint of the muon $g-2$~\cite{Altmannshofer:2019zhy,Biswas:2021dan}; see also, {\it e.g.}, Ref.~\cite{Bauer:2018onh} for various experimental constraints on $\zeta$. In this work, we shall ignore the gauge kinetic mixing term to ensure that the $U(1)_D$ gauge boson $W_{D}$ becomes a stable WIMP DM candidate.
One may alternatively impose an upper bound of $\zeta\lesssim 10^{-20}$ by requiring that the lifetime of $W_D$ is larger than the age of the Universe; see Appendix~\ref{apdx:DM-decay-width} for details.\footnote{In fact, a stronger bound, $\zeta \lesssim 10^{-26}$, exists when we take into account $\gamma$-ray observation \cite{Fermi-LAT:2015kyq}.}
Additionally, to consider $W_{D}$ as the WIMP DM, which is one of the main motivations of the present work, we consider all the particles to be neutral in $U(1)_D$ except the singlet scalar $\phi_D$ which is necessary for obtaining the $W_D$ mass. Introduction of $U(1)_{D}$ charges to any other fields would make $W_D$ unstable. 
Table~\ref{tab:contents} summarises the particle contents of the model under consideration and their charges.

The presence of an extra scalar that interact with the $\phi_h$ induces a mixing between the two. In unitary gauge, the expressions of $\phi_{h}$ and $\phi_{D}$, after the spontaneous breaking of the gauge symmetry, are given by
\begin{align}\label{eqn:phih}
\phi_{h}=
\begin{pmatrix}
0 \\
\dfrac{v+H}{\sqrt{2}}
\end{pmatrix}
\,,\qquad 
\phi_{D}=
\frac{v_{D}+H_{D}}{\sqrt{2}}
\,,
\end{align}
with the mass matrix
\begin{align}\label{eqn:mass-matrix}
\mathcal{M}^2_{scalar} = 
\left(\begin{array}{cc}
2\lambda_h\,v^2 ~~&~~ \lambda_{hD}\,v_{D}\,v \\
~~&~~\\
\lambda_{hD}\,v_{D}\,v ~~&~~ 2 \lambda_D\,v^2_{D}
\end{array}\right) 
\,.
\end{align}
Here, $v$ ($v_D$) denotes the VEV of the SM (dark) Higgs $\phi_h$ ($\phi_D$).
Diagonalisation of the mass matrix leads to the mass eigenstates
\begin{align}
H_{1}&= H \cos \theta - H_{D} \sin \theta \,, 
\nonumber \\
H_{2}&= H \sin \theta + H_{D} \cos \theta\,,
\end{align}
where $\theta$ is the mixing angle, given by
\begin{align}\label{eqn:scalarmix}
\tan 2\theta &= 
\frac{\lambda_{hD}\,v_{D}\,v}
{\lambda_h v^2 - \lambda_D v^2_{D}}
\,,
\end{align}
The mass eigenvalues are
\begin{align}
M^2_{H_1} &= \lambda_h v^2 + \lambda_D v^2_{D} + 
\sqrt{(\lambda_h v^2 - \lambda_D v^2_D)^2 + (\lambda_{hD}\,v\,v_D)^2} 
\,, \nonumber \\
M^2_{H_2} &= \lambda_h v^2 + \lambda_D v^2_D - 
\sqrt{(\lambda_h v^2 - \lambda_D v^2_D)^2 + (\lambda_{hD}\,v\,v_D)^2}
\,.
\label{eqn:massh2}
\end{align}
We consider the case where the dark Higgs is lighter than the SM Higgs. In other words, $M_{H_2}$ denotes the mass of the dark Higgs, and $M_{H_1}$ matches the SM Higgs mass. One may express the scalar quartic couplings in terms of the mixing angle and masses of the physical Higgses; we present the expressions in Appendix~\ref{apdx:quarticcouplings}.
When the dark Higgs acquires a VEV, the $U(1)_D$ gauge boson $W_D$ gets the mass of $M_{W_D} = g_D v_D$.

\subsection{Generation of neutrino masses with the extended seesaw mechanism}
\label{subsec:ESmechanism}
We consider the first two generations of the additional fermions, namely $N^{1}_{L}$, $N^{2}_{L}$, $S^{1}_{L}$, and $S^{2}_{L}$, to take part in the neutrino mass generation. The third generation is decoupled from the visible sector which is achieved by making them $Z_{2}$-odd. 
Therefore, the neutrino mass matrix can be expressed as
\begin{align}\label{eqn:neutrino-mass}
\mathcal{L}_{NM} = - \frac{1}{2} 
\begin{pmatrix}
\nu_L & S_{L} & N_{L} 
\end{pmatrix}
\begin{pmatrix}
0 & 0 & M^T_{D} \\
0 & \mu & M^T_{S} \\
M_{D} & M_{S} & M_{R}
\end{pmatrix}
\begin{pmatrix}
\nu_L \\
S_L \\
N_L
\end{pmatrix}
+ {\rm h.c.}
\end{align}
Here, $M_{D}$ is the $2 \times 3$ Dirac mass matrix,
\begin{align}\label{eqn:dirac-mass}
M_{D} =
\begin{pmatrix}
m^{e1}_{D} & m^{\mu1}_{D} & m^{\tau1}_{D} \\
m^{e2\,R}_{D} + i m^{e2\,I}_{D} & m^{\mu 2\,R}_{D} + i m^{\mu 2\,I}_{D}
& m^{\tau2\,R}_{D} + i m^{\tau2\,I}_{D}  
\end{pmatrix}
\,,
\end{align}
where $m^{ij}_{D} = y_{ij} v/\sqrt{2}$ and the superscript $R$ ($I$) stands for the real (imaginary) part. 
On the other hand, $M_{R}$ and $M_{S}$ in Eq.~\eqref{eqn:neutrino-mass} are $2 \times 2$ matrices which we choose to take as follows:
\begin{align}\label{eqn:righthanded-NM}
M_{R} = 
\begin{pmatrix}
M_{R}^{11} & 0\\
0 & M_{R}^{22}
\end{pmatrix}
\,,\qquad 
M_{S} = 
\begin{pmatrix}
M^{11}_{S} & 0\\
0 & M^{22}_{S}
\end{pmatrix}
\,.
\end{align}
Finally, we choose $\mu$ as a symmetric matrix. It is in general complex, and we parametrise it as
\begin{align}
\mu = 
\begin{pmatrix}
\mu^{R}_{11} + i \mu^{I}_{11} & \mu^{R}_{12} + i \mu^{I}_{12} \\
\mu^{R}_{12} + i \mu^{I}_{12} & \mu^{R}_{22} + i \mu^{I}_{22}
\end{pmatrix}\,.
\end{align}

To realise the non-zero neutrino masses in the extended seesaw framework, we consider the following hierarchy amongst the elements of $M_{D}$, $M_{R}$, and $M_{S}$ mass matrices \cite{Mitra:2011qr}:
\begin{align}
M_{R} > M_{S} > M_{D} \gg \mu 
\,,\qquad  
\mu < M^{T}_{S} M^{-1}_{R} M_{S}
\,.
\end{align}
With these assumptions, we can diagonalise the mass matrix shown in Eq.~\eqref{eqn:neutrino-mass} and
obtain the following set of mass matrices \cite{Mitra:2011qr}:
\begin{align}
m_{\nu} & \simeq M^T_{D} (M^T_{S})^{-1} \mu M^{-1}_{S} M_{D} 
\,,\nonumber \\
m_{S} & \simeq - M^T_{S} M^{-1}_{R} M_{S} 
\,, \label{eqn:first-diagonalisation} \\
m_{N} & \simeq M_{R}
\,.\nonumber
\end{align}
Once we diagonalise the $3\times 3$ matrix $m_{\nu}$, we get the masses of the active neutrinos.
The other two matrices give the masses of the sterile neutrinos.  
After the diagonalisation, we find the relation between the flavour basis $\left( \nu_L \; S_L \; N_L \right)^{T}$ and mass basis $\left( \nu_m \; S_m \; N_m \right)^{T}$ as
\begin{align}
\begin{pmatrix}
\nu^i_L \\ S^i_L \\ N^i_L
\end{pmatrix}
= \mathcal{U} 
\begin{pmatrix}
\nu^i_m \\ S^i_m \\ N^i_m
\end{pmatrix}
\,,
\end{align}
where the matrix $\mathcal{U} = \mathcal{U}_1 \mathcal{U}_2$. 
Note that $\mathcal{U}_1$ diagonalises the matrix in Eq.~\eqref{eqn:neutrino-mass}, while $\mathcal{U}_2$ diagonalises the mass matrices given in Eq.~\eqref{eqn:first-diagonalisation}. The expressions of $\mathcal{U}_1$ and $\mathcal{U}_2$ are given by \cite{Mitra:2011qr}
\begin{align}
\mathcal{U}_1 = \left(
\begin{smallmatrix}
1 - \frac{1}{2} M^{\dagger}_{D} 
(M^{-1}_{S})^{\dagger} 
M^{-1}_{S} M_{D}
& 
M^{\dagger}_{D} (M^{-1}_{S})^{\dagger}
&
M^{\dagger}_{D} M^{-1}_{R} 
\\
- M^{-1}_{S} M_{D} 
& 
1 - \frac{1}{2} (M^{-1}_{S} M_{D}) (M^{-1}_{S} M_{D})^{\dagger} - \frac{1}{2} M^{\dagger}_{S} M^{-2}_{R} M_{S}
& 
M^{\dagger}_{S} M^{-1}_{R} 
\\  
(M^{T})^{-1}_{S} \mu M^{-1}_{S} M_{D} 
& 
- M^{-1}_{R} M_{S} 
&
1 - \frac{1}{2} M^{-1}_{R} M_{S} M^{\dagger}_{S} M^{-1}_{R}
\end{smallmatrix}
\right)
\,,
\end{align}
and 
\begin{eqnarray}
\mathcal{U}_2 = \begin{pmatrix}
U & 0 & 0 \\
0 & W_S & 0 \\
0 & 0 & W_N
\end{pmatrix}\,,
\end{eqnarray}
where $U$, which is the Pontecorvo–Maki–Nakagawa–Sakata (PMNS) matrix \cite{Pontecorvo:1957qd,Maki:1962mu}, $W_{S}$, and $W_{N}$ diagonalise $m_{\nu}$, $m_{S}$, and $m_{N}$, respectively \cite{Mitra:2011qr}. 

\subsection{FIMP dark matter candidate}
\label{subsec:fimp-dm}
The remaining singlet neutrinos, $N^{3}_{L}$ and $S^{3}_{L}$, comprise a $2\times2$ mass matrix, and the lighter one is a good DM candidate. The mass matrix for the DM sector takes the following form:
\begin{align}\label{eqn:DM-mass}
\mathcal{L}_{\rm FIMP} = 
\begin{pmatrix}
S^3_{L} & N^3_{L}
\end{pmatrix}
\begin{pmatrix}
\mu^{\prime}_{33} & M^{\prime \, 33}_{SN} \\
M^{\prime \, 33}_{SN} & M^{\prime \, 33}_{R}
\end{pmatrix}
\begin{pmatrix}
S^3_{L}\\
N^3_{L}
\end{pmatrix}
\,,
\end{align}
where the elements are given by
\begin{align}
\mu^{\prime}_{33} &= 
\mu_{33} + \frac{\kappa v^2}{2 \Lambda} 
+ \frac{\kappa^{\prime} v_D^2}{2 \Lambda} 
\,,\nonumber \\
M^{\prime \, 33}_{SN} &= 
M_{S}^{33} + \frac{\alpha v^2}{2 \Lambda} 
+ \frac{\alpha^{\prime} v_D^2}{2 \Lambda} 
\,,\nonumber \\
M^{\prime \, 33}_{R} &= 
M_{R}^{33} + \frac{\xi v^2}{2 \Lambda} 
+ \frac{\xi^{\prime} v_D^2}{2 \Lambda}
\,.
\end{align}
In the limit $M^{\prime\,33}_{R} \gg M^{\prime\,33}_{SN}$, we diagonalise the mass matrix to obtain the eigenvalues expressed as
\begin{align}
M_{S_m} &= \mu^{\prime}_{33} - \frac{(M^{\prime\,33}_{SN})^2}
{M^{\prime\,33}_{R}} 
\,,\nonumber \\
M_{N_m} &= M^{\prime\,33}_{R}\,.
\end{align}
The relation between the mass eigenstates and the flavour eigenstates are given by
\begin{align}
\begin{pmatrix}
S_{m}^3 \\
N_{m}^3
\end{pmatrix}
\simeq 
\begin{pmatrix}
1 & \frac{M^{\prime\,33}_{SN}}{M^{\prime\,33}_{R}} \\
- \frac{M^{\prime\,33}_{SN}}{M^{\prime\,33}_{R}} & 1
\end{pmatrix}
\begin{pmatrix}
S^3_{L} \\
N^3_{L}
\end{pmatrix}
=
\begin{pmatrix}
1 & \delta \\
- \delta & 1
\end{pmatrix}
\begin{pmatrix}
S^3_{L} \\
N^3_{L}
\end{pmatrix}\,,
\end{align}
where $\delta \equiv M^{\prime\,33}_{SN}/M^{\prime\,33}_{R}$.
In our study, $S_{m}^3$ is the lighter one, becoming a good DM candidate, and $N_{m}^3$ is the next-to-lightest stable particle (NLSP). We shall drop the superscript `3' and use $S_m$ and $N_m$ to denote the lighter third-generation mass eigenstate, which becomes the FIMP DM, and the heavier third-generation mass eigenstate, which is the NLSP, respectively, hereinafter.

In Section~\ref{sec:DMpheno}, we shall explore the DM phenomenology in detail, focusing on the parameter space where $\Lambda \gtrsim 10^{14}$ GeV and $\alpha=\alpha^\prime = \xi=\xi^\prime = \kappa=\kappa^\prime = \mathcal{O}(1)$. In this case, both the DM candidate $S_m$ and the NLSP $N_m$ are produced out-of-equilibrium in the early Universe through the freeze-in processes of the type $H_i(+H_i) \rightarrow {\rm FIMP}+{\rm FIMP}$, where $i=1,2$. Thus, the effective couplings are in the ballpark of the FIMP-type DM. After the production, the NLSP decays into the lighter eigenstate; see Section~\ref{sec:DMpheno} for details.

\subsection{WIMP dark matter candidate}
The vector gauge boson $W_D$ associated with the extra dark $U(1)_D$ is, on the other hand, a good WIMP DM candidate in our model.
The vector boson $W_D$, being a WIMP, is produced via the standard freeze-out processes ${\rm SM}+{\rm SM} \leftrightarrow W_D + W_D$ which keep the WIMP in thermal equilibrium with the SM thermal bath.

As the model features both the FIMP DM, $S_m$, and the WIMP DM, $W_D$, a two-component DM scenario naturally arises in our model. We will present the detailed analysis in Section~\ref{sec:DMpheno}.

\section{Neutrino masses and lepton flavour violation bounds}
\label{sec:LFVnu}
The eigenvalues of the mass matrix $m_{\nu}$ represent the masses of the active neutrinos as we discussed in Section~\ref{subsec:ESmechanism}. Differences of their mass-squared will give us the solar mass difference $\Delta m^2_{12}$ and the atmospheric mass difference $\Delta m^2_{31}$. On the other hand, the elements of the PMNS matrix $U$ give us the oscillation angles $\theta_{12}$, $\theta_{13}$, and $\theta_{23}$.
In this work, we consider the recent bounds on the oscillation parameters~\cite{Esteban:2020cvm},
\begin{gather}
6.82  \leq  \frac{\Delta m^2_{21}}{10^{-5}\,\,{\rm eV^{2}}} \leq  8.04
\,,\qquad 
2.431  \leq  \frac{\Delta m^2_{31}}{10^{-3}\,\,{\rm eV^{2}}} \leq  2.599
\,,\nonumber\\
31.27  \leq  \theta_{12} \,\,[{}^{0}] \leq  35.86
\,,\qquad
8.20  \leq  \theta_{13} \,\,[{}^{0}] \leq  8.97
\,,\qquad
39.5  \leq  \theta_{23} \,\,[{}^{0}] \leq  52.00
\,.
\end{gather}
Additionally, we consider the bound on the sum of the non-decaying active neutrino masses from cosmology, {\it i.e.}, $\sum_{i} m_{\nu_i} < 0.23$ eV~\cite{Planck:2015fie,Planck:2018vyg}. 
Since we have additional sterile neutrinos, we also take into account the LFV processes. The most stringent bounds on the LFV processes come from $\mu \rightarrow e \gamma$, $\mu \rightarrow e e e$, and $\mu$-to-$e$ conversion CR$\left( \mu^{-} \,{\rm Ti} \rightarrow e^{-}\,{\rm Ti} \right)$. The recent bounds are given by Br$(\mu \rightarrow e \gamma) < 4.2\times10^{-13}$~\cite{MEG:2016leq}, Br$(\mu \rightarrow e\bar{e}e) < 1\times 10^{-12}$~\cite{SINDRUM:1987nra}, and CR$\left( \mu^{-} \,{\rm Ti} \rightarrow e^{-}\,{\rm Ti} \right) < 6.1 \times 10^{-13}$~\cite{Wintz:1998rp}. 
In determining the branching ratios for $\mu \rightarrow e \gamma$, $\mu \rightarrow e \bar{e} e$, and $\mu - e$ conversion rate, we follow Refs.~\cite{Ilakovac:1994kj,Lindner:2016bgg}. 
In order to satisfy all the aforementioned constraints, the elements shown in Eq.~\eqref{eqn:dirac-mass} and Eq.~\eqref{eqn:righthanded-NM} cannot take arbitrary values.
We thus vary the model parameters as below and obtain allowed parameter spaces by imposing the aforementioned constraints:
\begin{gather}
1 \leq M^{11}_{S}\,(= M^{22}_{S})\,\,[{\rm GeV}] \leq 1000
\,,\quad
10^{-5} \leq \frac{m^{\alpha 1}_{D}}{M^{11}_{S}}
\,,\quad
\frac{m^{\alpha 2\,R,I}_{D}}{M^{11}_{S}} \leq 10^{-1}
\,,\nonumber\\
10^{-9} \leq \mu^{R, I}_{ij}\,\,[{\rm GeV}] \leq 10^{-1}
\,.\label{eqn:parameters-range}
\end{gather}
where $\alpha = \{e,\mu,\tau\}$ and $i,j = 1,2$, and we have chosen $M_{N^1} = M_{N^2}= 2M^{11}_{S} $. The rest of the model parameters, which affect the DM relic density directly but do not take part in the neutrino mass, are fixed as
\begin{gather}
M_{S_m} = 20\;{\rm GeV}
\,,\quad 
M_{W_D} = 1.04628 \;{\rm GeV}
\,,\quad 
M_{N_m} = 300\;{\rm GeV}
\,,\quad
M_{H_2} = 2.2120 \; {\rm GeV}
\,,\nonumber\\ 
g_{D} = 3.1\times 10^{-4}
\,,\quad 
\sin\theta = 8.17\times 10^{-2}
\,,\quad 
\Lambda = 5.5 \times 10^{14}\,\,{\rm GeV}
\,,\nonumber\\ 
\kappa = \kappa^{\prime} = \alpha = \alpha^{\prime}
= \xi = \xi^{\prime} = 1
\,.\label{eqn:parameters-fixed}
\end{gather}
These fixed values are inspired by the DM studies as well as the FOPTs, as we will discuss later in the paper.

\begin{figure}[t!]
\centering
\includegraphics[scale=0.49]{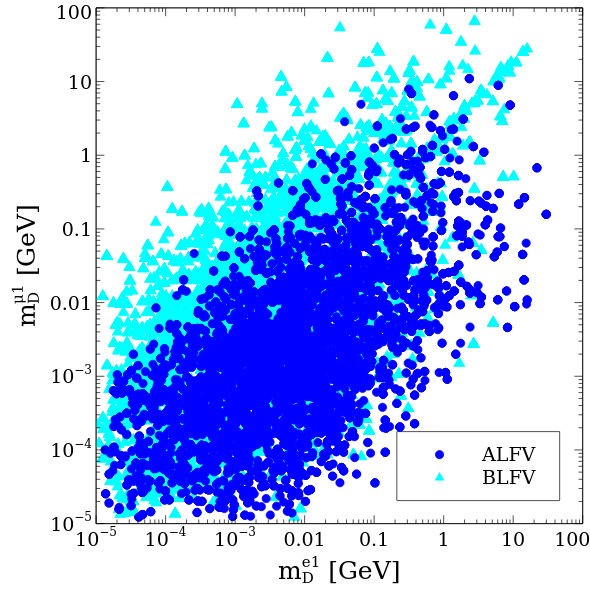}
\includegraphics[scale=0.49]{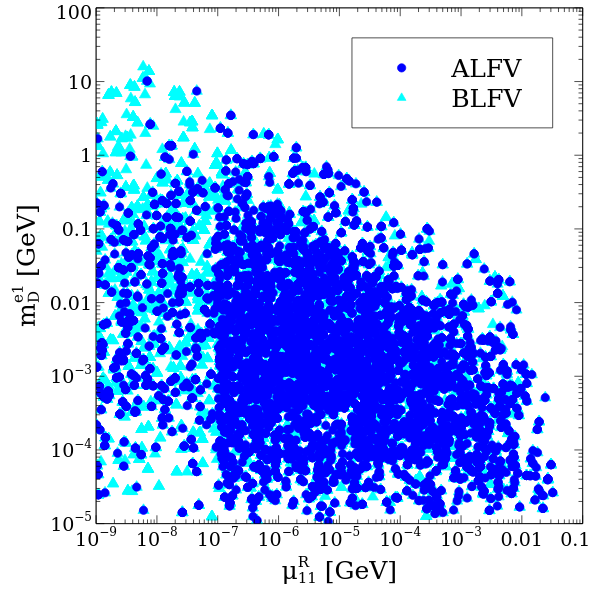}
\caption{Allowed parameter spaces after imposing the NOD constraints (cyan) and the NOD constraints plus LFV bounds (blue) are shown in the $m^{e1}_{D}$ -- $m^{\mu1}_{D}$ plane in the left panel and in the $\mu^{R}_{11}$ -- $m^{e1}_{D}$ plane in the right panel.
ALFV and BLFV respectively correspond to after and before imposing the LFV bounds.} 
\label{fig:neutrino-scatter-1}
\end{figure}

In the left panel (LP) and right panel (RP) of Fig.~\ref{fig:neutrino-scatter-1}, allowed parameter spaces are shown in the $m^{e1}_{D}$ -- $m^{\mu1}_{D}$ and $\mu^{R}_{11}$ -- $m^{e1}_{D}$ planes, respectively. The cyan points are obtained after imposing the NOD constraints. The blue points are obtained when we additionally impose the LFV bounds. From the LP of Fig.~\ref{fig:neutrino-scatter-1}, one may see a sharp correlation between $m^{e1}_{D}$ and $m^{\mu1}_{D}$. This is because both of them actively contribute to the neutrino mass, {\it i.e.}, they are the leading contributions in two different elements of the neutrino mass matrix $m_{\nu}$. Since we have taken the elements of $M_{S}$ to be equal, for a large value of $M_{S}$, we need a small value for $m^{e1}_{D}$ and $m^{\mu1}_{D}$, and similarly, for a small value of $M_{S}$, we need a large value for $m^{e1}_{D}$ and $m^{\mu1}_{D}$.  
Moreover, the LFV processes are mediated by the gauge bosons ($W^{\pm}, Z$), so those processes mainly depend on the active-sterile mixing terms, namely $M_{D}/M_{S}$ and $M_{D}/M_{R}$. Therefore, when we apply the LFV bounds, higher values of $m^{\mu1}_{D}$ get ruled out for each value of $m^{e1}_{D}$ .
On the other hand, in the RP of Fig.~\ref{fig:neutrino-scatter-1}, we see an anti-correlation between $m^{e1}_{D}$ and $\mu^{R}_{11}$, which is mainly due to the neutrino mass relation. Furthermore, elements of the matrix $\mu$ do not actively contribute to the LFV processes. Therefore, there is practically no shrink in the $\mu^{R}_{11}$ -- $m^{e1}_{D}$ plane after applying the LFV bounds. 

\begin{figure}[t!]
\centering
\includegraphics[scale=0.49]{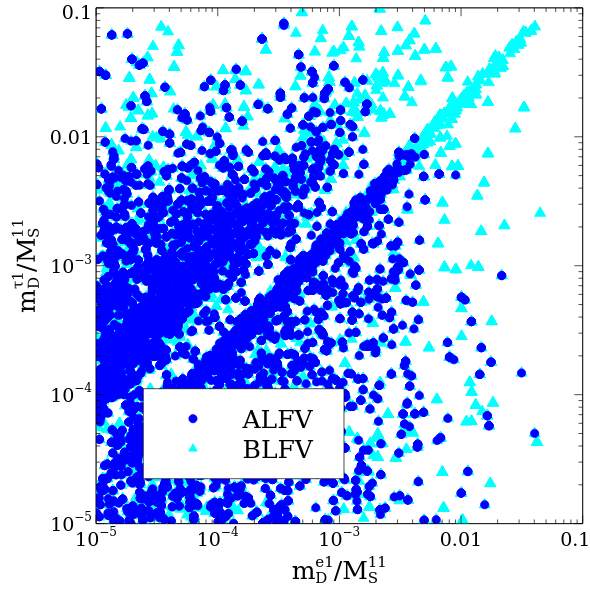}
\includegraphics[scale=0.49]{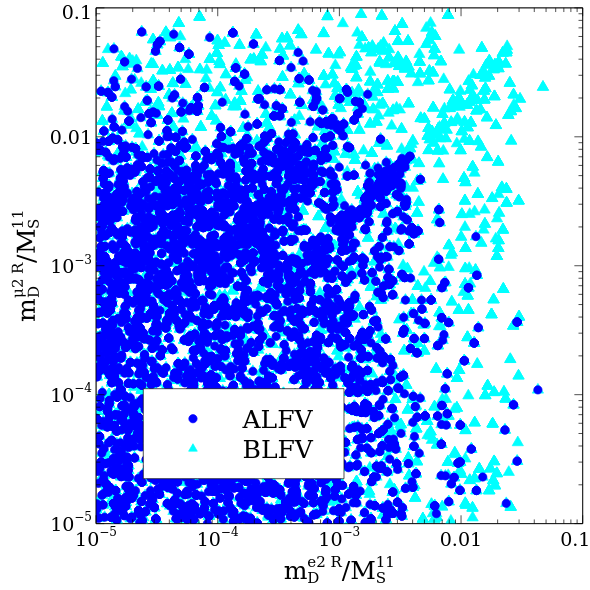}
\caption{
Allowed parameter spaces after imposing the NOD constraints (cyan) and the NOD constraints plus LFV bounds (blue) are shown in the $m^{e1}_{D}/M^{11}_{S}$ -- $m^{\tau1}_{D}/M^{11}_{S}$ plane in the left panel and in the $m^{e2\,R}_{D}/M^{11}_{S}$ -- $m^{\mu2\,R}_{D}/M^{11}_{S}$ plane in the right panel.
ALFV and BLFV respectively correspond to after and before imposing the LFV bounds.
}  
\label{fig:neutrino-scatter-2}
\end{figure}

\begin{figure}[t!]
\centering
\includegraphics[scale=0.49]{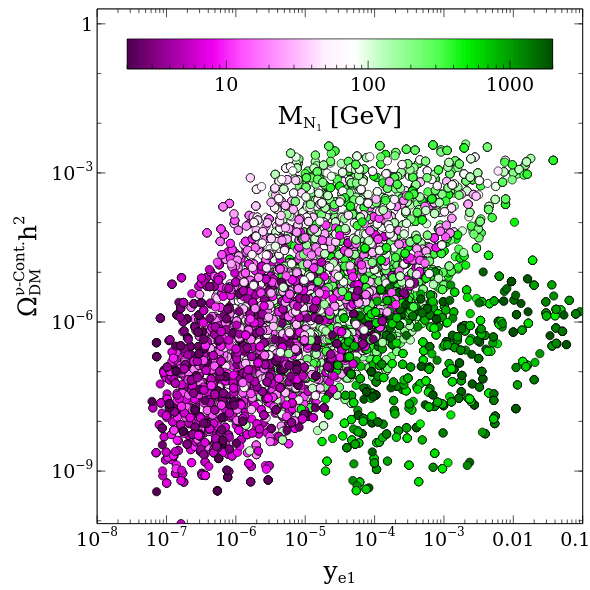}
\includegraphics[scale=0.49]{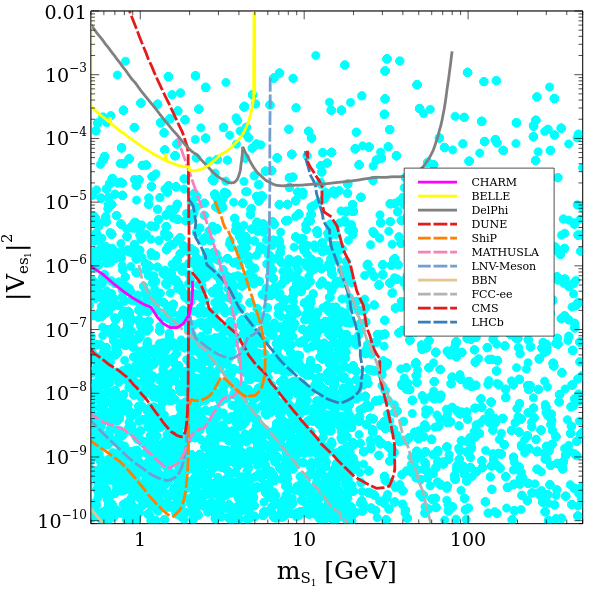}
\caption{
Allowed parameter spaces in the $y_{e1}$ -- $\Omega^{\nu-{\rm Cont.}}_{\rm DM}h^{2}$ (left) and $m_{S_1}$ -- $|V_{eS_{1}}|^2$ (right) planes after imposing the NOD constraints. 
Bounds coming from various ongoing (solid lines) and future (dashed lines) experiments are overlaid; see text for more details.
} 
\label{fig:neutrino-scatter-4}
\end{figure}

\begin{figure}[t!]
\centering
\includegraphics[scale=0.49]{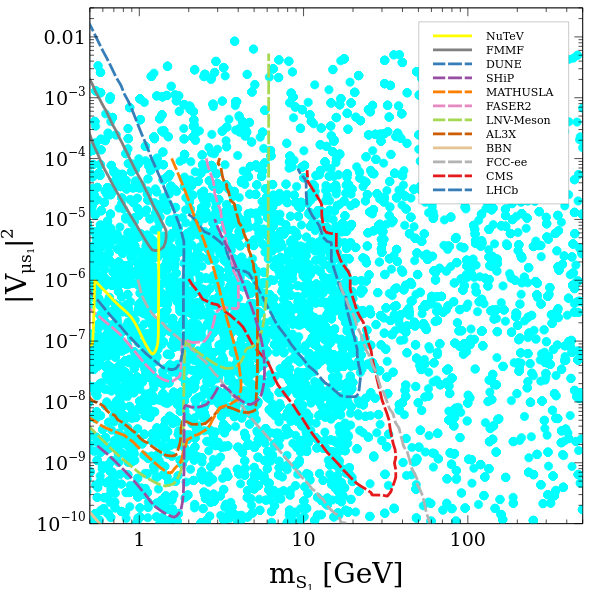}
\includegraphics[scale=0.49]{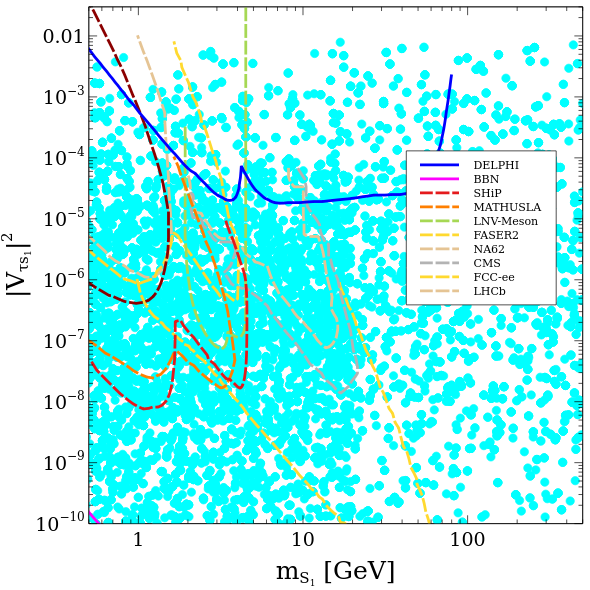}
\caption{
Allowed parameter spaces in the $m_{S_1}$ -- $|V_{\mu{}S_{1}}|^2$ (left) and $m_{S_1}$ -- $|V_{\tau{}S_{1}}|^2$ (right) planes after imposing the NOD constraints.
Bounds coming from various ongoing (solid lines) and future (dashed lines) experiments are overlaid; see text for more details.} 
\label{fig:neutrino-scatter-5}
\end{figure}

In Fig.~\ref{fig:neutrino-scatter-2}, we have shown the allowed parameter space in the $m^{e1}_{D}/M^{11}_{S}$ -- $m^{\tau1}_{D}/M^{11}_{S}$ and $m^{e2\,R}_{D}/M^{11}_{S}$ -- $m^{\mu2\,R}_{D}/M^{11}_{S}$ planes in the LP and RP, respectively, after imposing the NOD (cyan) and NOD plus LFV bounds (blue). The LFV bounds directly depend on the parameters $m^{\tau1}_{D}/M^{11}_{S}$, $m^{\tau1}_{D}/M^{11}_{S}$, $m^{e2\,R}_{D}/M^{11}_{S}$, and $m^{\mu2\,R}_{D}/M^{11}_{S}$ as they represent the strength of the active-sterile mixing.
Therefore, in the LP of Fig.~\ref{fig:neutrino-scatter-2}, we see that both parameters cannot take higher values simultaneously due to the LFV bounds. The same conclusion is also observed for the RP. 
Depending on the strength of the active-sterile mixing, we may detect the sterile neutrinos in many ongoing and future experiments which we shall discuss later in Fig.~\ref{fig:neutrino-scatter-4} and Fig.~\ref{fig:neutrino-scatter-5}.

\begin{figure}[t!]
\centering
\includegraphics[scale=0.49]{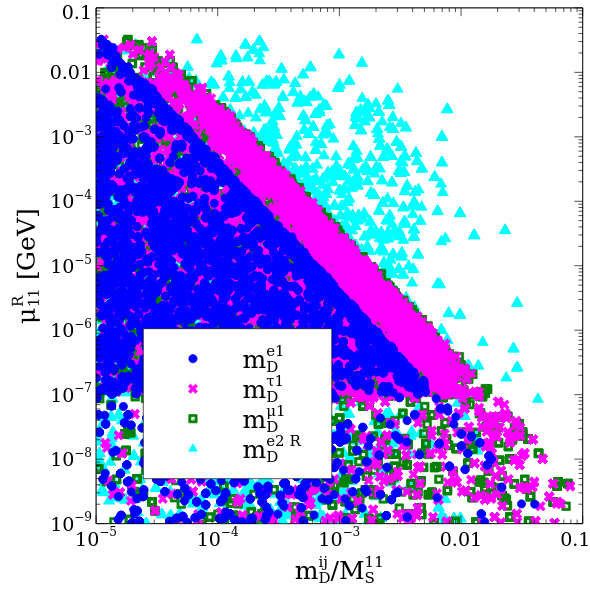}
\includegraphics[scale=0.49]{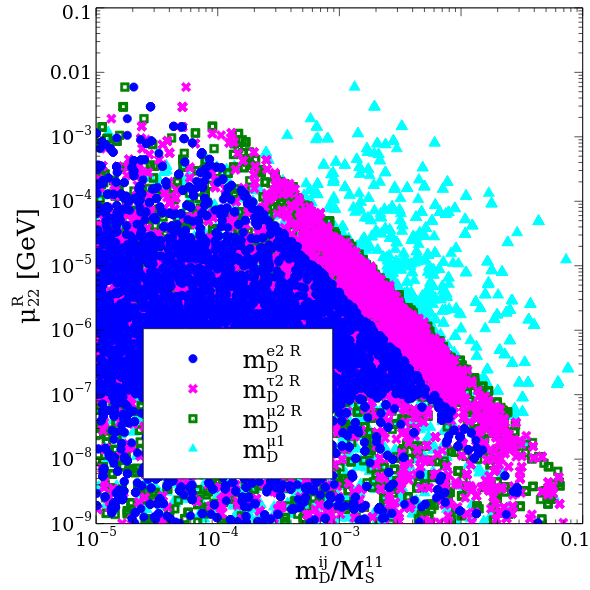}
\caption{
Allowed parameter spaces after imposing the NOD constraints and the LFV bounds are shown in the $m^{ij}_{D}/M^{11}_{S}$ -- $\mu^{R}_{11}$ ($ij = \{e1, \tau1, \mu1,e2\,R\}$) plane in the left panel and in the $m^{ij}_{D}/M^{11}_{S}$ -- $\mu^{R}_{22}$ ($ij = \{e2\,R, \tau2\,R, \mu2\,R,\mu1\}$) plane in the right panel.
} 
\label{fig:neutrino-scatter-3}
\end{figure}

In Fig.~\ref{fig:neutrino-scatter-3}, we have shown the scatter plots in the $M_{D}/M_{S}$ -- $\mu$ plane after imposing the NOD and LFV bounds.
In the LP of Fig.~\ref{fig:neutrino-scatter-3}, blue, magenta, green, and cyan points respectively correspond to $m^{e1}_{D}/M^{11}_{S}$ -- $\mu^{R}_{11}$, $m^{\tau1}_{D}/M^{11}_{S}$ -- $\mu^{R}_{11}$, $m^{\mu1}_{D}/M^{11}_{S}$ -- $\mu^{R}_{11}$, and $m^{e2\,R}_{D}/M^{11}_{S}$ -- $\mu^{R}_{11}$. 
In the RP, we have blue, magenta, green, and cyan points for $m^{e2\,R}_{D}/M^{11}_{S}$ -- $\mu^{R}_{22}$, $m^{\tau2\,R}_{D}/M^{11}_{S}$ -- $\mu^{R}_{22}$, $m^{\mu2\,R}_{D}/M^{11}_{S}$ -- $\mu^{R}_{22}$, and $m^{\mu1}_{D}/M^{11}_{S}$ -- $\mu^{R}_{22}$, respectively. 
One interesting point to note here is that there is a strong correlation amongst the blue, magenta, and green points in both the LP and RP, while we observe no relation amongst the cyan points. The points that exhibit the strong correlation strictly follow the relation $(M_{D}/M_{S})^{2} \mu < 10^{-11}$ GeV, which is the mass of the active neutrinos. 
Parameters denoted by the cyan points do not affect the neutrino mass directly; they either come with the multiplication of other terms or are absent in the neutrino mass matrix. Thus, in the end, their combinational effect never exceeds the light active 
neutrino mass.    


In the LP of Fig.~\ref{fig:neutrino-scatter-4}, we have shown the allowed parameter space in terms of the Yukawa coupling $y_{e1}$ and the DM relic density that is coming solely from the neutrino sector. 
$S_m$, which is a FIMP DM candidate as we discussed in Sec.~\ref{sec:model}, may be produced via annihilations of active neutrinos and extra heavy neutrinos, mediated by the Higgses, as $\nu_i + N_j \xrightarrow[]{H_{1,2}} S_m + S_m$ and $\nu_i + S_j \xrightarrow[]{H_{1,2}} S_m + S_m$, where $i=1,2,3$ and $j=1,2$. The LP of Fig.~\ref{fig:neutrino-scatter-4} indicates that this contribution is subdominant. One may understand the general behaviour as follows. When $M_{N_1}$ is smaller than 500 GeV, we have a linear relation between $y_{e1}$ and the DM relic density coming from the active and heavy neutrinos annihilations. It reflects the fact that $\Omega_{\rm DM}^{\nu-{\rm Cont.}}h^2 \propto y_{e1}^2$. When $M_{N_1}$ is larger than 1000 GeV, the contribution to the DM relic density is small as the mass is close to the chosen reheating temperature of $T_R = 3$ TeV; thus, a suppression occurs. We observe that, for the chosen range of parameter values \eqref{eqn:parameters-range}, the contribution of the active and extra heavy neutrinos to the total DM relic density is at most $\sim 3\%$.
The RP of Fig.~\ref{fig:neutrino-scatter-4} depicts the allowed region in the active-sterile mixing angle associated with electron $|V_{eS_{1}}|^{2}$ and sterile neutrino mass $m_{S_1}$ plane after imposing the NOD bounds.
The solid lines represent the present bounds which come from CHARM~\cite{CHARM:1985nku,CHARMII:1994jjr}, BELLE~\cite{Belle:2013ytx}, and DelPhi~\cite{DELPHI:1996qcc}, depending on the mass of the sterile neutrino. DelPhi demands the allowed range $|V_{eS_{1}}|^{2} < 10^{-4}$ for the sterile neutrino mass up to 100 GeV, whereas CHARM puts a bound on the active-sterile mixing angle $|V_{eS_{1}}|^{2} \lesssim 10^{-7}$ for the sterile neutrino mass less than 2 GeV. There are various proposed experiments, including DUNE~\cite{Krasnov:2019kdc,Ballett:2019bgd}, SHiP~\cite{SHiP:2018xqw}, MATHUSLA~\cite{Chou:2016lxi}, LNV-Meson~\cite{Chun:2019nwi}, FCC-ee~\cite{Blondel:2014bra,Alimena:2022hfr}, CMS~\cite{Drewes:2019fou}, and LHCb~\cite{Antusch:2017hhu,Drewes:2019fou}, which have the sensitivity reaching up to $|V_{eS_{1}}|^{2} \sim 10^{-10}$ for the sterile neutrino mass up to 100 GeV.

Figure~\ref{fig:neutrino-scatter-5} shows the allowed region in the active-sterile mixing associated with the muon $|V_{\mu{}S_{1}}|^2$ (LP) as well as tauon $|V_{\tau{}S_{1}}|^2$ (RP) and the sterile neutrino mass $m_{S_1}$ planes, after imposing the NOD bounds. 
In the LP, the recent bounds put by NuTeV~\cite{NuTeV:1999kej} and FMMF~\cite{FMMF:1994yvb} already rule out the sterile neutrino mass up to 2 GeV for the active-sterile neutrino mixing larger than $10^{-7}$. 
Various future experiments such as DUNE~\cite{Krasnov:2019kdc,Ballett:2019bgd}, SHiP~\cite{SHiP:2018xqw}, MATHUSLA~\cite{Chou:2016lxi}, FASER2~\cite{Feng:2017uoz}, LNV-Meson~\cite{Chun:2019nwi}, AL3X~\cite{Dercks:2018wum}, FCC-ee~\cite{Blondel:2014bra,Alimena:2022hfr}, CMS~\cite{Drewes:2019fou}, and LHCb~\cite{Antusch:2017hhu,Drewes:2019fou} are also presented by dashed lines which will probe the active-sterile mixing, $|V_{\mu{}S_{1}}|^2$, up to $10^{-10}$ for sterile neutrino mass as large as 100 GeV. 
On the other hand, from the RP of Fig.~\ref{fig:neutrino-scatter-5}, we see that the DELPHI experiment~\cite{DELPHI:1996qcc} already rules out $|V_{\tau{}S_{1}}|^2 > 3 \times 10^{-5}$ for the sterile neutrino mass up to 100 GeV. The future experiments shall cover the active-sterile mixing up to $|V_{\tau{}S_{1}}|^2 \sim 10^{-10}$ for the mass range up to 100 GeV.
For all the active-sterile mixing and sterile neutrino mass planes, there exits a bound coming from the Big Bang Nucleosynthesis (BBN) as well, if the sterile neutrino decays after the BBN. However, the BBN bound is weak for the parameter space we have considered.

\section{Dark matter phenomenology}
\label{sec:DMpheno}
We now discuss the production and detection prospects of the DM candidates in our model.\footnote{
We have utilised publicly available tools, including {\tt FeynRules}~\cite{Alloul:2013bka}, {\tt CalcHEP}~\cite{Belyaev:2012qa}, and {\tt micrOMEGAs}~\cite{Belanger:2006is}, for the DM studies.
}
Our model features both the WIMP and FIMP DM candidates as we discussed in Section~\ref{sec:model}. The dark gauge boson $W_D$ plays the WIMP role, and the lighter singlet neutrino of the third generation $S_m$ becomes the FIMP DM; the NLSP, $N_m$, will eventually decay to the FIMP DM $S_m$.
Thus, a two-component DM scenario naturally arises.
The WIMP part ensures the potential detectability in future, whereas the FIMP DM will be difficult to probe by the direct, indirect, or collider detection techniques. 

As we shall discuss in Section~\ref{sec:FOPTsGWs}, a low-mass BSM dark Higgs is favoured from the FOPT point of view~\cite{Carena:2019une}. Thus, in this section, we mainly focus on the range $1-200$ GeV for the dark Higgs. 
Furthermore, to avoid any potential problems with collider searches due to the low mass of the dark Higgs, we consider the mixing angle $\theta$ in Eq.~\eqref{eqn:scalarmix} to be small, focusing on $|\sin\theta| < 0.1$. In doing so, we may easily evade the Higgs signal strength bounds~\cite{CMS:2018uag,ATLAS:2016neq}. 

There are mainly five constraints that we have taken into account for the discussion of DM phenomenology: {\it i)} relic density, {\it ii)} direct detection bounds, {\it iii)} indirect detection bounds, {\it iv)} Higgs invisible decay, and {\it v)} Higgs signal strength bound.
We explain each category below before presenting the results.

\begin{itemize}
\item {\bf DM relic density:} 
We consider the bound on the total amount of DM relic density coming from the Planck experiment~\cite{Planck:2015fie,Planck:2018vyg}. 
Specifically, the following $3 \sigma$ bound is used, unless stated otherwise:
\begin{align}\label{eqn:DM-RD-3sigma}
0.1172 \leq 
\Omega_{\rm DM}h^{2} 
(=\Omega_{W_{D}}h^{2} + \Omega_{S_m}h^{2}) 
\leq 0.1226 
\,.
\end{align}
Here, $\Omega_{W_{D}}h^{2}$ ($\Omega_{S_m}h^{2}$) denotes the WIMP (FIMP) DM relic density.

\item {\bf Direct detection:} 
\begin{figure}[t!]
\centering
\includegraphics[scale=1.2]{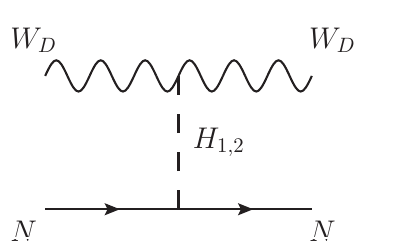}
\caption{DM direct detection diagram mediated by $H_{1,2}$.} 
\label{fig:DD-feynman}
\end{figure}
In our model, DM can have the elastic scattering with a nucleon $N$ as depicted in Fig.~\ref{fig:DD-feynman}. The analytical estimate for such a process takes the form~\cite{Berlin:2014tja},
\begin{align}
\sigma_{\rm SI} = \frac{\mu_*^{2}\,\sin^{2}2\theta\,g^2_{D}}{4 \pi v^{2}} 
\left(\frac{1}{M^2_{H_1}} - \frac{1}{M^2_{H_2}} \right)^{2} 
\left[ \frac{Z \tilde{f}_{p} + (A-Z)\tilde{f}_{n}}{A} \right]^{2}
\,,\label{eqn:DD-expression}
\end{align}
where $\mu_* = M_{W_D} M_{N}/(M_{W_D} + M_{N})$ is the reduced mass, with $M_{N}$ being the nucleon mass, $v$ is the electroweak VEV, $Z$ is the atomic number, $A$ is the atomic weight, and $\tilde{f}_{\alpha}$ $(\alpha = p,n)$ can be expressed as
\begin{align}
\frac{f_{\alpha}}{M_{N}} = \left( \frac{7}{9} \sum_{q=u,d,s} 
f^{\alpha}_{T_q} + \frac{2}{9}\right)\,,
\end{align}
with $f^{p(n)}_{T_u} = 0.020 (0.026)$, $f^{p(n)}_{T_d} = 0.026 (0.020)$, and $f^{p,n}_{T_s} = 0.043$~\cite{Junnarkar:2013ac}.
As we take the WIMP DM mass to be in the range of $1-100$ GeV, the DM may be detected by different experiments.
A part of the parameter space in the spin-independent direct detection (SIDD) cross-section, $\sigma_{\rm SI}$, and DM mass, $M_{W_{D}}$, plane is already ruled out by LUX-ZEPLIN-5.5T~\cite{LZ:2022ufs}, PandaX-4T~\cite{PandaX-II:2017hlx,Liu:2022zgu}, and Xenon-1T~\cite{XENON:2018voc} for the $10-100$ GeV DM mass range.
On the other hand, the mass range below 10 GeV will be explored by experiments such as DarkSide-50~\cite{DarkSide:2018kuk,DarkSide:2018bpj}, XENON-1T (M)~\cite{Ibe:2017yqa}, CDMSlite~\cite{SuperCDMS:2018gro}, and CRESST-III~\cite{CRESST:2017ues,CRESST:2019jnq}. 
Our SIDD cross-section is a few orders of magnitude below the current bound.    

\item {\bf Indirect Detection:} 
The WIMP DM can also be detected by observing the annihilation products, namely $b\bar{b}$, $\tau\bar\tau$, $\mu\bar\mu$, and $e\bar{e}$. When the WIMP DM mass is above the $b$-quark mass, the bound from the $b\bar{b}$ final state dominates. Fermi-LAT + MAGIC Segue 1~\cite{MAGIC:2016xys} puts the stringent bound on the $\langle\sigma v\rangle_{b\bar{b}}$ -- $M_{W_D}$ plane. On the other hand, when the DM mass is smaller than the $b$-quark mass, the DM annihilates to $\tau\bar{\tau}$, $\mu\bar{\mu}$, and $e\bar{e}$ dominantly. The bounds come from the study of FERMI-LAT~\cite{Fermi-LAT:2015att,Leane:2018kjk}, CMB~\cite{Leane:2018kjk}, and AMS~\cite{Bergstrom:2013jra,Leane:2018kjk}. 
We shall discuss the details of the indirect detection bound when we present our resultant plots.

\item {\bf Invisible decay:} 
When the DM mass is below half of the SM Higgs mass, there is a possibility that the SM Higgs will have an invisible decay, $\Gamma^{\rm inv}_{H_1}$. Thus, one needs to make sure that the invisible decay is always smaller than the current bound~\cite{ATLAS:2019cid},
\begin{align}
\frac{\Gamma^{\rm inv}_{H_1}}{\Gamma_{H_1}} < 0.26\,.
\end{align}
The decay width of the SM Higgs to the WIMP DM in the present case takes the form,
\begin{align}
\Gamma^{\rm inv}_{H_1} = \frac{M^3_{H_1} g^2_{H_{1}W_{D}W_{D}}}
{128\,\pi M^4_{W_D}} \sqrt{1 - \frac{4 M_{W^2_{D}}}{M^2_{H_1}}}
\left( 1 - \frac{4 M^2_{W_D}}{M^2_{H_1}} + \frac{12 M^4_{W_D}}{M^4_{H_1}}\right)
\,,
\end{align}
where $g_{H_{1}W_{D}W_{D}} = - 2 g_{D} M_{W_D} \sin \theta$. 
The allowed parameter region in the $M_{W_D}$ -- $\sigma_{\rm SI}$ plane after imposing the invisible decay constraints shall be presented in Section~\ref{subsec:DMParamScans} with the SM Higgs decay width $\Gamma_{H_1} = 4.156$ MeV.

\item {\bf Higgs signal strength:} 
The Higgs signal strength can be estimated by measuring its production and decay ratio with the SM values. It can be defined as
\begin{eqnarray}
\tilde{\mu} =  \mu_{H_1} \mu_f
= \frac{\sigma_{H_1}}{\sigma^{\rm SM}_{H_1}} \frac{\mathcal{B}_{f}}{\mathcal{B}^{\rm SM}_{f}}
\,,
\end{eqnarray}
where $\mu_{H_1} = \sigma_{H_1}/\sigma^{\rm SM}_{H_1}$ is the ratio of the Higgs production in the new model and the SM, and $\mu_f = \mathcal{B}_{f}/\mathcal{B}^{\rm SM}_{f}$ is the ratio of the branchings of the Higgs to a channel $f$. The current bound on $\tilde{\mu}$ after a combined analysis is given by~\cite{CMS:2018uag}
\begin{align}
\tilde{\mu} = 1.17 \pm 0.10\,.
\end{align}
Assuming that the Higgs boson has the same kind of branchings as the SM case, we find that $\tilde{\mu} \sim \cos^{2}\theta$. By taking the $3\sigma$ range, we obtain that $\sin\theta < 0.36$. Since we consider a small mixing angle, namely $\sin\theta < 0.1$, we thus always satisfy the bound from the Higgs signal strength. 

\end{itemize}

\subsection{Dark matter production}
\label{subsec:DMproduction}
Let us temporarily consider a regime where only the coupling $\kappa$ is active and the mixing between $S_L^3$ and $N_L^3$ is negligible, in which case $S_m \simeq S_L^3$. Let us also consider the case where the mixing between the Higgses is small, {\it i.e.}, $\cos\theta \simeq 1$. In this case, the FIMP DM $S_m$ is produced dominantly by the Higgs scattering process, and the analytical solution for the yield is given by~\cite{Hall:2009bx} 
\begin{align}\label{eqn:YSmapprox}
Y_{S_m}	= \int^{T_{R}}_{T_{\rm end}} \frac{1}{S\mathcal{H}T}\left(\frac{4 \kappa}{\Lambda}\right)^2 \frac{1}{16 \pi^5} \mathrm{~T}^6
\,,
\end{align}
where $T_{R}$ is the reheating temperature, $T_{\rm end} \simeq 1$ MeV is the temperature after which we may safely assume that no DM production occurs, and the entropy $S$ and the Hubble parameter $\mathcal{H}$ are given by
\begin{align}
S=\frac{2 \pi^2 g_S T^3}{45}
\,, \qquad
\mathcal{H}=\frac{1.66\sqrt{g_{\rho}} T^2}{M_{\rm Pl}}
\,,
\end{align}
with $M_{\rm Pl} = 1.22 \times 10^{19}$ GeV, and $g_S$ and $g_{\rho}$ being respectively the entropy and the energy density degrees of freedom of the Universe; we take $g_{S,\rho} \sim 100$. 
In achieving Eq.~\eqref{eqn:YSmapprox}, it is assumed that the masses of the associated particles in the production may be neglected compared to the temperature at which DM production happens which, for the process under consideration, is the reheating temperature. Thus, the production depends on the highest temperature, obtaining the UV freeze-in contribution. Consideration of masses of the associated particles has therefore a negligible effect in the DM production. Moreover, as in the IR freeze-in case, it is assumed that one may safely ignore the back-reaction of the DM in the Boltzmann equations since the number density is always smaller than the equilibrium number density.
With these assumptions, the final relic density for the $T_{R}$-dominated regime is given by
\begin{align}\label{eqn:key}
\Omega_{S_m} h^2 = M_{S_m} \frac{S_0}{\rho_c}  Y_{S_m}\,,
\end{align}
where $S_0/\rho_c \simeq 2.74 \times 10^8$ is the ratio of the entropy today and the critical energy density.
\begin{figure}[t!]
\centering
\includegraphics[scale=0.8]{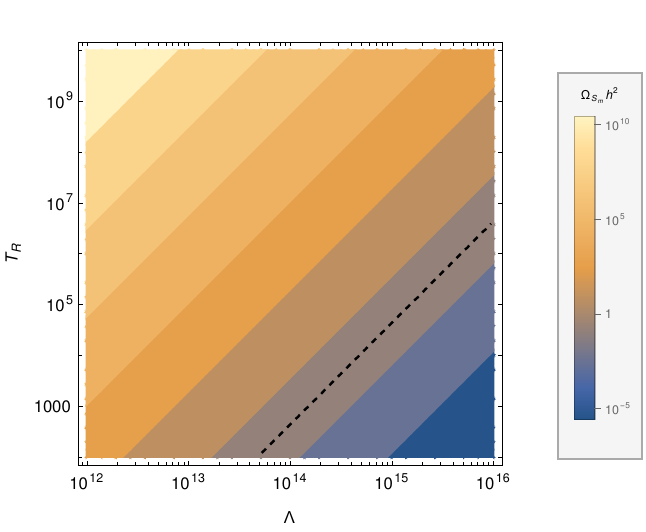}
\caption{FIMP DM relic density in terms of $\Lambda$ and the reheating temperature $T_R$ in units of GeV. The black dashed line indicates the correct relic abundance. Here, $M_{S_m}=100$ GeV is chosen. One may notice that low values of $T_R$ are preferred to obtain a correct relic abundance.} 
\label{fig:TRlambda-plot}
\end{figure}
We have checked that the result from numerical analyses performed by using {\tt micrOMEGAs}~\cite{Belanger:2006is} for the $T_R$-dominated regime matches well with the analytical expression \eqref{eqn:YSmapprox}.
Figure~\ref{fig:TRlambda-plot} shows the relic density of the FIMP DM, $\Omega_{S_m} h^2$, in the $T_R-\Lambda$ plane, using the analytical result. The black dashed line indicates the correct relic abundance. We see that low values of the reheating temperature are preferred. For the rest of the work, we will therefore concentrate on the low reheating temperature. In particular, we shall choose $T_R = 3$ TeV throughout this section. 

With the knowledge obtained above, we now re-introduce all the couplings and numerically evolve the full Boltzmann equations using {\tt micrOMEGAs} to obtain the DM relic densities. The relevant Boltzmann equations are
\begin{align}
\frac{d Y_{W_D}}{dz} &= -\frac{2 \pi^2}{45} \frac{M_{\rm Pl} M_{H_1} 
\sqrt{g_\rho(z)}}{1.66 z^2} \sum_{A,B \in {\rm SM}}\langle \sigma v \rangle_{W_{D}W_{D} \rightarrow
A B} \left( Y^2_{W_D} - Y^{{\rm eq}\,2}_{W_D} \right)\, \nonumber \\
\frac{d Y_{N_m}}{dz} &= \frac{4 \pi^2}{45} \frac{M_{\rm Pl} M_{H_1} \sqrt{g_\rho(z)}}{1.66 z^2} \sum_{i,j \in {\rm SM}, W_{D}, H_2} 
\langle \sigma v \rangle_{ij} \left(Y^{{\rm eq}}_{i} Y^{{\rm eq}}_{j} - Y^2_{N_m}\right)
\nonumber \\
& - \frac{M_{\rm Pl} z \sqrt{g_\rho(z)}}{1.66 M^2_{H_1} g_{S}(z)} 
 \sum_{f_{1}, f_{2} \in {\rm SM}} \langle \Gamma_{N_{m} \rightarrow S_{m} f_{1} f_{2}} \rangle \left( Y_{N_{m}} - Y_{S_{m}} Y_{f_1} Y_{f_2} \right)
 \theta \left(1 - \frac{M_{S_m} + M_{f_1} + M_{f_2}}{M_{N_m}} \right)
 \nonumber \\
& + \frac{2 M_{\rm Pl} z \sqrt{g_\rho(z)}}{1.66 M_{H_1}^2 g_{S}(z)} 
 \sum_{i=1,2} \langle \Gamma_{H_{i} \rightarrow N_{m} N_{m}} \rangle \left( Y_{H_{i}} - Y_{N_{m}} Y_{N_{m}} \right)
 \theta \left(1 - \frac{2 M_{N_m}}{M_{H_i}} \right)
 \, \nonumber \\
\frac{d Y_{S_m}}{dz} &= \frac{4 \pi^2}{45} \frac{M_{\rm Pl} M_{H_1} \sqrt{g_\rho(z)}}{1.66 z^2} \sum_{i,j \in {\rm SM}, W_{D}, H_2} 
\langle \sigma v \rangle_{ij} \left(Y^{{\rm eq}}_{i} Y^{{\rm eq}}_{j} - Y^2_{S_m}\right)
\nonumber \\
& + \frac{M_{\rm Pl} z \sqrt{g_\rho(z)}}{1.66 M^2_{H_1} g_{S}(z)} 
 \sum_{f_{1}, f_{2} \in {\rm SM}} \langle \Gamma_{N_{m} \rightarrow S_{m} f_{1} f_{2}} \rangle \left( Y_{N_{m}} - Y_{S_{m}} Y_{f_1} Y_{f_2} \right)
 \theta \left(1 - \frac{M_{S_m} + M_{f_1} + M_{f_2}}{M_{N_m}} \right)
  \nonumber \\
& + \frac{2 M_{\rm Pl} z \sqrt{g_\rho(z)}}{1.66 M^2_{H_1} g_{S}(z)} 
 \sum_{i=1,2} \langle \Gamma_{H_{i} \rightarrow S_{m} S_{m}} \rangle \left( Y_{H_{i}} - Y_{S_{m}} Y_{S_{m}} \right)
 \theta \left(1 - \frac{2 M_{S_m}}{M_{H_i}} \right)\,,
 \label{BE-DM}
\end{align}
where $z \equiv M_{H_1}/T$, $\langle \sigma v \rangle_{AB \rightarrow CD}$ is the thermally-averaged cross section times velocity, $\langle \Gamma_{A \rightarrow BCD} \rangle$ is the thermally-averaged decay width, and $\theta(x)$ is the Heaviside step function. 
We present the relevant Feynman diagrams in Appendix~\ref{apdx:FeynmanDiagrams}.

\begin{figure}[t!]
\centering
\includegraphics[scale=0.6]{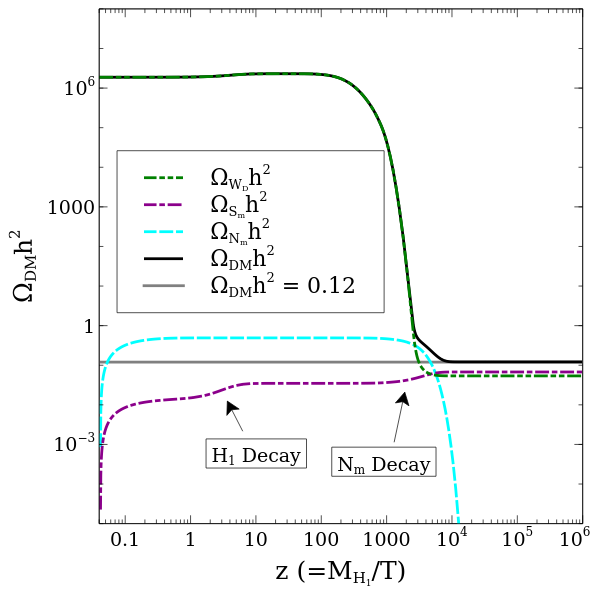}
\caption{DM production by the freeze-out and freeze-in mechanisms and its evolution in terms of $z\equiv M_{H_1}/T$. 
The model parameters are chosen as $M_{N_{m}} = 300$ GeV, $M_{S_{m}} = 20$ GeV, $M_{W_D} = 1.04628$ GeV, $\Lambda = 5.5 \times 10^{14}$ GeV, $\kappa = \kappa^{\prime} = \xi = \xi^{\prime} = \alpha = \alpha^{\prime} = 1$, $M_{H_2} = 2.212$ GeV, $g_{D} = 3.1 \times 10^{-4}$, $\delta =10^{-2}$, and $\sin\alpha = 8.17 \times 10^{-2}$. For the reheating temperature, we have used $T_R = 3$ TeV.
The green double-dot-dashed (purple dot-dashed) line corresponds to the WIMP (FIMP) DM relic density. The cyan dashed line represents the NLSP relic density. The sum of the WIMP and FIMP DM relic densities is depicted by the black solid line, while the grey solid line shows the present DM relic density measured by the Planck, $\Omega_{\rm DM}h^2 = \Omega_{\rm Tot}h^2 = 0.12$.
} 
\label{fig:bp-line-plot}
\end{figure}

In Fig.~\ref{fig:bp-line-plot}, the DM production by freeze-out and freeze-in mechanisms are shown. The green double-dot-dashed line corresponds to the WIMP DM production by the freeze-out mechanism. It freezes out at $T \simeq M_{W_D}/20$ which corresponds to $z \simeq 2500$. The cyan dashed line represents the production of the NLSP $N_m$, which later decays to the FIMP DM $S_m$ at $z \simeq 3500$. The NLSP is produced in the early Universe at $T \simeq 3000$ GeV, {\it i.e.}, $z \simeq 0.03$ through $2 \rightarrow 2$ processes present in our model.
The purple dot-dashed line indicates the FIMP DM production via the freeze-in mechanism. At its initial production, we see a sharp rise at $z = 0.03$ which represents the production by the $2 \rightarrow 2$ processes like the NLSP case. There exists a second rise in the production shortly after $z = 1$ which is due to the decay of the SM-like Higgs, $H_1$. Finally, a third rise happens at $z \simeq 3500$ when the NLSP decays to the FIMP DM. 
The sum of the WIMP and FIMP DM relic densities is depicted by the black solid line, which coincides with the Planck measurement of total DM relic density $\Omega_{\rm Tot}h^{2} = 0.12$ today which is represented by the grey solid line. We have chosen the parameter values in such a way that the WIMP and FIMP DM contribute equally, namely $\Omega_{S_m}h^2 \approx \Omega_{W_D}h^2 \approx \Omega_{\rm Tot}h^{2}/2$.

\begin{figure}[t!]
\centering
\includegraphics[scale=0.49]{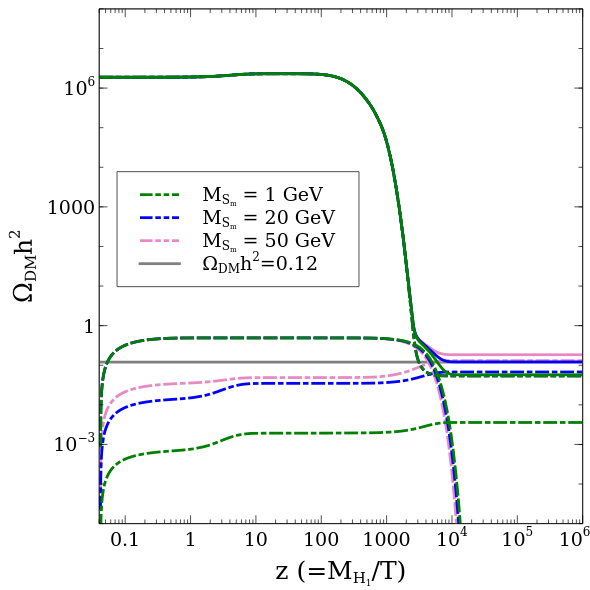}
\includegraphics[scale=0.49]{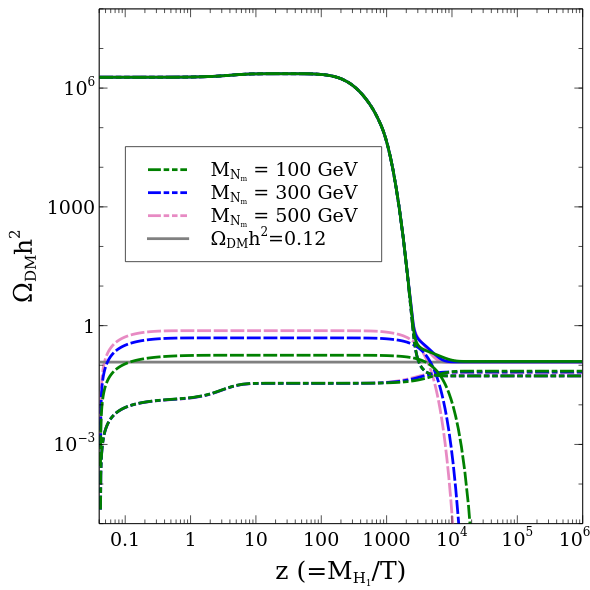}
\caption{Dependence of the DM relic densities on three different values of the FIMP DM mass (left) and the NLSP mass (right). The other parameters are chosen as $M_{W_D} = 1.04628$ GeV, $\Lambda = 5.5 \times 10^{14}$ GeV, $\kappa = \kappa^{\prime} = \xi = \xi^{\prime} = \alpha = \alpha^{\prime} = 1$, $M_{H_2} = 2.212$ GeV, $g_{D} = 3.1 \times 10^{-4}$, $\delta =10^{-2}$, and $\sin\alpha = 8.17 \times 10^{-2}$. For the reheating temperature, we have used $T_R = 3$ TeV.
For the left plot, $M_{N_{m}} = 300$ GeV is chosen, while $M_{S_{m}} = 20$ GeV is used for the right plot. The double-dot-dashed (dot-dashed) lines represent the WIMP (FIMP) DM relic densities, while the dashed lines indicate the NLSP relic densities. The solid lines are the sum of the FIMP and WIMP DM relic densities, while the grey solid line denotes the current DM relic density of 0.12.
} 
\label{fig:line-plot-1}
\end{figure}

Dependence of the DM relic densities on the FIMP DM mass is shown in the LP of Fig.~\ref{fig:line-plot-1}. 
One may see that the variation of the FIMP DM mass does not alter the WIMP DM relic density, which is depicted by double-dot-dashed lines. 
The dashed lines correspond to the NLSP ($N_m$) relic densities. The decay length of NLSP is not affected by the DM mass unless we choose $M_{N_m} \simeq M_{S_m} + M_{H_{1,2}}$. The dot-dashed lines below $\Omega_{\rm DM} h^{2} \simeq 10^{-1}$ are the FIMP DM evolutions. We see that, for $M_{S_m} = 1$ and 20 GeV, there is a slight rise in the DM density which corresponds to the FIMP DM production from the SM Higgs decay at around $z=1$. This rise is, however, negligible for the $M_{S_m} = 50$ GeV case due to the phase space suppression from the SM Higgs decay. The second rise at $z \sim 3500$ happens when NLSP decays to the FIMP.
Total DM relic density, which is the sum of the WIMP and FIMP relic densities, is represented by the solid lines. We see that the total DM relic density mainly follows the WIMP DM relic density. 
The RP of Fig.~\ref{fig:line-plot-1}, shows the dependence of the DM relic densities on the NLSP mass. The NLSP masses are all above 100 GeV, so its production happens through $2 \rightarrow 2$ processes. The NLSP relic density varies linearly with its mass, and its contribution to the DM relic density is given by $(M_{N_m}/M_{S_m})\Omega_{N_m}h^{2}$. This is similar to the SuperWIMP mechanism \cite{Covi:1999ty} and associated with the conservation of the comoving number densities between two out-of-equilibrium species. For example, for a process $A \rightarrow B+\cdots$, if $A$ and $B$ are out of equilibrium, then one has $Y_{A} = Y_{B}$, and thus, $\Omega_{A}h^{2}/M_{A} = \Omega_{B}h^{2}/M_{B}$.

\begin{figure}[t!]
\centering
\includegraphics[scale=0.49]{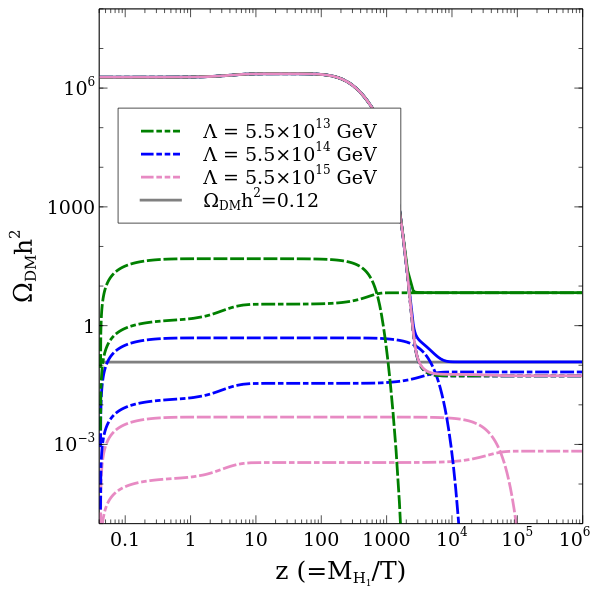}
\includegraphics[scale=0.49]{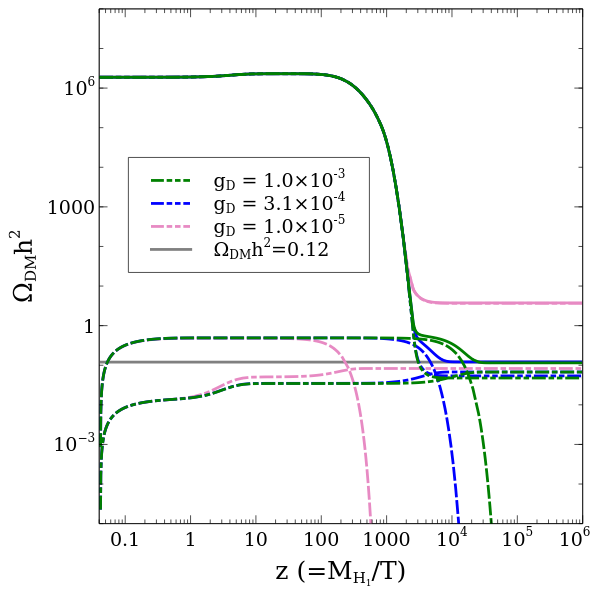}
\caption{Dependence of the DM relic densities on three different values of $\Lambda$ (left) and the dark gauge coupling $g_D$ (right). The other parameters are kept the same as those given in Fig.~\ref{fig:bp-line-plot}.
The double-dot-dashed (dot-dashed) lines represent the WIMP (FIMP) DM relic densities, while the dashed lines indicate the NLSP relic densities. The solid lines are the sum of the FIMP and WIMP DM relic densities, while the grey solid line denotes the current DM relic density of 0.12.
} 
\label{fig:line-plot-2}
\end{figure}

Figure~\ref{fig:line-plot-2} presents dependence of the DM relic densities on three different values $\Lambda$ (LP) and $g_{D}$ (RP). From the LP, one may see the significant changes in the FIMP DM and the NLSP relic densities with the variation of $\Lambda$.
This behaviour is due to the fact that their production strength is inversely proportional to $\Lambda$. We see that the green lines, which correspond to $\Lambda = 5.5 \times 10^{13}$ GeV, have larger relic densities than the blue and pink lines, which respectively correspond to $\Lambda = 5.5 \times 10^{14}$ GeV and $\Lambda = 5.5 \times 10^{15}$ GeV. 
The RP of Fig.~\ref{fig:line-plot-2} shows the variation of the DM relic densities for three different values of the $U(1)_{D}$ gauge coupling $g_D$. The green lines are for $g_{D} = 10^{-3}$, the blue lines are for $g_{D} = 3.1 \times 10^{-4}$, and the pink lines are for $g_{D} = 10^{-5}$. 
We see that the $g_{D} = 10^{-5}$ case, which is depicted by the pink lines, has the largest WIMP DM relic density. It is because a smaller value of $g_D$ reduces the WIMP DM annihilation cross-section ($W_{D}+W_{D} \rightarrow {\rm SM} + {\rm SM}$) which affects inversely the WIMP DM relic density. 
On the other hand, the FIMP DM relic density is controlled by the strength of the dark Higgs VEV $v_{D} = M_{W_D}/g_D$. The VEV $v_D$ is linearly proportional to the coupling strength responsible for the FIMP DM production through the BSM Higgs decay. One interesting thing we may note is that $g_D$ does not affect the NLSP production as its production is governed by the $2 \rightarrow 2$ processes. However, the NLSP decay depends on the VEV $v_D$; the NLSP decays faster with the increment of $v_{D}$ or decrement of $g_D$. Moreover, FIMP DM production from the decays of the SM Higgs has an effect only when $v_{D}$ is large enough; otherwise, the $v_D$-associated part in $H_{1} \rightarrow S_{m} + S_{m}$ production is suppressed. 

\begin{figure}[t!]
\centering
\includegraphics[scale=0.49]{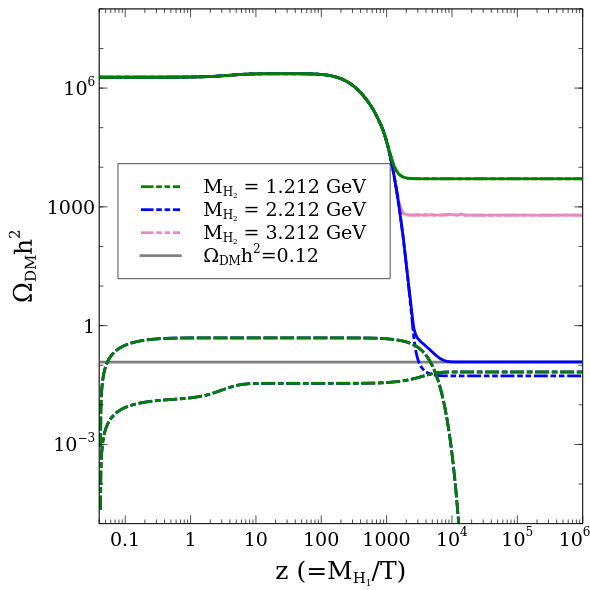}
\includegraphics[scale=0.49]{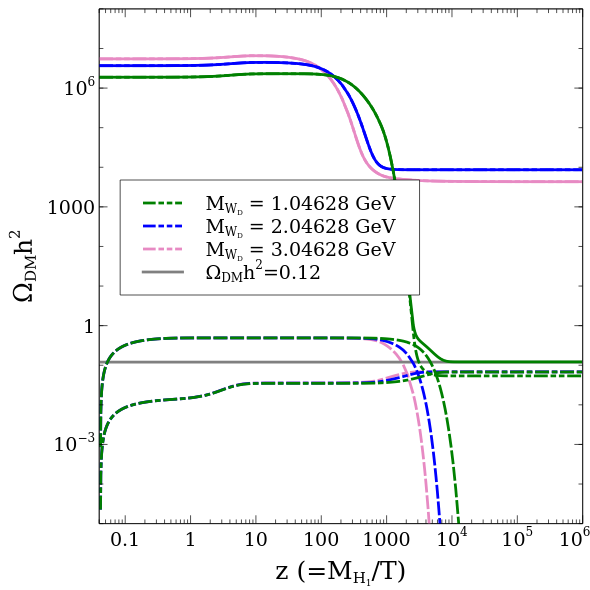}
\caption{Dependence of the DM relic densities on three different values of the dark Higgs mass $M_{H_2}$ (left) and the dark gauge boson WIMP mass $M_{W_D}$ (right). The other parameters are kept the same as those given in Fig.~\ref{fig:bp-line-plot}.
The double-dot-dashed (dot-dashed) lines represent the WIMP (FIMP) DM relic densities, while the dashed lines indicate the NLSP relic densities. The solid lines are the sum of the FIMP and WIMP DM relic densities, while the grey solid line denotes the current DM relic density of 0.12.} 
\label{fig:line-plot-3}
\end{figure}

In Fig.~\ref{fig:line-plot-3}, we show the dependence of the DM relic densities on three different values of $M_{H_2}$ (LP) and $M_{W_D}$ (RP).
From the LP, we see that there is no effect of $M_{H_2}$ on the FIMP DM production, while the effect on the WIMP DM production is significant. It is the case since the WIMP DM relic density is mainly controlled by how far we are from the resonance region of the second Higgs $H_2$. 
On the other hand, from the RP of Fig.~\ref{fig:line-plot-3}, we see that changing the WIMP DM mass affects both the WIMP and FIMP DM productions. The effect on the WIMP DM is due to the fact that, with the change of $M_{W_D}$, we are moving away from the resonance region of the second Higgs $H_2$, and thus we have more production of the WIMP DM. Additionally, the NLSP decay is proportional to the VEV $v_{D} = M_{W_D}/g_D$. Therefore, by increasing the value of $M_{W_D}$, the value of $v_D$ increases as well, which triggers an early decay of NLSP.

\subsection{Exploration of allowed parameter spaces}
\label{subsec:DMParamScans}
With the understandings we have acquired in Section~\ref{subsec:DMproduction} by studying the behaviours of the DM relic densities near the point \eqref{eqn:parameters-fixed}, we attempt to obtain allowed parameter regions amongst the different parameters after imposing that the DM relic density satisfies the range $0.01 \leq \Omega_{\rm DM} h^{2} \leq 0.12$.
The lower limit of 0.01 is to ensure more allowed points. Note, however, that our conclusion remains unchanged even if the $3\sigma$ range shown in Eq.~\eqref{eqn:DM-RD-3sigma} is considered.
We also discuss bounds on the WIMP DM parameters coming from the direct and indirect detections of DM. 
We perform parameter scans with the following parameter ranges:
\begin{gather}
1 \leq M_{S_m}\,\,[{\rm GeV}] \leq 100
\,,\quad
100 \leq M_{N_m}\,\,[{\rm GeV}] \leq 1000
\,,\quad 
1 \leq M_{W_D}\,\,[{\rm GeV}] \leq 100
\,,\nonumber \\
5.5\times 10^{13} \leq \Lambda\,\,[{\rm GeV}] \leq 5.5\times 10^{15}
\,,\quad 
10^{-4} \leq g_{D} \leq 10^{-1}
\,, \quad 
1.5 \leq \frac{M_{H_2}}{M_{W_D}}
\leq 2.5
\,, \label{eqn:DM-range-scatter}
\end{gather}
with the rest of the model parameters being fixed as those given in Eq.~\eqref{eqn:parameters-fixed}. We have taken $T_R = 3$ TeV for the reheating temperature.
The chosen range of the ratio $M_{H_2}/M_{W_D}$ is due to the observation that the WIMP DM mass needs to be close to the resonance region, as shown in Fig.~\ref{fig:line-plot-3}.

\begin{figure}[t!]
\centering
\includegraphics[scale=0.49]{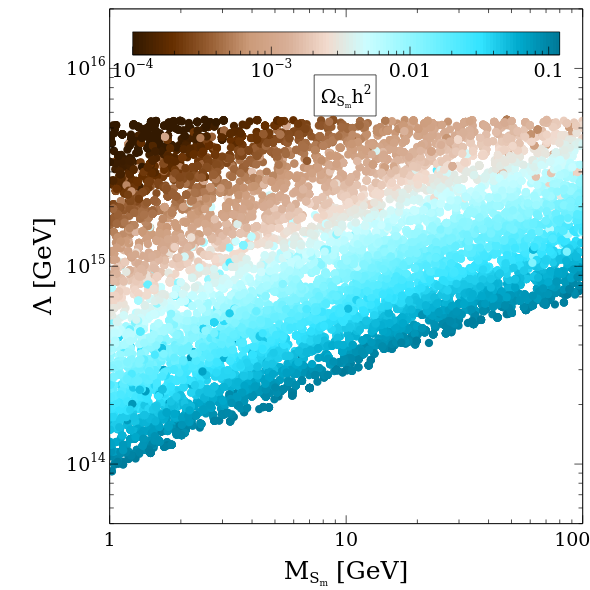}
\includegraphics[scale=0.49]{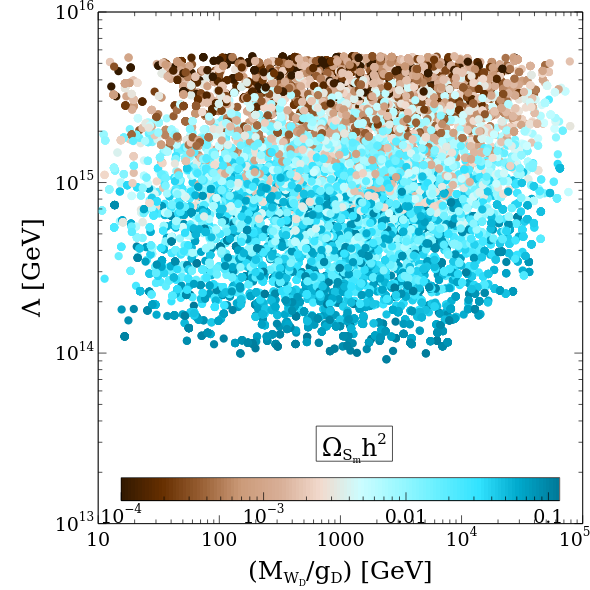}
\caption{Allowed parameter space satisfying $0.01 \leq \Omega_{\rm DM}h^2 \leq 0.12$ in the $M_{S_m}$ -- $\Lambda$ (left) and $M_{W_D}/g_D$ -- $\Lambda$ (right) planes. The colour of the points represents the FIMP DM relic density.} 
\label{fig:DM-scatter-plot-1}
\end{figure}

The LP of Fig.~\ref{fig:DM-scatter-plot-1} shows the allowed region in the $M_{S_m}$ -- $\Lambda$ plane where the colour represents the FIMP DM relic density. One may easily see that, for a fixed value of the FIMP DM mass, increasing the value of $\Lambda$ makes the FIMP contribution to DM relic density decrease, as the FIMP DM production is inversely proportional to $\Lambda$. 
The lower limit in the $\Lambda$ value comes from the maximum allowed range for DM relic density, $\Omega_{\rm DM}h^2 = 0.12$, since the relic density is proportional to $M_{S_m}/\Lambda^2$. 
In the RP of Fig.~\ref{fig:DM-scatter-plot-1}, we present the allowed range in the $M_{W_D}/g_D$ -- $\Lambda$ plane. One may again observe that, as we go to a higher value of $\Lambda$, we have a smaller FIMP DM contribution. The VEV $v_D = M_{W_D}/g_D$ linearly contributes to the FIMP DM relic density, and thus, for a higher value of $v_D$, we need a higher value of $\Lambda$ to get the correct DM relic density value; we notice this in particular in the region $g_D < 10^{-3}$ and $10<M_{W_D} \; [{\rm GeV}]<100$. This correlation between $v_D$ and $\Lambda$ is observed for higher values of $v_D$, whereas we do not see such a correlation for lower values of $v_D$.
In both the LP and RP of Fig.~\ref{fig:DM-scatter-plot-1}, we see that there is no upper bound on $\Lambda$. This is because such higher values of $\Lambda$ reduce the FIMP DM relic density, and the DM relic density bound can be satisfied from the contribution of the WIMP DM. 

\begin{figure}[t!]
\centering
\includegraphics[scale=0.49]{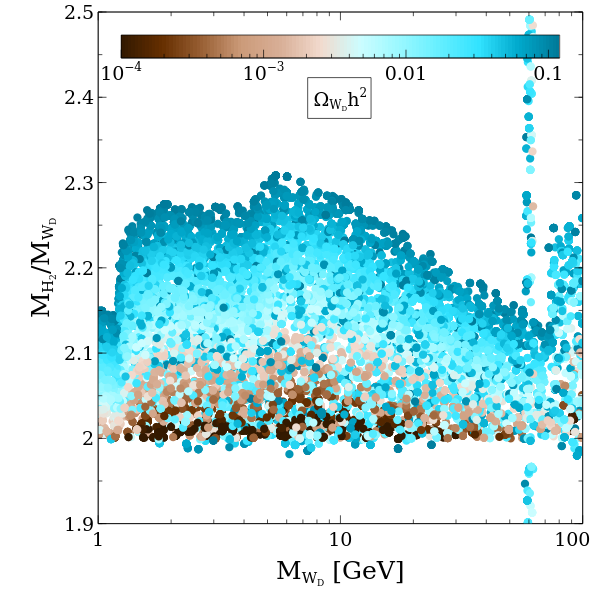}
\includegraphics[scale=0.49]{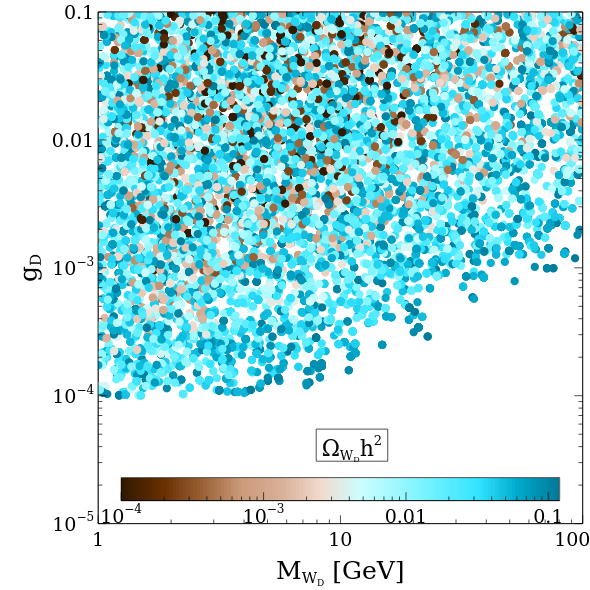}
\caption{Allowed parameter space satisfying $0.01 \leq \Omega_{\rm DM}h^2 \leq 0.12$ in the $M_{W_D}$ -- $M_{H_2}/M_{W_D}$ (left) and $M_{W_D}$ -- $g_D$ (right) planes. The colour of the points represents the WIMP DM relic density.} 
\label{fig:DM-scatter-plot-2}
\end{figure}

The LP of Fig.~\ref{fig:DM-scatter-plot-2} shows the allowed range in the $M_{W_D}$ -- $M_{H_2}/M_{W_D}$ plane after imposing $0.01\leq \Omega_{\rm DM}h^2 \leq 0.12$. It is clearly shown in the figure that, to obtain the WIMP DM relic density below 0.12, we need to stay near the resonance region, {\it i.e.}, $M_{H_2} \sim 2M_{W_D}$. It is also clear that, when we are very close to the resonance region, we have a smaller WIMP contribution in the DM relic density, while a larger WIMP contribution is obtained as we depart from the resonance region. 
Moreover, we see that, for $M_{W_D} \simeq 62.5$ GeV, the dark Higgs mass $M_{H_2}$ may take any value. This is due to the fact that the dominating contribution comes from the SM Higgs resonance. 
The RP of Fig.~\ref{fig:DM-scatter-plot-2} shows the allowed region in the $M_{W_D}$ -- $g_{D}$ plane after imposing $0.01\leq \Omega_{\rm DM}h^2 \leq 0.12$. One may see from the figure that, if we increase the value of $g_D$, one may have a smaller contribution of WIMP DM. This happens because the annihilation cross section increases with $g_D$, and the WIMP DM relic density is inversely proportional to annihilation cross section.
For higher values of $M_{W_D} \gtrsim 10 \;{\rm GeV}$, we see no allowed point in the range $10^{-4} \lesssim g_D \lesssim 10^{-3}$ as the region has a dominating FIMP DM contribution due to the high value of the VEV $v_D$.

\begin{figure}[t!]
\centering
\includegraphics[scale=0.85]{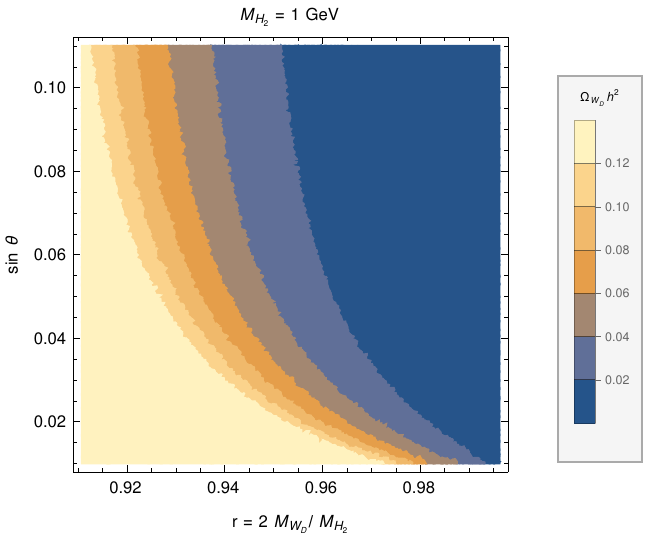}
\caption{WIMP DM relic density as a function of the mixing angle $\sin\theta$ and $r\equiv 2M_{W_D}/M_{H_2}$ that quantifies the closeness to the resonance region. The exact resonance point is where $r=1$. For this example, we have set $M_{H_2} = 1$ GeV. The other relevant parameter is $g_D$, and it is fixed as $g_D = 10^{-4}$. In order to have a significant amount of WIMP relic density while not overproducing, we need to be near the range of $0.92 \lesssim r \lesssim 0.98$ for $\sin\theta < 0.1$.}
\label{fig:sinthetavsr}
\end{figure}

How close one needs to be to the resonance region is studied in Fig.~\ref{fig:sinthetavsr} where the WIMP DM relic density is shown in terms of the mixing angle $\sin\theta$ and $r\equiv 2M_{W_D}/M_{H_2}$. The parameter $r$ quantifies the closeness to the resonance region, and $r=1$ corresponds to the exact resonance point. We find that $r$ typically takes a value between $0.92$ and $0.98$, depending on the value of the mixing angle, if we ask for the WIMP DM component to be a significant part of the total relic density. 
We observe that the window for a WIMP DM relic density of at least $10 \%$ of the total DM relic density is narrower for smaller values of the mixing angle. At a fixed value of $r$, the relic density decreases as the value of $\sin \theta$ increases. One may understand this as follows: The process keeping the WIMP DM in thermal equilibrium is ${\rm DM} + {\rm DM} \leftrightarrow {\rm SM} + {\rm SM}$, and it is mediated by $H_2$. We thus find that the cross section is proportional to $\cos^2 \theta \sin^2 \theta$. 
Higher values of $r$ mean being closer to the resonant point where the cross section increases. Therefore, the relic density decreases following the standard behaviour, $\Omega_{\rm WIMP} h^2 \sim 1/ \langle \sigma v \rangle$. Once we depart too much from the resonance region, we may overproduce the WIMP DM and overclose the Universe.

\begin{figure}[t!]
\centering
\includegraphics[scale=0.49]{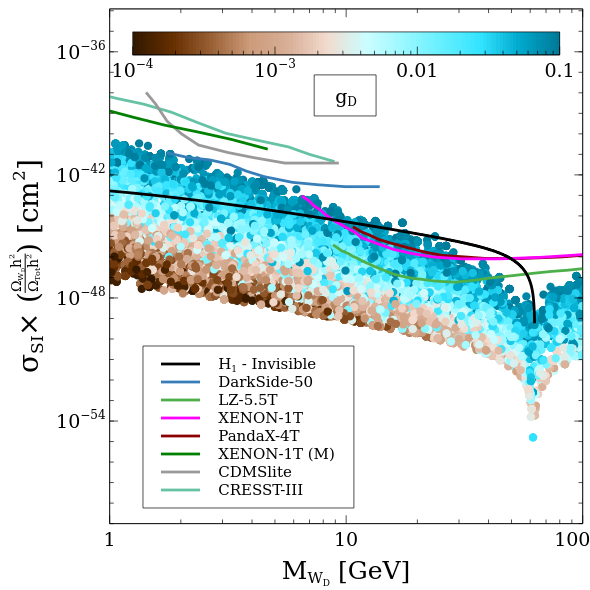}
\includegraphics[scale=0.49]{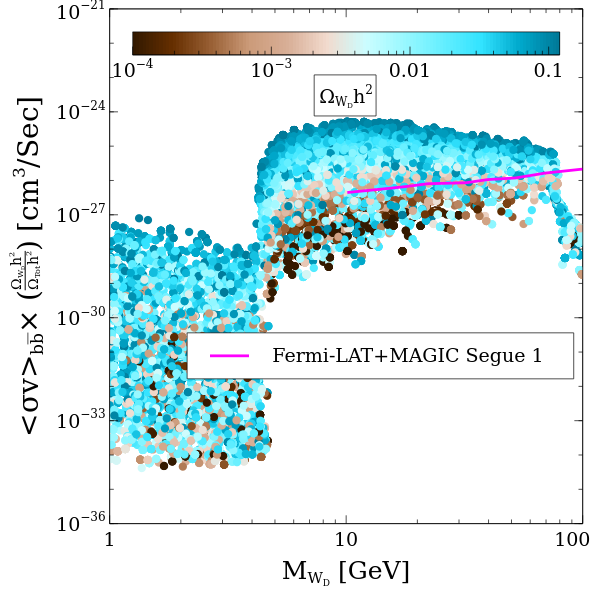}
\caption{
Allowed parameter space satisfying $0.01 \leq \Omega_{\rm DM}h^2 \leq 0.12$ in the $M_{W_D}$ -- $(\Omega_{W_D}/\Omega_{\rm Tot})\sigma_{\rm SI}$ (left) and $M_{W_D}$ -- $(\Omega_{W_D}/\Omega_{\rm Tot})\langle\sigma v\rangle_{b\bar{b}}$ (right) planes. Here, $\Omega_{\rm Tot}h^2 = 0.12$ is total DM relic density today. The black solid line in the left panel indicates the Higgs invisible decay constraint. Various direct and indirect detection bounds are also overlaid with coloured solid lines; see text for detailed explanation. The colour of the points represents the value of the dark gauge coupling $g_D$ (left) and the WIMP DM relic density (right).
} 
\label{fig:DM-scatter-plot-3}
\end{figure}

Figure~\ref{fig:DM-scatter-plot-3} shows the allowed region in the $M_{W_D}$ -- $(\Omega_{W_D}/\Omega_{\rm Tot})\sigma_{\rm SI}$ (LP) and $M_{W_D}$ -- $(\Omega_{W_D}/\Omega_{\rm Tot})\langle\sigma v\rangle_{b\bar{b}}$ (RP) planes, together with various direct and indirect detection bounds that are depicted by solid lines.
Note that we have rescaled the $y$-axes by the amount of the WIMP DM relic density compared to the total DM in the Universe $\Omega_{\rm Tot} h^{2} = 0.12$. 
The LP of Fig.~\ref{fig:DM-scatter-plot-3} may be easily understood with the direct detection expression given by Eq.~\eqref{eqn:DD-expression}, which states that $\sigma_{\rm SI}$ is proportional to $g^2_D$. 
One may estimate the percentage of the WIMP DM relic that each sample point corresponds to with the help of the RP of Fig.~\ref{fig:DM-scatter-plot-2}. Comparing the LP of Fig.~\ref{fig:DM-scatter-plot-3} and RP of Fig.~\ref{fig:DM-scatter-plot-2}, one can easily see that lower values of $g_{D}$ correspond to lower values of $\sigma_{SI} (\propto g^2_{D})$ and higher values of the WIMP DM relic density as the density is inversely proportional to $g^2_{D}$.
A sharp dip at $M_{W_D} \simeq 62.5$ GeV happens because of the mutual cancellation between the SM Higgs- and the BSM Higgs-mediated processes as one may see from Eq.~\ref{eqn:DD-expression}. 
A part of the $M_{W_D} > 7$ GeV region is already ruled out by the different direct detection experiments such as XENON-1T~\cite{XENON:2018voc}, PandaX-4T~\cite{PandaX-II:2017hlx,Liu:2022zgu}, and LUX-ZEPLIN-5.5T~\cite{LZ:2022ufs}. 
The region of DM mass below 7 GeV will be explored by DarkSide-50~\cite{DarkSide:2018kuk,DarkSide:2018bpj}, XENON-1T(M)~\cite{Ibe:2017yqa}, CDMSlite~\cite{SuperCDMS:2018gro}, and CRESST-III~\cite{CRESST:2017ues,CRESST:2019jnq}. 
The black solid line corresponds to the bound from the Higgs invisible decay which is obtained by staying near the dark Higgs resonance region, {\it i.e.}, $M_{H_2} \sim 2M_{W_D}$, so that the WIMP DM never becomes over-abundant. The region above the black solid line is already ruled out by the current bound on the branching of the Higgs invisible decay mode.
We note that our model predicts much lower values for $\sigma_{\rm SI}$ compared to the aforementioned bounds.
From the RP of Fig.~\ref{fig:DM-scatter-plot-3}, we see that there is a dip in $\langle\sigma v\rangle_{b\bar{b}}$ for the WIMP DM mass below 5 GeV. This is due to the fact that, for this range, the channel $W_{D}W_{D} \rightarrow b \bar{b}$ is not active. The region of $M_{W_D}\gtrsim 10$ GeV is constrained by the Fermi-LAT + MAGIC Segue 1 data~\cite{MAGIC:2016xys}. We observe that most of the parameter space which contributes dominantly to the DM relic is already ruled out by the indirect detection bound. 
We have also checked the present bounds on the DM annihilation to $\mu^+\mu^-$ and $\tau^+\tau^-$, and we present the results in Appendix~\ref{apdx:mumutautaubounds}.

\section{First-order phase transitions and associated gravitational waves}
\label{sec:FOPTsGWs}
The extra dark $U(1)_D$ Higgs field not only gives a mass to the WIMP DM $W_D$, but it also changes the vacuum evolution. We study the evolution of the vacuum state and the dynamics of the phase transition in this section.
We first compute the one-loop finite-temperature effective potential,
\begin{align}
V_{\rm eff} = V^{(0)} + V^{(1)}_{\rm CW} + V^{(1)}_T
\,,
\end{align}
where $V^{(0)}$ is the tree-level scalar potential, $V^{(1)}_{\rm CW}$ is the one-loop Coleman-Weinberg potential, and $V^{(1)}_T$ is the finite-temperature correction.
In terms of the background fields $\bar{H}$ and $\bar{H}_D$ of the SM Higgs and the $U(1)_D$ dark Higgs, the tree-level scalar potential is given by
\begin{align}
V^{(0)} =
\frac{1}{2}\mu_h^2\bar{H}^2
+\frac{1}{2}\mu_D^2\bar{H}_D^2
+\frac{1}{4}\lambda_h^2\bar{H}^4
+\frac{1}{4}\lambda_D^2\bar{H}_D^4
+\frac{1}{4}\lambda_{hD}^2\bar{H}^2\bar{H}_D^2
\,.
\end{align}
The Coleman-Weinberg potential can generically be written as
\begin{align}\label{eqn:CWpot}
V^{(1)}_{\rm CW} &=
\pm \sum_i n_i \frac{M_i^4(\bar{H},\bar{H}_D)}{64\pi^2}\left[
\ln\frac{M_i^2(\bar{H},\bar{H}_D)}{\bar \mu^2} - c_i
\right]
\,,
\end{align}
where the $+$ ($-$) sign is for bosons (fermions), $n_i$ is the number of degrees of freedom of the species $i$, $M_i$ is the field-dependent mass, the constants $c_i$ are 1/2 for transverse gauge bosons and 2/3 for the rest, and $\bar \mu$ is the renormalisation scale.
The expressions for the field-dependent masses $M_i$ and $n_i$ are summarised in Appendix \ref{apdx:CWmass}. Depending on the choice of the renormalisation scale $\bar{\mu}$, the effective potential $V_{\rm eff}$ changes, and hence one may arrive at different results. This renormalisation scale dependence has been explored in {\it e.g.} Refs.~\cite{Chiang:2018gsn,Croon:2020cgk}\footnote{
See also {\it e.g.} Refs.~\cite{Nielsen:1975fs,Fukuda:1975di,Patel:2011th,Chiang:2017zbz,Chiang:2018gsn,Croon:2020cgk,Schicho:2022wty} for the gauge dependence issue.
}. Together with the gauge dependence issue, we do not attempt to address the issue of the renormalisation scale dependence as it goes beyond the scope of the current work. We thus ignore the one-loop Coleman-Weinberg corrections in the followings by assuming that the renormalisation scale $\bar{\mu}$ is chosen in such a way that the Coleman-Weinberg corrections are minimised. 
The finite-temperature correction is given by~\cite{Dolan:1973qd}
\begin{align}\label{eqn:TEMPpot}
V_T^{(1)} = \frac{T^4}{2\pi^2}\sum_i n_i I_\pm
\left(
\frac{M_i^2(\bar{H},\bar{H}_D)}{T^2}
\right)\,,
\end{align}
with
\begin{align}
I_\pm(x^2) = \pm \int_0^\infty dy \, y^2 \,
\ln\left(
1 \mp e^{-\sqrt{y^2+x^2}}
\right)\,,
\end{align}
where $I_+$ is for bosons and $I_-$ is for fermions.
The re-summed ring diagrams are taken into account by replacing the field-dependent masses as~\cite{Parwani:1991gq}
\begin{align}
M_i^2 \rightarrow \widetilde{M}_i^2 = M_i^2 + \Pi_i(T)\,,
\end{align}
where $\Pi_i(T)$ are the thermal masses~\cite{Carrington:1991hz}. We present the thermal mass expressions in Appendix~\ref{apdx:THERMmass}.
Up to the leading $\mathcal{O}(T)$ order, the effective potential is thus given by
\begin{align}
V_{\rm eff} &= 
\frac{1}{2}\left(
\mu_h^2 + \Pi_H 
\right) \bar{H}^2
+\frac{1}{2}\left( 
\mu_D^2 + \Pi_{H_D} 
\right) \bar{H}_D^2 
\nonumber\\
&\quad 
+\frac{1}{4}\lambda_h^2 \bar{H}^4
+\frac{1}{4}\lambda_D^2 \bar{H}_D^4
+\frac{1}{4}\lambda_{hD}^2\bar{H}^2\bar{H}_D^2
-E^{\rm SM}\bar{H}^3 T
+\cdots 
\,,\label{eqn:Veff}
\end{align}
where $E^{\rm SM} \equiv (2 g_2^3+{\sqrt{g_1^2+g_2^{2}}}^3)/(32 \pi)$ and `$\cdots$' include sub-leading, negligible terms. Note that we have assumed that the dark gauge coupling $g_D$ is small enough not to affect the leading-order terms.

\begin{figure}
\centering
\includegraphics[scale=1]{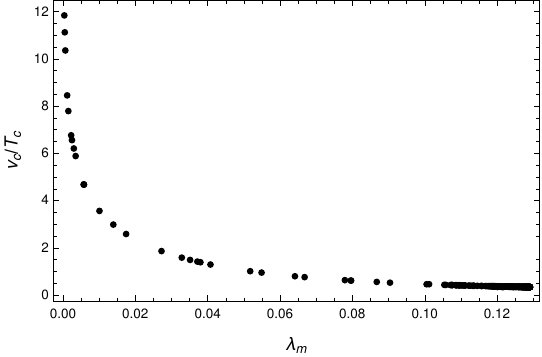}
\caption{Numerically computed $v_c/T_c$ values as a function of $\lambda_m \equiv \lambda_h-\lambda_{hD}^2/(4\lambda_D)$. Being in agreement with the analytical expression \eqref{eqn:vcTcExpr}, strong FOPTs, $v_c/T_c\gtrsim 1$, are achieved for small values of $\lambda_m$, or equivalently, small values of the dark $U(1)_D$ Higgs mass. We note that our result matches with the one presented in Ref.~\cite{Carena:2019une}.}
\label{fig:vcTcPlot}
\end{figure}

In the presence of the extra Higgs field, FOPTs may arise. FOPTs with a dark $U(1)_D$ have been studied in \textit{e.g.} Refs.~\cite{Chao:2014ina,Hashino:2018zsi,Breitbach:2018ddu,Borah:2021ocu}. For studies of the phase transition with an extra scalar field, see, \textit{e.g.}, Refs.~\cite{Cline:2012hg,Vaskonen:2016yiu,Kurup:2017dzf,Chiang:2018gsn,Ellis:2018mja,Carena:2019une,Biondini:2022ggt,Schicho:2022wty}.
In particular, in Ref.~\cite{Carena:2019une} where the studied scalar potential has the same form as Eq.~\eqref{eqn:Veff}, it was shown both analytically and numerically that the FOPTs could be strong, characterised by $v_c/T_c \gtrsim 1$. Here, $T_c$ is the critical temperature at which the potential minima become degenerate, and $v_c$ is the VEV of the SM Higgs field at $T_c$. It indicates that one of the Sakharov conditions for successful electroweak baryogenesis can be fulfilled~\cite{Sakharov:1967dj}.
Since the scalar field space is now two-dimensional due to the extra Higgs field, one may achieve either one-step or two-step phase transitions. Noting that the VEV of the $U(1)_D$ Higgs is non-zero at zero temperature, the one-step phase transition has the pattern $(\langle H \rangle, \langle H_D \rangle) = (0,0) \rightarrow (v,v_D)$, while the two-step phase transition may occur via $(\langle H \rangle, \langle H_D \rangle) = (0,0) \rightarrow (0,v_D^\prime) \rightarrow (v,v_D)$ or $(\langle H \rangle, \langle H_D \rangle) = (0,0) \rightarrow (v^\prime,0) \rightarrow (v,v_D)$.
For the two-step phase transition of the pattern $(\langle H \rangle, \langle H_D \rangle) = (0,0) \rightarrow (0,v_D^\prime) \rightarrow (v,v_D)$, the second step breaks the electroweak symmetry, giving~\cite{Carena:2019une}
\begin{align}\label{eqn:vcTcExpr}
\frac{v_c}{T_c} = 
\frac{2E^{\rm SM}}{\lambda_h - \lambda_{hD}^2/(4\lambda_D)}
= \frac{4 E^{\rm SM} v^2}{M_{H_1}^2}\left(1+\sin ^2 \theta\, \frac{M_{H_1}^2-M_{H_2}^2}{M_{H_2}^2}\right)\,.
\end{align}
One may see that strongly FOPTs, $v_c/T_c \gtrsim 1$, can be achieved when the dark Higgs is lighter than the SM Higgs. The one-step phase transition shows a similar behaviour~\cite{Carena:2019une}.
Utilising the publicly available tool {\tt CosmoTransitions}~\cite{Wainwright:2011kj}, we numerically compute $v_c/T_c$ for a wide range of the parameter space and present the result in Fig.~\ref{fig:vcTcPlot}. The explored parameter range is as follows:
\begin{gather}
0.01 \leq M_{H_2} \; [{\rm GeV}] \leq 100
\,,\quad 
-0.1 \leq \sin\theta \leq 0.1 
\,,\nonumber\\
10 \leq v_D \; [{\rm GeV}] \leq 10^{4}
\,,\quad 
10^{-5} \leq g_D \leq 10^{-2}
\,.\label{eqn:PTGWparams}
\end{gather}
We note that these four input parameters are the only relevant model parameters. The other model parameters can be derived from the above input parameters. 
The $x$-axis of Fig.~\ref{fig:vcTcPlot} is defined as $\lambda_m \equiv \lambda_h-\lambda_{hD}^2/(4\lambda_D)$.
We observe that strong FOPTs could be achieved for small values of $\lambda_m$, which is in good agreement with both the analytical estimate \eqref{eqn:vcTcExpr} and the results of Ref.~\cite{Carena:2019une}.

FOPTs may produce observable stochastic GWs~\cite{Kamionkowski:1993fg}. There are three main contributions to the GWs from the FOPT: bubble wall collisions $\Omega_{\rm col}h^{2}$, sound wave in plasma $\Omega_{\rm sw}h^{2}$, and the magneto-hybrodynamic turbulence $\Omega_{\rm turb}h^{2}$. The total GWs are then $\Omega_{\rm GW}h^{2}  \simeq \Omega_{\rm col}h^{2} +\Omega_{\rm sw}h^{2}+\Omega_{\rm turb}h^{2}$.
The GWs coming from the bubble wall collisions are given by~\cite{Caprini:2015zlo}
\begin{align}\label{eqn:Oh2col}
\Omega_{\rm col} h^{2} =
1.67 \times 10^{-5}
\left(\frac{\mathcal{H}_{*}}{\beta}\right)^{2}
\left(\frac{\kappa_{\phi} \alpha}{1+\alpha}\right)^{2}
\left(\frac{100}{g_{*}}\right)^{\frac{1}{3}}
\left(\frac{0.11 v_{w}^{3}}{0.42+v_{w}^{2}}\right)
\left( \frac{3.8\left(f / f_{\rm col}\right)^{2.8}}{1+2.8\left(f / f_{\rm col}\right)^{3.8}} \right)
\,,
\end{align}
while the turbulence contribution is~\cite{Caprini:2015zlo}
\begin{align}
\Omega_{\rm turb} h^{2} =
3.35 \times 10^{-4}
\left(\frac{\mathcal{H}_{*}}{\beta}\right)
\left(\frac{\kappa_{\rm turb} \alpha}{1+\alpha}\right)^{\frac{3}{2}}
\left(\frac{100}{g_{*}}\right)^{\frac{1}{3}}   
\left( \frac{v_{w} \left(f / f_{\rm turb}\right)^{3}}{\left[1+\left(f / f_{\rm turb}\right)\right]^{\frac{11}{3}}\left(1+8 \pi f / h_{*}\right)} \right)
\,,
\end{align}
where
\begin{align}
h_{*}=1.65 \times 10^{-5} \,{\rm Hz}
\left(\frac{T_{*}}{100 {\rm GeV}}\right)
\left(\frac{g_{*}}{100}\right)^{\frac{1}{6}}
\,.
\end{align}
Finally, the sound-wave contribution to the GW signal can be expressed as~\cite{Ellis:2018mja,Ellis:2019oqb,Ellis:2020awk,Guo:2020grp}
\begin{align}
\Omega_{\rm sw} h^{2} &=
4.80 \times 10^{-6}
\;{\rm min}
\left\{
1, \frac{2(8\pi)^{1/3}}{\sqrt{3}}v_w\left(\frac{\mathcal{H}_*}{\beta}\right)
\sqrt{\frac{1+\alpha}{\kappa_v \alpha}}
\right\}
\nonumber\\
&\qquad
\times
\left(\frac{\mathcal{H}_{*}}{\beta}\right)
\left(\frac{\kappa_{v} \alpha}{1+\alpha}\right)^{2}
\left(\frac{100}{g_{*}}\right)^{\frac{1}{3}}
v_{w} 
\left(f / f_{\rm sw}\right)^{3}
\left(\frac{7}{4+3\left(f / f_{\rm sw}\right)^{2}}\right)^{\frac{7}{2}}
\,.
\end{align}
For the expressions for $f_{\rm col}$, $f_{\rm sw}$, $f_{\rm turb}$, $v_w$, $\kappa$, $\kappa_v$, and $\kappa_{\rm trub}$, see Appendix \ref{apdx:GWexpr}.

\begin{figure}[t!]
\centering
\includegraphics[scale=1]{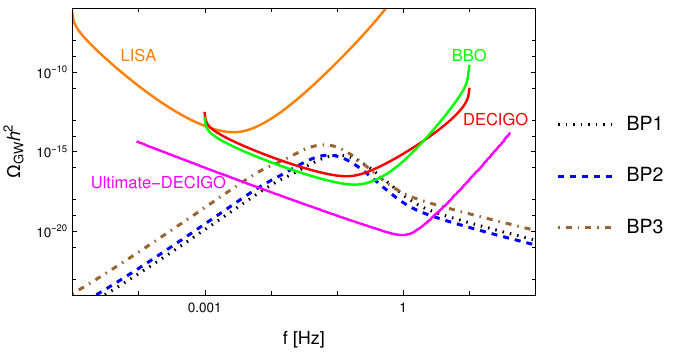}
\caption{
FOPT-associated GW spectra for our three BPs summarised in Table~\ref{tab:BPs}.
The black dotted line corresponds to the first BP, the blue dashed line depicts the second BPs, and the brown dot-dashed line represents the third BP. The sensitivity curves of future space-base GW experiments, including LISA, BBO, DECIGO, and Ultimate-DECIGO, are shown as well. We consulted Ref.~\cite{Schmitz:2020syl} for the data for LISA, BBO, and DECIGO, and Ref.~\cite{Ringwald:2020vei} for the Ultimate-DECIGO.
}
\label{fig:GWsBPs}
\end{figure}

Three parameters that play the key roles in the GW signal are
\begin{align}
\alpha=\frac{\rho_{\rm vac}}{\rho_{\rm rad}^{*}}
\,,\qquad
\frac{\beta}{\mathcal{H}_{*}}=
T_{*} \frac{d S_{\rm E}}{d T} \bigg\vert_{T_{*}}
\,,\quad {\rm and} \quad
T_*
\,,
\end{align}
where $S_{\rm E} = S_3/T$ is the Euclidean action of a bubble with $S_3$ being the three-dimensional action, $\rho_{\text{vac}}$ the released energy density during the phase transition, and $\rho_{\rm rad}^{*}=g_{*} \pi^{2} T_{*}^{4} / 30$, with $g_{*}$ being the number of effective degrees of freedom at $T=T_*$.
We take $T_*$ to be the nucleation temperature $T_n$.
We employ {\tt CosmoTransitions} \cite{Wainwright:2011kj} to numerically compute the three key parameters, $\alpha$, $\beta/\mathcal{H}_*$, and the nucleation temperature $T_n$.
In Fig.~\ref{fig:GWsBPs}, we present the FOPT-associated GW signals for three BPs together with the sensitivity curves of future space-based GW experiments such as LISA, DECIGO, and BBO. The three BPs, that account for not only the neutrino masses and the correct DM relic density, but also the strong FOPTs, are summarised in Table~\ref{tab:BPs}. One may notice that the three BPs have different DM compositions. In the case of the first BP (BP1), both the WIMP and FIMP contribute equally to the total DM relic density, while the BP2 (BP3) is mostly composed of the FIMP (WIMP) DM.
We see from Fig.~\ref{fig:GWsBPs} that the GW signals for all the BPs are well within the reach of the detectability threshold of BBO, DECIGO, and Ultimate-DECIGO. 

\begin{table}[t!]
\centering
\renewcommand{\arraystretch}{1.2}
\tabcolsep=0.1cm
\begin{tabular}{||c|c|c|c|c|c|c|c|c|c|c||}
\hline \hline
BPs & $v_D$ [TeV] & $M_{H_2}$ [GeV] &
$\sin \theta$ & 
$g_D$ [$10^{-4}$]  &
$\alpha$ & $\frac{\beta}{\mathcal{H}_*}$ & $T_n$ [GeV] &
$\frac{v_c}{T_c}$ &
$\frac{\Omega_{\rm WIMP}}{\Omega_{\rm Tot}}$ &
$\frac{\Omega_{\rm FIMP}}{\Omega_{\rm Tot}}$ 
\\ \hline
BP1 & 3.37 & 2.21 & 0.082 & 3.1 
& 0.238 & 13671 & 34.43 & 4.67 &
0.46 & 0.54 
\\ \hline
BP2 & 0.673& 2.77 & -0.076 & 19.7 
&0.139 & 6760.0 & 46.67 & 3.56 &
0.044 & 0.956
\\ \hline
BP3 & 4.63 & 1.0 & 0.060 & 1.0
& 0.461 & 13820 & 21.58 & 6.76 &
0.87 & 0.13\\
\hline \hline
\end{tabular}
\caption{Three BPs. The first four columns represent the input model parameters, the fifth, sixth, and the seventh columns are GW-related quantities, the eighth column shows the strength of the FOPT. The last two columns denote the WIMP and FIMP contributions to the total DM relic density $\Omega_{\rm Tot}h^2=0.12$; for the first BP, both the WIMP and FIMP equally contribute to the total DM relic density, while the second (third) BP is mostly composed of FIMP (WIMP) DM. In all the three cases, the LFV bounds are satisfied, and the neutrino masses can successfully be generated. The GW signals corresponding to the three BPs are shown in Fig.~\ref{fig:GWsBPs}.
}
\label{tab:BPs}
\end{table}

\section{Collider Searches}
\label{sec:collider}
The present work deals with the WIMP and FIMP-type DMs. Due to the feeble interaction of the FIMP, it is difficult to probe it at collider experiments.
We can, however, focus on general search strategies for BSM particles in the context of the present work. In particular, we may study the production of the second Higgs $H_2$ at the $pp$ or $e^{+}e^{-}$ colliders and look for its subsequent decay. Suitably adjusting the WIMP DM mass allows the second Higgs to decay mainly to the WIMP DM, and we may look for the missing energy with mono-jet or di-jet signals in the final state. Otherwise, if $H_2$ does not dominantly decay to the WIMP DM, then it will decay to the SM particles such as the SM Higgs. See, {\it e.g.}, Refs.~\cite{Banerjee:2015gca,Belanger:2021slj}. The relevant signal channels for our current study at the $pp$ collider would be
\begin{align}
pp \rightarrow p j + H_{2} &\rightarrow  p j + \cancel{E}_T\,\,(p \geq 1), \nonumber \\
&\rightarrow n j + m l\,\,(n,m \geq 1)\,,
\end{align}
where $j$ corresponds to the initial or final state jets, $l$ is associated with the SM lepton, and $\cancel{E}_{T}$ is the transverse missing energy.
Similar to the SM Higgs searches, we can also investigate 
\begin{align}
e^{+}e^{-} \rightarrow Z H_{2} &\rightarrow nj + pl\,\,(n, p\rightarrow 1)
\nonumber \\
&\rightarrow nj + pl + \cancel{E}_{T}\,.
\end{align}
at the $e^{+}e^{-}$ collider.
The exact values of the integers, $n$, $m$, and $p$, depend on the production cross section and dominance of the associated backgrounds.
Moreover, exploring the singlet fermions ($S_{L}, N_{L}$) of our model at different colliders is an interesting direction; see, {\it e.g.}, Ref.~\cite{Banerjee:2015gca}.
Further comments require a full-fledged collider study which is out of the scope of the current work, and we leave it for future study.

\section{Conclusion}
\label{sec:conc}
We have studied an extension of the Standard Model that accounts for the dark matter and the smallness of the neutrino masses under the extended seesaw framework. In our model, two sets of three-generation neutrinos are introduced; the first two generations provide the light neutrinos with a mass, and the third-generation neutrinos become FIMP-like particles. Amongst these FIMP-like particles, the heavier one eventually decays into the lighter one, and thus, we have the lighter third-generation neutrino as the FIMP dark matter candidate. Our model also contains a WIMP dark matter candidate, namely the dark $U(1)_D$ gauge boson. Thus, a two-component WIMP-FIMP dark matter scenario naturally arises in our model.

We have explored allowed parameter spaces by using the lepton flavour violating bounds as well as the neutrino oscillation data. Much of the parameter spaces are already tightly constrained, but we have shown that there are viable parameter regions which are free from the constraints. Prospects of various future experiments have been discussed as well. Interestingly, the contribution to the FIMP dark matter relic density coming from neutrinos scattering is found to be up to a $3\%$ of the total relic density for the range of the model parameters considered in our study.
We have also discussed the dependence of the relic density on the model parameters. Utilising publicly available tools, we have performed extensive numerical parameter scans in order to study the evolutions of the dark matter candidates. Parameter spaces compatible with the bounds from (in-)direct detection and collider searches are presented. In particular, we have showed regions where a two-component dark matter scenario is realised and testable by future (in-)direct experiments.

The dark $U(1)_D$ Higgs field plays a major role in the FIMP and WIMP dark matter productions. In addition, the extra scalar field also changes the evolution of the vacuum state in the scalar sector, making a first-order phase transition possible. We have demonstrated that the strength of the electroweak first-order phase transition, quantified by the quantity $v_c/T_c$, where $T_c$ is the critical temperature and $v_c$ is the SM Higgs vacuum expectation value at $T_c$, may become larger than unity for small values of the dark $U(1)_D$ Higgs mass.
Therefore, one of the essential ingredients for a successful electroweak baryogenesis is achieved in our model. We have also studied stochastic gravitational waves associated with the first-order phase transitions and showed that the gravitational wave signals are strong enough to be detectable by future experiments such as BBO and DECIGO.

Three benchmark points, that explicitly demonstrate the capability of {\it i)} having a correct dark matter relic density, {\it ii)} generating the non-zero neutrino masses with the extended seesaw mechanism, {\it iii)} achieving a strongly first-order phase transition, and {\it iv)} emitting stochastic gravitational waves detectable by future experiments, are presented. Thus, the model studied in this work has an exciting potential detectability not only with future (in-)direct detection experiments and collider searches, but also with future gravitational wave experiments.

\acknowledgments
J.K. would like to thank Yikun Wang for useful discussions on the phase transition and the use of {\tt CosmoTransitions}.
S.K. would like to acknowledge Genevi\`{e}ve B\'{e}langer for the help related with {\tt micrOMEGAs}.
The work of F.C. is supported by the European Union’s Horizon 2020 research and innovation programme under the Marie Sk\l{}odowska-Curie grant agreement No 860881-HIDDeN. This work used the Scientific Compute Cluster at GWDG, the joint data center of Max Planck Society for the Advancement of Science (MPG) and University of G\"ottingen.

\appendix

\section{WIMP DM decay width through kinetic mixing}
\label{apdx:DM-decay-width}
In the presence of the mixing between the WIMP DM and the $U(1)_Y$ gauge boson, the DM may decay to, {\it e.g.}, electrons, through the coupling between the DM and SM fermions \cite{Biswas:2021dan}. For simplicity, we consider the decay of the DM to electrons. The decay width is then given by
\begin{align}
\Gamma_{W_{D} \rightarrow ee} = \frac{M_{W_{D}} g^2_{W_{D}ee}}{12 \pi}
\left(1 + \frac{2 m^2_{e}}{M^2_{W_D}} \right) 
\sqrt{1 - \frac{4 m^2_{e}}{M^2_{W_D}}}\,,
\end{align}
where $g_{W_{D}ee} = 3 e \zeta/(4 \cos\theta_w)$, $e = \sqrt{4 \pi \alpha}$, $\alpha$ is the fine-structure constant, $\theta_w$ is the weak angle, and $\zeta$ is the gauge kinetic mixing parameter introduced in \eqref{eqn:lag}. Considering the DM mass of 10 GeV, and requiring the lifetime of the DM to be is larger than the age of the universe, we get an upper bound on the gauge kinetic mixing parameter as $\zeta < 10^{-20}$.
When the decay of the DM to the SM fermions is open, the $\gamma$-ray observation may become relevant \cite{Fermi-LAT:2015kyq}. In this case, the DM lifetime should be greater than $10^{29}$s \cite{Fermi-LAT:2015kyq}, which puts an even stronger bound of $\zeta < 10^{-26}$.

\section{Quartic couplings}
\label{apdx:quarticcouplings}
The scalar quartic couplings may be written in terms of the mixing angle and masses of the physical Higgses as follows:
\begin{align}
\lambda_{h} &= \frac{M^2_{H_1}\cos^2\theta + M^2_{H_2}\sin^2\theta}{2 v^2}\,, \nonumber \\
 \lambda_{D} &= \frac{M^2_{H_1}\sin^2\theta + M^2_{H_2}\cos^2\theta}{2 v_D^2}\,,\\
\lambda_{hD} &= \frac{(M^2_{H_1} - M^2_{H_2})
\sin2\theta}{2 v v_D}\,.\nonumber
\end{align} 

\section{Feynman Diagrams}
\label{apdx:FeynmanDiagrams}
\begin{figure}
\centering
\includegraphics[scale=0.5]{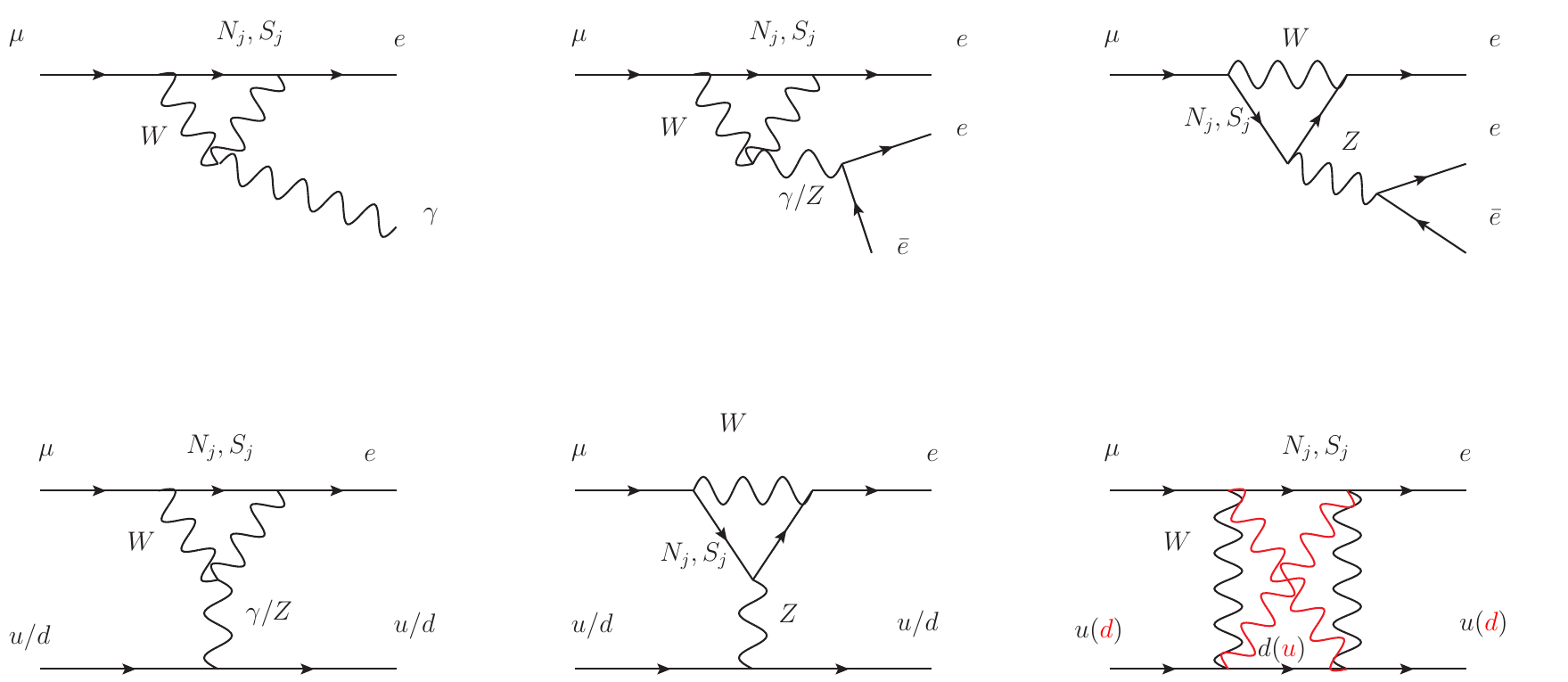}
\caption{Feynman diagrams for LFV processes.}
\label{fig:lfv}
\end{figure}

\begin{figure}
\centering
\includegraphics[scale=0.5]{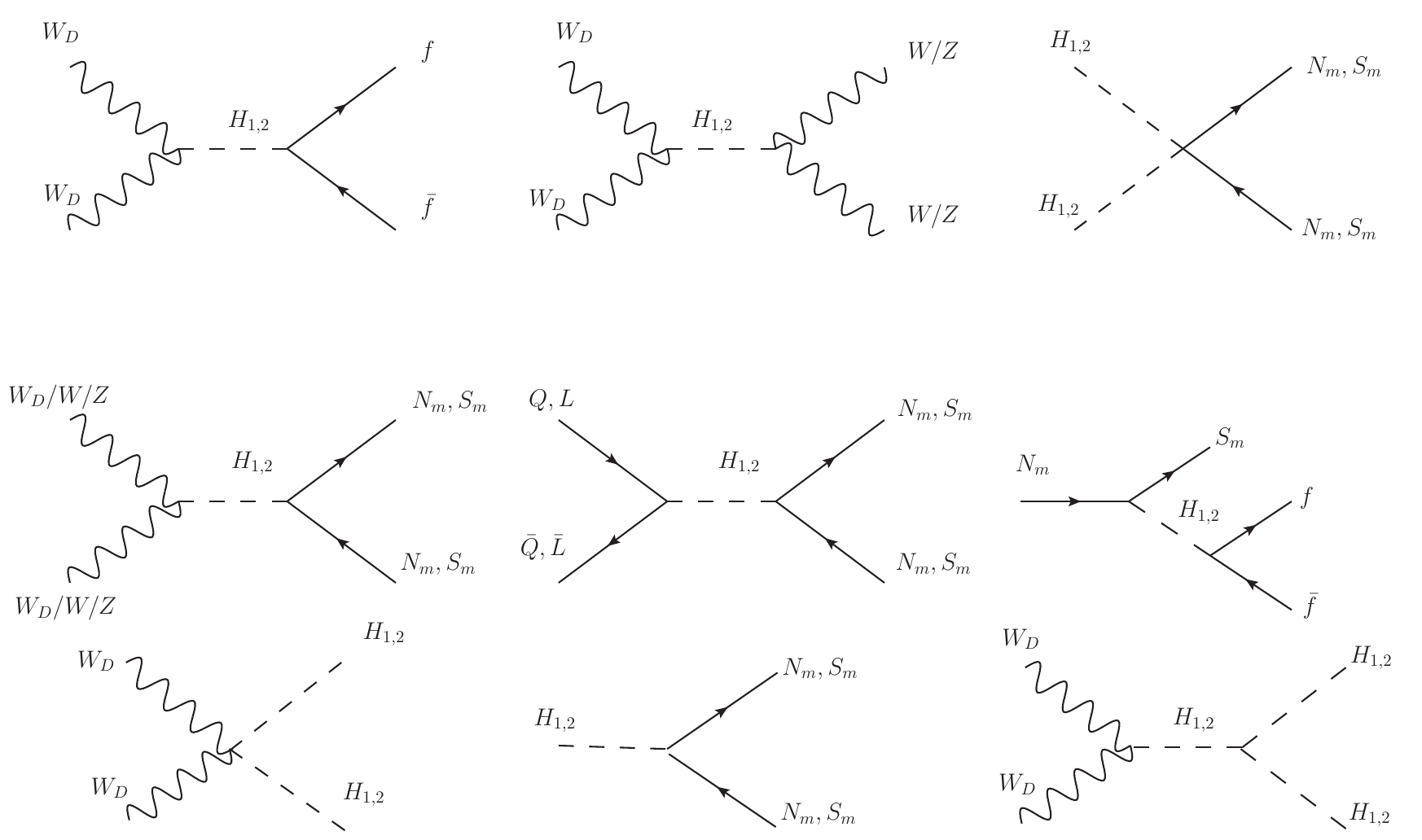}
\caption{Feynman diagrams relevant for the DM analysis.}
\label{fig:dm-feynman}
\end{figure}

Figure~\ref{fig:lfv} shows the diagrams which contribute to the processes $\mu \rightarrow e \gamma$, $\mu \rightarrow e e \bar{e}$, and $\mu$-to-$e$ conversion. We have considered these diagrams for the discussion of the LFV bounds.
The Feynman diagrams relevant for our DM analysis are shown in Fig.~\ref{fig:dm-feynman}.

\section{DM annihilation to $\tau^{+}\tau^{-}$ and $\mu^{+}\mu^{-}$}
\label{apdx:mumutautaubounds}
\begin{figure}[t!]
\centering
\includegraphics[scale=0.49]{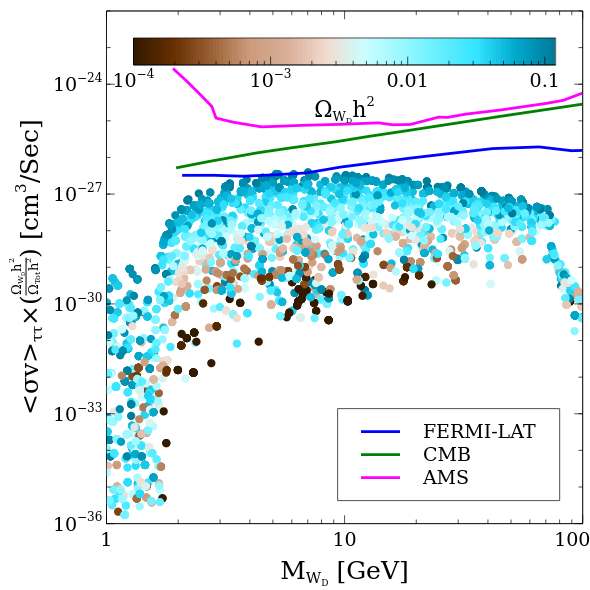}
\includegraphics[scale=0.49]{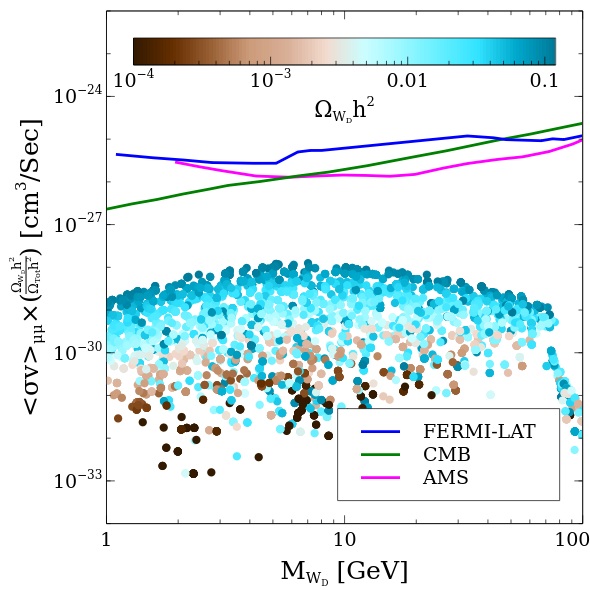}
\caption{
Allowed parameter space satisfying $0.01 \leq \Omega_{\rm DM}h^2 \leq 0.12$ in the $M_{W_D}$ -- $(\Omega_{W_D}/\Omega_{\rm Tot})\langle\sigma v\rangle_{\tau\tau}$ (left) and $M_{W_D}$ -- $(\Omega_{W_D}/\Omega_{\rm Tot})\langle\sigma v\rangle_{\mu\mu}$ (right) planes. Here, $\Omega_{\rm Tot}h^2 = 0.12$ is total DM relic density today. Various indirect detection bounds are overlaid with solid lines; see text for detailed explanation. The colour of the points represents the WIMP DM relic density.
} 
\label{fig:DM-scatter-plot-4}
\end{figure}
The LP and RP of Fig.~\ref{fig:DM-scatter-plot-4} present the DM annihilation to $\tau^{+}\tau^{-}$ and $\mu^{+}\mu^{-}$, respectively, together with the bounds from FERMI-LAT~\cite{Fermi-LAT:2015att, Leane:2018kjk}, CMB~\cite{Leane:2018kjk}, and AMS~\cite{Bergstrom:2013jra,Leane:2018kjk} data. 
We find that our model predicts $\langle \sigma v \rangle$ orders of magnitude lower than the current bound. We expect that our model parameter space may be explored in future by different ongoing indirect detection experiments. Finally, we note that the DM annihilation to $e^{+}e^{-}$ is many orders below than the current bound as well.

\section{Field-dependent masses}
\label{apdx:CWmass}
We summarise the expressions for the field-dependent masses $M_i$ as well as the number of degrees of freedom $n_i$ that appear in the one-loop Coleman-Weinberg potential \eqref{eqn:CWpot}:
\begin{align*}
M_{H,H_D}^2 &= 
\left(\begin{array}{cc}
\mu_h^2 + 3\lambda_h \bar{H}^2 + \frac{1}{2}\lambda_{hD}\bar{H}_D^2 &
\lambda_{hD}\bar{H}\bar{H}_D \\
\lambda_{hD}\bar{H}\bar{H}_D &
\mu_D^2 + 3\lambda_D \bar{H}_D^2 + \frac{1}{2}\lambda_{hD}\bar{H}^2
\end{array}\right)
\,,\\
M_{W^{3},B}^2 &= 
\left(\begin{array}{cc}
\frac{1}{4}g_2^2\bar{H}^2 & \frac{1}{4}g_1 g_2\bar{H}^2 \\
\frac{1}{4}g_1 g_2\bar{H}^2 & \frac{1}{4}g_1^2\bar{H}^2
\end{array}\right)
\,,\\
M_{H^\pm,H^0}^2 &= \mu_h^2 + \lambda_h \bar{H}^2 + \frac{1}{2}\lambda_{hD}\bar{H}_D^2
\,,\qquad
M_{H_D^0}^2 = \mu_D^2 + \lambda_D \bar{H}_D^2 + \frac{1}{2}\lambda_{hD}\bar{H}^2
\,,\\
M_{W^{1,2}} &= \frac{1}{4}g_2^2\bar{H}^2
\,,\qquad
M_{Z_D}^2 = g_D^2\bar{H}_D^2
\,,\qquad
M_t^2 = \frac{1}{2}y_t^2\bar{H}^2
\,,
\end{align*}
and
\begin{gather*}
n_{H} = n_{H_D} = n_{H^\pm} = n_{H^0} = n_{H_D^0} = 1 \,,\quad
n_{W^{1,2,3}} = 6 \,,\quad
n_B = 3 \,,\quad
n_t = 12 \,,\quad
n_{Z_D} = 3
\,.
\end{gather*}
Note that we have considered only the most dominant SM top quark for the fermionic states.

\section{Thermal masses}
\label{apdx:THERMmass}
We summarise the expressions for the thermal masses $\Pi_i$ that enter the one-loop temperature-dependent potential \eqref{eqn:TEMPpot}:
\begin{align*}
\Pi_H &= \Pi_{H^\pm,H^0} = \frac{T^2}{48}\left(
3g_1^2 + 9g_2^2 + 12y_t^2 + 24\lambda_h + 4\lambda_{hD}
\right)
\,,\\
\Pi_{H_D} &= \Pi_{H_D^0} = \frac{T^2}{12}\left(
3g_D^2 + 4\lambda_D + 2\lambda_{hD}
\right)
\,,\\
\Pi_{W_L^{1,2,3}} &= \frac{11}{6}g_2^2T^2
\,,\qquad
\Pi_{B_L} = \frac{11}{6}g_1^2T^2
\,,\qquad
\Pi_{Z_{D L}} = \frac{1}{3}g_D^2T^2
\,,
\end{align*}
where we have considered only the most dominant SM top quark for the fermionic states.
Note that fermions and transverse modes of the gauge bosons do not receive any thermal correction.

\section{Gravitational wave-related expressions}
\label{apdx:GWexpr}
The quantities $f_{\rm col}$, $f_{\rm sw}$, and $f_{\rm turb}$ that appear in $\Omega_{\rm col}h^2$, $\Omega_{\rm sw}h^2$, and $\Omega_{\rm turb}h^2$ are given as follows \cite{Caprini:2015zlo}:
\begin{align*}
f_{\rm col} &=
1.65 \times 10^{-5} \, {\rm Hz}
\left(\frac{0.62}{1.8-0.1 v_{w}+v_{w}^{2}}\right)
\left(\frac{\beta}{\mathcal{H}_{*}}\right)
\left(\frac{T_{*}}{100 {\rm GeV}}\right)
\left(\frac{g_{*}}{100}\right)^{\frac{1}{6}}
\,,\\
f_{\rm sw} &=
1.9 \times 10^{-5} \, {\rm Hz} 
\left(\frac{1}{v_{w}}\right)
\left(\frac{\beta}{\mathcal{H}_{*}}\right)
\left(\frac{T_{*}}{100 {\rm GeV}}\right)
\left(\frac{g_{*}}{100}\right)^{\frac{1}{6}}
\,,\\
f_{\rm turb} &=
2.7 \times 10^{-5} \, {\rm Hz} 
\left(\frac{1}{v_{w}}\right)
\left(\frac{\beta}{\mathcal{H}_{*}}\right)
\left(\frac{T_{*}}{100 {\rm GeV}}\right)
\left(\frac{g_{*}}{100}\right)^{\frac{1}{6}}
\,.
\end{align*}
The bubble wall velocity $v_w$ is given by \cite{Steinhardt:1981ct}
\begin{align*}
v_{w}=\frac{\sqrt{1 / 3}+\sqrt{\alpha^{2}+2 \alpha / 3}}{1+\alpha}
\,,
\end{align*}
and we adopt \cite{Kamionkowski:1993fg}
\begin{align*}
\kappa=
\frac{0.715 \alpha+(4 / 27) \sqrt{3 \alpha / 2}}{1+0.715 \alpha}
\,, \qquad 
\kappa_{v}=
\frac{\alpha}{0.73+0.083 \sqrt{\alpha}+\alpha}
\,,\qquad
\kappa_{\rm turb} = 0.1 \kappa_{v}
\,,
\end{align*}


\bibliographystyle{JHEP}
\input{U1DWFVFESPTGW-arXiv-version2.bbl}

\end{document}

%% file: U1DWFVFESPTGW-arXiv-version2.bbl
\providecommand{\href}[2]{#2}\begingroup\raggedright\endgroup

%% file: U1DWFVFESPTGW-arXiv-version2.bbl
\begin{thebibliography}{100}

\bibitem{Super-Kamiokande:1998kpq}
T.~S.-K. Collaboration and Y.~F. et~{al}, \emph{Evidence for oscillation of
  atmospheric neutrinos},
  \href{https://doi.org/10.1103/PhysRevLett.81.1562}{\emph{Physical Review
  Letters} {\bfseries 81} (1998) 1562}
  [\href{https://arxiv.org/abs/hep-ex/9807003}{{\ttfamily hep-ex/9807003}}].

\bibitem{Gonzalez-Garcia:2002bkq}
M.~C. {Gonzalez-Garcia} and Y.~Nir, \emph{Neutrino {{Masses}} and {{Mixing}}:
  {{Evidence}} and {{Implications}}},
  \href{https://doi.org/10.1103/RevModPhys.75.345}{\emph{Reviews of Modern
  Physics} {\bfseries 75} (2003) 345}
  [\href{https://arxiv.org/abs/hep-ph/0202058}{{\ttfamily hep-ph/0202058}}].

\bibitem{Esteban:2020cvm}
I.~Esteban, M.~C. {Gonzalez-Garcia}, M.~Maltoni, T.~Schwetz and A.~Zhou,
  \emph{The fate of hints: Updated global analysis of three-flavor neutrino
  oscillations}, \href{https://doi.org/10.1007/JHEP09(2020)178}{\emph{Journal
  of High Energy Physics} {\bfseries 2020} (2020) 178}
  [\href{https://arxiv.org/abs/2007.14792}{{\ttfamily 2007.14792}}].

\bibitem{Planck:2015fie}
{\scshape Planck} collaboration, \emph{Planck 2015 results. {{XIII}}.
  {{Cosmological}} parameters},
  \href{https://doi.org/10.1051/0004-6361/201525830}{\emph{Astronomy \&
  Astrophysics} {\bfseries 594} (2016) A13}
  [\href{https://arxiv.org/abs/1502.01589}{{\ttfamily 1502.01589}}].

\bibitem{Planck:2018vyg}
{\scshape Planck} collaboration, \emph{Planck 2018 results. {{VI}}.
  {{Cosmological}} parameters},
  \href{https://doi.org/10.1051/0004-6361/201833910}{\emph{Astronomy \&
  Astrophysics} {\bfseries 641} (2020) A6}
  [\href{https://arxiv.org/abs/1807.06209}{{\ttfamily 1807.06209}}].

\bibitem{Minkowski:1977sc}
P.~Minkowski, \emph{{{$\mu\rightarrow$e$\gamma$}} at a rate of one out of
  10{$^9$} muon decays?},
  \href{https://doi.org/10.1016/0370-2693(77)90435-X}{\emph{Physics Letters B}
  {\bfseries 67} (1977) 421}.

\bibitem{Gell-Mann:1979vob}
M.~{Gell-Mann}, P.~Ramond and R.~Slansky, \emph{Complex {{Spinors}} and
  {{Unified Theories}}},  \href{https://arxiv.org/abs/1306.4669}{{\ttfamily
  1306.4669}}.

\bibitem{Kang:2006sn}
S.~K. Kang and C.~S. Kim, \emph{Extended double seesaw model for neutrino mass
  spectrum and low scale leptogenesis},
  \href{https://doi.org/10.1016/j.physletb.2006.12.071}{\emph{Physics Letters
  B} {\bfseries 646} (2007) 248}
  [\href{https://arxiv.org/abs/hep-ph/0607072}{{\ttfamily hep-ph/0607072}}].

\bibitem{Mitra:2011qr}
M.~Mitra, G.~Senjanovic and F.~Vissani, \emph{Neutrinoless {{Double Beta
  Decay}} and {{Heavy Sterile Neutrinos}}},
  \href{https://doi.org/10.1016/j.nuclphysb.2011.10.035}{\emph{Nuclear Physics
  B} {\bfseries 856} (2012) 26}
  [\href{https://arxiv.org/abs/1108.0004}{{\ttfamily 1108.0004}}].

\bibitem{Majee:2008mn}
S.~K. Majee, M.~K. Parida and A.~Raychaudhuri, \emph{Neutrino mass and
  low-scale leptogenesis in a testable {{SUSY SO}}(10) model},
  \href{https://arxiv.org/abs/0807.3959}{{\ttfamily 0807.3959}}.

\bibitem{Kawasaki:2004qu}
M.~Kawasaki, K.~Kohri and T.~Moroi, \emph{Big-{{Bang}} nucleosynthesis and
  hadronic decay of long-lived massive particles},
  \href{https://doi.org/10.1103/PhysRevD.71.083502}{\emph{Physical Review D:
  Particles and Fields} {\bfseries 71} (2005) 083502}
  [\href{https://arxiv.org/abs/astro-ph/0408426}{{\ttfamily
  astro-ph/0408426}}].

\bibitem{Hook:2018sai}
A.~Hook, R.~McGehee and H.~Murayama, \emph{Cosmologically viable low-energy
  supersymmetry breaking},
  \href{https://doi.org/10.1103/PhysRevD.98.115036}{\emph{Physical Review D:
  Particles and Fields} {\bfseries 98} (2018) 115036}
  [\href{https://arxiv.org/abs/1801.10160}{{\ttfamily 1801.10160}}].

\bibitem{Ostriker:1973uit}
J.~P. Ostriker and P.~J.~E. Peebles, \emph{A {{Numerical Study}} of the
  {{Stability}} of {{Flattened Galaxies}}: Or, can {{Cold Galaxies Survive}}?},
  \href{https://doi.org/10.1086/152513}{\emph{The Astrophysical Journal}
  {\bfseries 186} (1973) 467}.

\bibitem{Corbelli:1999af}
E.~Corbelli and P.~Salucci, \emph{The {{Extended Rotation Curve}} and the
  {{Dark Matter Halo}} of {{M33}}},
  \href{https://doi.org/10.1046/j.1365-8711.2000.03075.x}{\emph{Monthly Notices
  of the Royal Astronomical Society} {\bfseries 311} (2000) 441}
  [\href{https://arxiv.org/abs/astro-ph/9909252}{{\ttfamily
  astro-ph/9909252}}].

\bibitem{Gunn:1978gr}
J.~E. Gunn, B.~W. Lee, I.~Lerche, D.~N. Schramm and G.~Steigman, \emph{Some
  astrophysical consequences of the existence of a heavy stable neutral
  lepton.}, \href{https://doi.org/10.1086/156335}{\emph{The Astrophysical
  Journal} {\bfseries 223} (1978) 1015}.

\bibitem{Hut:1977zn}
P.~Hut, \emph{Limits on masses and number of neutral weakly interacting
  particles}, \href{https://doi.org/10.1016/0370-2693(77)90139-3}{\emph{Physics
  Letters B} {\bfseries 69} (1977) 85}.

\bibitem{Lee:1977ua}
B.~W. Lee and S.~Weinberg, \emph{Cosmological {{Lower Bound}} on
  {{Heavy-Neutrino Masses}}},
  \href{https://doi.org/10.1103/PhysRevLett.39.165}{\emph{Physical Review
  Letters} {\bfseries 39} (1977) 165}.

\bibitem{Bertone:2004pz}
G.~Bertone, D.~Hooper and J.~Silk, \emph{Particle dark matter: Evidence,
  candidates and constraints},
  \href{https://doi.org/10.1016/j.physrep.2004.08.031}{\emph{Physics Reports}
  {\bfseries 405} (2005) 279}
  [\href{https://arxiv.org/abs/hep-ph/0404175}{{\ttfamily hep-ph/0404175}}].

\bibitem{XENON:2018voc}
{\scshape XENON} collaboration, \emph{Dark {{Matter Search Results}} from a
  {{One Ton-Year Exposure}} of {{XENON1T}}},
  \href{https://doi.org/10.1103/PhysRevLett.121.111302}{\emph{Physical Review
  Letters} {\bfseries 121} (2018) 111302}
  [\href{https://arxiv.org/abs/1805.12562}{{\ttfamily 1805.12562}}].

\bibitem{CMS:2016lcl}
{\scshape CMS} collaboration, \emph{Phenomenological {{MSSM}} interpretation of
  {{CMS}} searches in pp collisions at sqrt(s) = 7 and 8 {{TeV}}},
  \href{https://doi.org/10.1007/JHEP10(2016)129}{\emph{Journal of High Energy
  Physics} {\bfseries 10} (2016) 129}
  [\href{https://arxiv.org/abs/1606.03577}{{\ttfamily 1606.03577}}].

\bibitem{MAGIC:2016xys}
{\scshape MAGIC, Fermi-LAT} collaboration, \emph{Limits to dark matter
  annihilation cross-section from a combined analysis of {{MAGIC}} and
  {{Fermi-LAT}} observations of dwarf satellite galaxies},
  \href{https://doi.org/10.1088/1475-7516/2016/02/039}{\emph{Journal of
  Cosmology and Astroparticle Physics} {\bfseries 02} (2016) 039}
  [\href{https://arxiv.org/abs/1601.06590}{{\ttfamily 1601.06590}}].

\bibitem{Arcadi:2017kky}
G.~Arcadi, M.~Dutra, P.~Ghosh, M.~Lindner, Y.~Mambrini, M.~Pierre et~al.,
  \emph{The waning of the {{WIMP}}? {{A}} review of models, searches, and
  constraints},
  \href{https://doi.org/10.1140/epjc/s10052-018-5662-y}{\emph{European Physical
  Journal C} {\bfseries 78} (2018) 203}
  [\href{https://arxiv.org/abs/1703.07364}{{\ttfamily 1703.07364}}].

\bibitem{PandaX-II:2016vec}
{\scshape PandaX-II} collaboration, \emph{Dark {{Matter Results}} from
  {{First}} 98.7 {{Days}} of {{Data}} from the {{PandaX-II Experiment}}},
  \href{https://doi.org/10.1103/PhysRevLett.117.121303}{\emph{Physical Review
  Letters} {\bfseries 117} (2016) 121303}
  [\href{https://arxiv.org/abs/1607.07400}{{\ttfamily 1607.07400}}].

\bibitem{LUX:2016ggv}
{\scshape LUX} collaboration, \emph{Results from a {{Search}} for {{Dark
  Matter}} in the {{Complete LUX Exposure}}},
  \href{https://doi.org/10.1103/PhysRevLett.118.021303}{\emph{Physical Review
  Letters} {\bfseries 118} (2017) 021303}
  [\href{https://arxiv.org/abs/1608.07648}{{\ttfamily 1608.07648}}].

\bibitem{McDonald:2001vt}
J.~McDonald, \emph{Thermally {{Generated Gauge Singlet Scalars}} as
  {{Self-Interacting Dark Matter}}},
  \href{https://doi.org/10.1103/PhysRevLett.88.091304}{\emph{Physical Review
  Letters} {\bfseries 88} (2002) 091304}
  [\href{https://arxiv.org/abs/hep-ph/0106249}{{\ttfamily hep-ph/0106249}}].

\bibitem{Choi:2005vq}
K.-Y. Choi and L.~Roszkowski, \emph{E-{{WIMPs}}},
  \href{https://arxiv.org/abs/hep-ph/0511003}{{\ttfamily hep-ph/0511003}}.

\bibitem{Kusenko:2006rh}
A.~Kusenko, \emph{Sterile {{Neutrinos}}, {{Dark Matter}}, and {{Pulsar
  Velocities}} in {{Models}} with a {{Higgs Singlet}}},
  \href{https://doi.org/10.1103/PhysRevLett.97.241301}{\emph{Physical Review
  Letters} {\bfseries 97} (2006) 241301}
  [\href{https://arxiv.org/abs/hep-ph/0609081}{{\ttfamily hep-ph/0609081}}].

\bibitem{Hall:2009bx}
L.~J. Hall, K.~Jedamzik, J.~{March-Russell} and S.~M. West, \emph{Freeze-in
  production of {{FIMP}} dark matter},
  \href{https://doi.org/10.1007/JHEP03(2010)080}{\emph{Journal of High Energy
  Physics} {\bfseries 03} (2010) 080}
  [\href{https://arxiv.org/abs/0911.1120}{{\ttfamily 0911.1120}}].

\bibitem{Cheung:2011nn}
C.~Cheung, G.~Elor and L.~Hall, \emph{Gravitino freeze-in},
  \href{https://doi.org/10.1103/PhysRevD.84.115021}{\emph{Physical Review D}
  {\bfseries 84} (2011) 115021}
  [\href{https://arxiv.org/abs/1103.4394}{{\ttfamily 1103.4394}}].

\bibitem{Elahi:2014fsa}
F.~Elahi, C.~Kolda and J.~Unwin, \emph{{{UltraViolet}} freeze-in},
  \href{https://doi.org/10.1007/JHEP03(2015)048}{\emph{Journal of High Energy
  Physics} {\bfseries 03} (2015) 048}
  [\href{https://arxiv.org/abs/1410.6157}{{\ttfamily 1410.6157}}].

\bibitem{Arcadi:2015ffa}
G.~Arcadi, L.~Covi and M.~Nardecchia, \emph{Gravitino dark matter and low-scale
  baryogenesis},
  \href{https://doi.org/10.1103/PhysRevD.92.115006}{\emph{Physical Review D}
  {\bfseries 92} (2015) 115006}
  [\href{https://arxiv.org/abs/1507.05584}{{\ttfamily 1507.05584}}].

\bibitem{Bernal:2017kxu}
N.~Bernal, M.~Heikinheimo, T.~Tenkanen, K.~Tuominen and V.~Vaskonen, \emph{The
  dawn of {{FIMP Dark Matter}}: {{A}} review of models and constraints},
  \href{https://doi.org/10.1142/S0217751X1730023X}{\emph{International Journal
  of Modern Physics A} {\bfseries 32} (2017) 1730023}
  [\href{https://arxiv.org/abs/1706.07442}{{\ttfamily 1706.07442}}].

\bibitem{Benakli:2017whb}
K.~Benakli, Y.~Chen, E.~Dudas and Y.~Mambrini, \emph{Minimal model of gravitino
  dark matter},
  \href{https://doi.org/10.1103/PhysRevD.95.095002}{\emph{Physical Review D}
  {\bfseries 95} (2017) 095002}
  [\href{https://arxiv.org/abs/1701.06574}{{\ttfamily 1701.06574}}].

\bibitem{Bernal:2018qlk}
N.~Bernal, M.~Dutra, Y.~Mambrini, K.~Olive, M.~Peloso and M.~Pierre,
  \emph{Spin-2 portal dark matter},
  \href{https://doi.org/10.1103/PhysRevD.97.115020}{\emph{Physical Review D}
  {\bfseries 97} (2018) 115020}
  [\href{https://arxiv.org/abs/1803.01866}{{\ttfamily 1803.01866}}].

\bibitem{Bernal:2019mhf}
N.~Bernal, F.~Elahi, C.~Maldonado and J.~Unwin, \emph{Ultraviolet {{Freeze-in}}
  and {{Non-Standard Cosmologies}}},
  \href{https://doi.org/10.1088/1475-7516/2019/11/026}{\emph{Journal of
  Cosmology and Astroparticle Physics} {\bfseries 11} (2019) 026}
  [\href{https://arxiv.org/abs/1909.07992}{{\ttfamily 1909.07992}}].

\bibitem{Barman:2019lvm}
B.~Barman, S.~Bhattacharya and M.~Zakeri, \emph{Non-abelian vector boson as
  {{FIMP}} dark matter},
  \href{https://doi.org/10.1088/1475-7516/2020/02/029}{\emph{JCAP} {\bfseries
  02} (2020) 029} [\href{https://arxiv.org/abs/1905.07236}{{\ttfamily
  1905.07236}}].

\bibitem{Covi:2020pch}
L.~Covi, A.~Ghosh, T.~Mondal and B.~Mukhopadhyaya, \emph{Models of decaying
  {{FIMP Dark Matter}}: Potential links with the {{Neutrino Sector}}},
  \href{https://arxiv.org/abs/2008.12550}{{\ttfamily 2008.12550}}.

\bibitem{Khan:2020pso}
S.~Khan, \emph{Explaining {{Xenon-1T}} signal with {{FIMP}} dark matter and
  neutrino mass in a {{U}}(1){{{\textsubscript{X}}}} extension},
  \href{https://doi.org/10.1140/epjc/s10052-021-09397-x}{\emph{The European
  Physical Journal C: Particles and Fields} {\bfseries 81} (2021) 598}
  [\href{https://arxiv.org/abs/2007.13008}{{\ttfamily 2007.13008}}].

\bibitem{Garcia:2020hyo}
M.~A.~G. Garcia, Y.~Mambrini, K.~A. Olive and S.~Verner, \emph{The case for
  decaying spin-3/2 dark matter},
  \href{https://doi.org/10.1103/PhysRevD.102.083533}{\emph{Physical Review D}
  {\bfseries 102} (2020) 083533}
  [\href{https://arxiv.org/abs/2006.03325}{{\ttfamily 2006.03325}}].

\bibitem{Bernal:2020qyu}
N.~Bernal, J.~Rubio and H.~Veerm{\"a}e, \emph{{{UV Freeze-in}} in {{Starobinsky
  Inflation}}},
  \href{https://doi.org/10.1088/1475-7516/2020/10/021}{\emph{Journal of
  Cosmology and Astroparticle Physics} {\bfseries 10} (2020) 021}
  [\href{https://arxiv.org/abs/2006.02442}{{\ttfamily 2006.02442}}].

\bibitem{Barman:2020plp}
B.~Barman, D.~Borah and R.~Roshan, \emph{Effective theory of freeze-in dark
  matter}, \href{https://doi.org/10.1088/1475-7516/2020/11/021}{\emph{JCAP}
  {\bfseries 11} (2020) 021}
  [\href{https://arxiv.org/abs/2007.08768}{{\ttfamily 2007.08768}}].

\bibitem{Barman:2020ifq}
B.~Barman, S.~Bhattacharya and B.~Grzadkowski, \emph{Feebly coupled vector
  boson dark matter in effective theory},
  \href{https://doi.org/10.1007/JHEP12(2020)162}{\emph{JHEP} {\bfseries 12}
  (2020) 162} [\href{https://arxiv.org/abs/2009.07438}{{\ttfamily
  2009.07438}}].

\bibitem{Barman:2021yaz}
B.~Barman, P.~Ghosh, A.~Ghoshal and L.~Mukherjee, \emph{Shedding flavor on dark
  via freeze-in: {{U}}(1){{{\textsubscript{B-3L i}}}} gauged extensions},
  \href{https://doi.org/10.1088/1475-7516/2022/08/049}{\emph{JCAP} {\bfseries
  08} (2022) 049} [\href{https://arxiv.org/abs/2112.12798}{{\ttfamily
  2112.12798}}].

\bibitem{Barman:2021lot}
B.~Barman and A.~Ghoshal, \emph{Scale invariant {{FIMP}} miracle},
  \href{https://doi.org/10.1088/1475-7516/2022/03/003}{\emph{JCAP} {\bfseries
  03} (2022) 003} [\href{https://arxiv.org/abs/2109.03259}{{\ttfamily
  2109.03259}}].

\bibitem{Belanger:2021slj}
G.~B{\'e}langer, S.~Khan, R.~Padhan, M.~Mitra and S.~Shil, \emph{Right handed
  neutrinos, {{TeV}} scale {{BSM}} neutral {{Higgs}} boson, and {{FIMP}} dark
  matter in an {{EFT}} framework},
  \href{https://doi.org/10.1103/PhysRevD.104.055047}{\emph{Physical Review D:
  Particles and Fields} {\bfseries 104} (2021) 055047}
  [\href{https://arxiv.org/abs/2104.04373}{{\ttfamily 2104.04373}}].

\bibitem{Barman:2022njh}
B.~Barman and A.~Ghoshal, \emph{Probing pre-{{BBN}} era with scale invarint
  {{FIMP}}},  \href{https://arxiv.org/abs/2203.13269}{{\ttfamily 2203.13269}}.

\bibitem{Choi:2015fiu}
K.~Choi and S.~H. Im, \emph{Realizing the relaxion from multiple axions and its
  {{UV}} completion with high scale supersymmetry},
  \href{https://doi.org/10.1007/JHEP01(2016)149}{\emph{JHEP} {\bfseries 01}
  (2016) 149} [\href{https://arxiv.org/abs/1511.00132}{{\ttfamily
  1511.00132}}].

\bibitem{Kaplan:2015fuy}
D.~E. Kaplan and R.~Rattazzi, \emph{Large field excursions and approximate
  discrete symmetries from a clockwork axion},
  \href{https://doi.org/10.1103/PhysRevD.93.085007}{\emph{Physical Review D:
  Particles and Fields} {\bfseries 93} (2016) 085007}
  [\href{https://arxiv.org/abs/1511.01827}{{\ttfamily 1511.01827}}].

\bibitem{Giudice:2016yja}
G.~F. Giudice and M.~McCullough, \emph{A clockwork theory},
  \href{https://doi.org/10.1007/JHEP02(2017)036}{\emph{JHEP} {\bfseries 02}
  (2017) 036} [\href{https://arxiv.org/abs/1610.07962}{{\ttfamily
  1610.07962}}].

\bibitem{Kim:2017mtc}
J.~Kim and J.~McDonald, \emph{Clockwork {{Higgs}} portal model for freeze-in
  dark matter},
  \href{https://doi.org/10.1103/PhysRevD.98.023533}{\emph{Physical Review D:
  Particles and Fields} {\bfseries 98} (2018) 023533}
  [\href{https://arxiv.org/abs/1709.04105}{{\ttfamily 1709.04105}}].

\bibitem{Kim:2018xsp}
J.~Kim and J.~Mcdonald, \emph{Freeze-in dark matter from a sub-{{Higgs}} mass
  clockwork sector via the higgs portal},
  \href{https://doi.org/10.1103/PhysRevD.98.123503}{\emph{Physical Review D:
  Particles and Fields} {\bfseries 98} (2018) 123503}
  [\href{https://arxiv.org/abs/1804.02661}{{\ttfamily 1804.02661}}].

\bibitem{Goudelis:2018xqi}
A.~Goudelis, K.~A. Mohan and D.~Sengupta, \emph{Clockworking {{FIMPs}}},
  \href{https://doi.org/10.1007/JHEP10(2018)014}{\emph{JHEP} {\bfseries 10}
  (2018) 014} [\href{https://arxiv.org/abs/1807.06642}{{\ttfamily
  1807.06642}}].

\bibitem{Zurek:2008qg}
K.~M. Zurek, \emph{Multi-component dark matter},
  \href{https://doi.org/10.1103/PhysRevD.79.115002}{\emph{Physical Review D:
  Particles and Fields} {\bfseries 79} (2009) 115002}
  [\href{https://arxiv.org/abs/0811.4429}{{\ttfamily 0811.4429}}].

\bibitem{Profumo:2009tb}
S.~Profumo, K.~Sigurdson and L.~Ubaldi, \emph{Can we discover multi-component
  {{WIMP}} dark matter?},
  \href{https://doi.org/10.1088/1475-7516/2009/12/016}{\emph{JCAP} {\bfseries
  12} (2009) 016} [\href{https://arxiv.org/abs/0907.4374}{{\ttfamily
  0907.4374}}].

\bibitem{Feldman:2010wy}
D.~Feldman, Z.~Liu, P.~Nath and G.~Peim, \emph{Multicomponent dark matter in
  supersymmetric hidden sector extensions},
  \href{https://doi.org/10.1103/PhysRevD.81.095017}{\emph{Physical Review D:
  Particles and Fields} {\bfseries 81} (2010) 095017}
  [\href{https://arxiv.org/abs/1004.0649}{{\ttfamily 1004.0649}}].

\bibitem{Ko:2010at}
P.~Ko and Y.~Omura, \emph{Supersymmetric {{U}}(1){{B X U}}(1){{L}} model with
  leptophilic and leptophobic cold dark matters},
  \href{https://doi.org/10.1016/j.physletb.2011.06.009}{\emph{Physics Letters
  B} {\bfseries 701} (2011) 363}
  [\href{https://arxiv.org/abs/1012.4679}{{\ttfamily 1012.4679}}].

\bibitem{Drozd:2011aa}
A.~Drozd, B.~Grzadkowski and J.~Wudka, \emph{Multi-scalar-singlet extension of
  the standard model - the case for dark matter and an invisible higgs boson},
  \href{https://doi.org/10.1007/JHEP04(2012)006}{\emph{JHEP} {\bfseries 04}
  (2012) 006} [\href{https://arxiv.org/abs/1112.2582}{{\ttfamily 1112.2582}}].

\bibitem{Aoki:2012ub}
M.~Aoki, M.~Duerr, J.~Kubo and H.~Takano, \emph{Multi-component dark matter
  systems and their observation prospects},
  \href{https://doi.org/10.1103/PhysRevD.86.076015}{\emph{Physical Review D:
  Particles and Fields} {\bfseries 86} (2012) 076015}
  [\href{https://arxiv.org/abs/1207.3318}{{\ttfamily 1207.3318}}].

\bibitem{Bhattacharya:2013hva}
S.~Bhattacharya, A.~Drozd, B.~Grzadkowski and J.~Wudka, \emph{Two-component
  dark matter}, \href{https://doi.org/10.1007/JHEP10(2013)158}{\emph{JHEP}
  {\bfseries 10} (2013) 158} [\href{https://arxiv.org/abs/1309.2986}{{\ttfamily
  1309.2986}}].

\bibitem{Baek:2013dwa}
S.~Baek, P.~Ko and W.-I. Park, \emph{Hidden sector monopole, vector dark matter
  and dark radiation with {{Higgs}} portal},
  \href{https://doi.org/10.1088/1475-7516/2014/10/067}{\emph{JCAP} {\bfseries
  10} (2014) 067} [\href{https://arxiv.org/abs/1311.1035}{{\ttfamily
  1311.1035}}].

\bibitem{Esch:2014jpa}
S.~Esch, M.~Klasen and C.~E. Yaguna, \emph{A minimal model for two-component
  dark matter}, \href{https://doi.org/10.1007/JHEP09(2014)108}{\emph{JHEP}
  {\bfseries 09} (2014) 108} [\href{https://arxiv.org/abs/1406.0617}{{\ttfamily
  1406.0617}}].

\bibitem{Ko:2014bka}
P.~Ko and Y.~Tang, \emph{{{$\nu\Lambda$MDM}}: {{A}} model for sterile neutrino
  and dark matter reconciles cosmological and neutrino oscillation data after
  {{BICEP2}}},
  \href{https://doi.org/10.1016/j.physletb.2014.10.035}{\emph{Physics Letters
  B} {\bfseries 739} (2014) 62}
  [\href{https://arxiv.org/abs/1404.0236}{{\ttfamily 1404.0236}}].

\bibitem{Bian:2014cja}
L.~Bian, T.~Li, J.~Shu and X.-C. Wang, \emph{Two component dark matter with
  multi-{{Higgs}} portals},
  \href{https://doi.org/10.1007/JHEP03(2015)126}{\emph{JHEP} {\bfseries 03}
  (2015) 126} [\href{https://arxiv.org/abs/1412.5443}{{\ttfamily 1412.5443}}].

\bibitem{Karam:2015jta}
A.~Karam and K.~Tamvakis, \emph{Dark matter and neutrino masses from a
  scale-invariant multi-{{Higgs}} portal},
  \href{https://doi.org/10.1103/PhysRevD.92.075010}{\emph{Physical Review D:
  Particles and Fields} {\bfseries 92} (2015) 075010}
  [\href{https://arxiv.org/abs/1508.03031}{{\ttfamily 1508.03031}}].

\bibitem{Arcadi:2016kmk}
G.~Arcadi, C.~Gross, O.~Lebedev, Y.~Mambrini, S.~Pokorski and T.~Toma,
  \emph{Multicomponent dark matter from gauge symmetry},
  \href{https://doi.org/10.1007/JHEP12(2016)081}{\emph{JHEP} {\bfseries 12}
  (2016) 081} [\href{https://arxiv.org/abs/1611.00365}{{\ttfamily
  1611.00365}}].

\bibitem{DuttaBanik:2016jzv}
A.~Dutta~Banik, M.~Pandey, D.~Majumdar and A.~Biswas, \emph{Two component
  {{WIMP}}\textendash{{FImP}} dark matter model with singlet fermion, scalar
  and pseudo scalar},
  \href{https://doi.org/10.1140/epjc/s10052-017-5221-y}{\emph{The European
  Physical Journal C: Particles and Fields} {\bfseries 77} (2017) 657}
  [\href{https://arxiv.org/abs/1612.08621}{{\ttfamily 1612.08621}}].

\bibitem{Karam:2016rsz}
A.~Karam and K.~Tamvakis, \emph{Dark matter from a classically scale-invariant
  {{SU}}(3){{{\textsubscript{X}}}}},
  \href{https://doi.org/10.1103/PhysRevD.94.055004}{\emph{Physical Review D:
  Particles and Fields} {\bfseries 94} (2016) 055004}
  [\href{https://arxiv.org/abs/1607.01001}{{\ttfamily 1607.01001}}].

\bibitem{Bhattacharya:2016ysw}
S.~Bhattacharya, P.~Poulose and P.~Ghosh, \emph{Multipartite interacting scalar
  dark matter in the light of updated {{LUX}} data},
  \href{https://doi.org/10.1088/1475-7516/2017/04/043}{\emph{JCAP} {\bfseries
  04} (2017) 043} [\href{https://arxiv.org/abs/1607.08461}{{\ttfamily
  1607.08461}}].

\bibitem{Ko:2016fcd}
P.~Ko and Y.~Tang, \emph{Residual non-abelian dark matter and dark radiation},
  \href{https://doi.org/10.1016/j.physletb.2017.02.033}{\emph{Physics Letters
  B} {\bfseries 768} (2017) 12}
  [\href{https://arxiv.org/abs/1609.02307}{{\ttfamily 1609.02307}}].

\bibitem{Aoki:2016glu}
M.~Aoki and T.~Toma, \emph{Implications of two-component dark matter induced by
  forbidden channels and thermal freeze-out},
  \href{https://doi.org/10.1088/1475-7516/2017/01/042}{\emph{JCAP} {\bfseries
  01} (2017) 042} [\href{https://arxiv.org/abs/1611.06746}{{\ttfamily
  1611.06746}}].

\bibitem{Ahmed:2017dbb}
A.~Ahmed, M.~Duch, B.~Grzadkowski and M.~Iglicki, \emph{Multi-{{Component Dark
  Matter}}: The vector and fermion case},
  \href{https://doi.org/10.1140/epjc/s10052-018-6371-2}{\emph{The European
  Physical Journal C: Particles and Fields} {\bfseries 78} (2018) 905}
  [\href{https://arxiv.org/abs/1710.01853}{{\ttfamily 1710.01853}}].

\bibitem{Aoki:2018gjf}
M.~Aoki and T.~Toma, \emph{Boosted self-interacting dark matter in a
  multi-component dark matter model},
  \href{https://doi.org/10.1088/1475-7516/2018/10/020}{\emph{JCAP} {\bfseries
  10} (2018) 020} [\href{https://arxiv.org/abs/1806.09154}{{\ttfamily
  1806.09154}}].

\bibitem{Chakraborti:2018lso}
S.~Chakraborti and P.~Poulose, \emph{Interplay of scalar and fermionic
  components in a multi-component dark matter scenario},
  \href{https://doi.org/10.1140/epjc/s10052-019-6933-y}{\emph{The European
  Physical Journal C: Particles and Fields} {\bfseries 79} (2019) 420}
  [\href{https://arxiv.org/abs/1808.01979}{{\ttfamily 1808.01979}}].

\bibitem{Poulin:2018kap}
A.~Poulin and S.~Godfrey, \emph{Multicomponent dark matter from a hidden gauged
  {{SU}}(3)}, \href{https://doi.org/10.1103/PhysRevD.99.076008}{\emph{Physical
  Review D: Particles and Fields} {\bfseries 99} (2019) 076008}
  [\href{https://arxiv.org/abs/1808.04901}{{\ttfamily 1808.04901}}].

\bibitem{YaserAyazi:2018lrv}
S.~Yaser~Ayazi and A.~Mohamadnejad, \emph{Scale-invariant two component dark
  matter}, \href{https://doi.org/10.1140/epjc/s10052-019-6651-5}{\emph{The
  European Physical Journal C: Particles and Fields} {\bfseries 79} (2019) 140}
  [\href{https://arxiv.org/abs/1808.08706}{{\ttfamily 1808.08706}}].

\bibitem{Chakraborti:2018aae}
S.~Chakraborti, A.~Dutta~Banik and R.~Islam, \emph{Probing multicomponent
  extension of inert doublet model with a vector dark matter},
  \href{https://doi.org/10.1140/epjc/s10052-019-7165-x}{\emph{The European
  Physical Journal C: Particles and Fields} {\bfseries 79} (2019) 662}
  [\href{https://arxiv.org/abs/1810.05595}{{\ttfamily 1810.05595}}].

\bibitem{Bhattacharya:2019fgs}
S.~Bhattacharya, P.~Ghosh, A.~K. Saha and A.~Sil, \emph{Two component dark
  matter with inert {{Higgs}} doublet: Neutrino mass, high scale validity and
  collider searches},
  \href{https://doi.org/10.1007/JHEP03(2020)090}{\emph{JHEP} {\bfseries 03}
  (2020) 090} [\href{https://arxiv.org/abs/1905.12583}{{\ttfamily
  1905.12583}}].

\bibitem{Chen:2019pnt}
C.-R. Chen, Y.-X. Lin, C.~S. Nugroho, R.~Ramos, Y.-L.~S. Tsai and T.-C. Yuan,
  \emph{Complex scalar dark matter in the gauged two-{{Higgs-doublet}} model},
  \href{https://doi.org/10.1103/PhysRevD.101.035037}{\emph{Physical Review D:
  Particles and Fields} {\bfseries 101} (2020) 035037}
  [\href{https://arxiv.org/abs/1910.13138}{{\ttfamily 1910.13138}}].

\bibitem{Yaguna:2019cvp}
C.~E. Yaguna and {\'O}.~Zapata, \emph{Multi-component scalar dark matter from a
  {{Z}}{{{\textsubscript{N}}}} symmetry: A systematic analysis},
  \href{https://doi.org/10.1007/JHEP03(2020)109}{\emph{JHEP} {\bfseries 03}
  (2020) 109} [\href{https://arxiv.org/abs/1911.05515}{{\ttfamily
  1911.05515}}].

\bibitem{Bhattacharya:2019tqq}
S.~Bhattacharya, N.~Chakrabarty, R.~Roshan and A.~Sil, \emph{Multicomponent
  dark matter in extended {{U}}(1){{{\textsubscript{B-L}}}}: Neutrino mass and
  high scale validity},
  \href{https://doi.org/10.1088/1475-7516/2020/04/013}{\emph{JCAP} {\bfseries
  04} (2020) 013} [\href{https://arxiv.org/abs/1910.00612}{{\ttfamily
  1910.00612}}].

\bibitem{Betancur:2020fdl}
A.~Betancur, G.~Palacio and A.~Rivera, \emph{Inert doublet as multicomponent
  dark matter},
  \href{https://doi.org/10.1016/j.nuclphysb.2020.115276}{\emph{Nuclear Physics
  B} {\bfseries 962} (2021) 115276}
  [\href{https://arxiv.org/abs/2002.02036}{{\ttfamily 2002.02036}}].

\bibitem{Belanger:2020hyh}
G.~B{\'e}langer, A.~Pukhov, C.~E. Yaguna and {\'O}.~Zapata, \emph{The
  {{Z}}{$_5$} model of two-component dark matter},
  \href{https://doi.org/10.1007/JHEP09(2020)030}{\emph{JHEP} {\bfseries 09}
  (2020) 030} [\href{https://arxiv.org/abs/2006.14922}{{\ttfamily
  2006.14922}}].

\bibitem{Belanger:2021lwd}
G.~Belanger, A.~Mjallal and A.~Pukhov, \emph{Two dark matter candidates:
  {{The}} case of inert doublet and singlet scalars},
  \href{https://doi.org/10.1103/PhysRevD.105.035018}{\emph{Physical Review D:
  Particles and Fields} {\bfseries 105} (2022) 035018}
  [\href{https://arxiv.org/abs/2108.08061}{{\ttfamily 2108.08061}}].

\bibitem{Bhattacharya:2021rwh}
S.~Bhattacharya, S.~Chakraborti and D.~Pradhan, \emph{Electroweak symmetry
  breaking and {{WIMP-FIMP}} dark matter},
  \href{https://doi.org/10.1007/JHEP07(2022)091}{\emph{JHEP} {\bfseries 07}
  (2022) 091} [\href{https://arxiv.org/abs/2110.06985}{{\ttfamily
  2110.06985}}].

\bibitem{Das:2021zea}
P.~Das, M.~K. Das and N.~Khan, \emph{The {{FIMP-WIMP}} dark matter in the
  extended singlet scalar model},
  \href{https://doi.org/10.1016/j.nuclphysb.2022.115677}{\emph{Nuclear Physics
  B} {\bfseries 975} (2022) 115677}
  [\href{https://arxiv.org/abs/2104.03271}{{\ttfamily 2104.03271}}].

\bibitem{Betancur:2021ect}
A.~Betancur, A.~Castillo, G.~Palacio and J.~Suarez, \emph{Multicomponent scalar
  dark matter at high-intensity proton beam experiments},
  \href{https://doi.org/10.1088/1361-6471/ac65a6}{\emph{Journal of Physics G:
  Nuclear and Particle Physics} {\bfseries 49} (2022) 075003}
  [\href{https://arxiv.org/abs/2109.11586}{{\ttfamily 2109.11586}}].

\bibitem{Chakrabarty:2021kmr}
N.~Chakrabarty, R.~Roshan and A.~Sil, \emph{Two-component doublet-triplet
  scalar dark matter stabilizing the electroweak vacuum},
  \href{https://doi.org/10.1103/PhysRevD.105.115010}{\emph{Physical Review D:
  Particles and Fields} {\bfseries 105} (2022) 115010}
  [\href{https://arxiv.org/abs/2102.06032}{{\ttfamily 2102.06032}}].

\bibitem{Mohamadnejad:2021tke}
A.~Mohamadnejad, \emph{Electroweak phase transition and gravitational waves in
  a two-component dark matter model},
  \href{https://doi.org/10.1007/JHEP03(2022)188}{\emph{Journal of High Energy
  Physics} {\bfseries 03} (2022) 188}
  [\href{https://arxiv.org/abs/2111.04342}{{\ttfamily 2111.04342}}].

\bibitem{DiazSaez:2021pfw}
B.~D{\'i}az~S{\'a}ez, K.~M{\"o}hling and D.~St{\"o}ckinger, \emph{Two real
  scalar {{WIMP}} model in the assisted freeze-out scenario},
  \href{https://doi.org/10.1088/1475-7516/2021/10/027}{\emph{JCAP} {\bfseries
  10} (2021) 027} [\href{https://arxiv.org/abs/2103.17064}{{\ttfamily
  2103.17064}}].

\bibitem{Choi:2021yps}
S.-M. Choi, J.~Kim, P.~Ko and J.~Li, \emph{A multi-component {{SIMP}} model
  with {{U}}(1){{{\textsubscript{X}}}}\textrightarrow{} {{Z}}{$_2$}
  \texttimes{} {{Z}}{$_{3}$}},
  \href{https://doi.org/10.1007/JHEP09(2021)028}{\emph{JHEP} {\bfseries 09}
  (2021) 028} [\href{https://arxiv.org/abs/2103.05956}{{\ttfamily
  2103.05956}}].

\bibitem{Belanger:2022qxt}
G.~Belanger, A.~Mjallal and A.~Pukhov, \emph{{{WIMP}} and {{FIMP}} dark matter
  in the inert doublet plus singlet model},
  \href{https://arxiv.org/abs/2205.04101}{{\ttfamily 2205.04101}}.

\bibitem{Das:2022oyx}
A.~Das, S.~Gola, S.~Mandal and N.~Sinha, \emph{Two-component scalar and
  fermionic dark matter candidates in a generic {{U}}(1){{X}} model},
  \href{https://doi.org/10.1016/j.physletb.2022.137117}{\emph{Physics Letters
  B} {\bfseries 829} (2022) 137117}
  [\href{https://arxiv.org/abs/2202.01443}{{\ttfamily 2202.01443}}].

\bibitem{Ho:2022erb}
S.-Y. Ho, P.~Ko and C.-T. Lu, \emph{Scalar and fermion two-component {{SIMP}}
  dark matter with an accidental {{Z}}{$_4$} symmetry},
  \href{https://doi.org/10.1007/JHEP03(2022)005}{\emph{JHEP} {\bfseries 03}
  (2022) 005} [\href{https://arxiv.org/abs/2201.06856}{{\ttfamily
  2201.06856}}].

\bibitem{Costa:2022oaa}
F.~Costa, S.~Khan and J.~Kim, \emph{A {{Two-Component Dark Matter Model}} and
  its {{Associated Gravitational Waves}}},
  \href{https://doi.org/10.1007/JHEP06(2022)026}{\emph{Journal of High Energy
  Physics} {\bfseries 06} (2022) 026}
  [\href{https://arxiv.org/abs/2202.13126}{{\ttfamily 2202.13126}}].

\bibitem{Kamionkowski:1993fg}
M.~Kamionkowski, A.~Kosowsky and M.~S. Turner, \emph{Gravitational
  {{Radiation}} from {{First-Order Phase Transitions}}},
  \href{https://doi.org/10.1103/PhysRevD.49.2837}{\emph{Physical Review D}
  {\bfseries 49} (1994) 2837}
  [\href{https://arxiv.org/abs/astro-ph/9310044}{{\ttfamily
  astro-ph/9310044}}].

\bibitem{Baker:2019nia}
J.~Baker, J.~Bellovary, P.~L. Bender, E.~Berti, R.~Caldwell, J.~Camp et~al.,
  \emph{The {{Laser Interferometer Space Antenna}}: {{Unveiling}} the
  {{Millihertz Gravitational Wave Sky}}},
  \href{https://arxiv.org/abs/1907.06482}{{\ttfamily 1907.06482}}.

\bibitem{Seto:2001qf}
N.~Seto, S.~Kawamura and T.~Nakamura, \emph{Possibility of {{Direct
  Measurement}} of the {{Acceleration}} of the {{Universe Using}} 0.1 {{Hz Band
  Laser Interferometer Gravitational Wave Antenna}} in {{Space}}},
  \href{https://doi.org/10.1103/PhysRevLett.87.221103}{\emph{Physical Review
  Letters} {\bfseries 87} (2001) 221103}
  [\href{https://arxiv.org/abs/astro-ph/0108011}{{\ttfamily
  astro-ph/0108011}}].

\bibitem{Kawamura:2006up}
S.~Kawamura, T.~Nakamura, M.~Ando, N.~Seto, K.~Tsubono, K.~Numata et~al.,
  \emph{The {{Japanese}} space gravitational wave
  antenna\textemdash{{DECIGO}}},
  \href{https://doi.org/10.1088/0264-9381/23/8/S17}{\emph{Classical and Quantum
  Gravity} {\bfseries 23} (2006) S125}.

\bibitem{Sato:2017dkf}
S.~Sato, S.~Kawamura, M.~Ando, T.~Nakamura, K.~Tsubono, A.~Araya et~al.,
  \emph{The status of {{DECIGO}}},
  \href{https://doi.org/10.1088/1742-6596/840/1/012010}{\emph{Journal of
  Physics: Conference Series} {\bfseries 840} (2017) 012010}.

\bibitem{Isoyama:2018rjb}
S.~Isoyama, H.~Nakano and T.~Nakamura, \emph{Multiband {{Gravitational-Wave
  Astronomy}}: {{Observing}} binary inspirals with a decihertz detector,
  {{B-DECIGO}}}, \href{https://doi.org/10.1093/ptep/pty078}{\emph{Progress of
  Theoretical and Experimental Physics} {\bfseries 2018} (2018) 073E01}
  [\href{https://arxiv.org/abs/1802.06977}{{\ttfamily 1802.06977}}].

\bibitem{Kawamura:2020pcg}
S.~Kawamura, M.~Ando, N.~Seto, S.~Sato, M.~Musha, I.~Kawano et~al.,
  \emph{Current status of space gravitational wave antenna {{DECIGO}} and
  {{B-DECIGO}}},  \href{https://arxiv.org/abs/2006.13545}{{\ttfamily
  2006.13545}}.

\bibitem{Corbin:2005ny}
V.~Corbin and N.~J. Cornish, \emph{Detecting the {{Cosmic Gravitational Wave
  Background}} with the {{Big Bang Observer}}},
  \href{https://doi.org/10.1088/0264-9381/23/7/014}{\emph{Classical and Quantum
  Gravity} {\bfseries 23} (2006) 2435}
  [\href{https://arxiv.org/abs/gr-qc/0512039}{{\ttfamily gr-qc/0512039}}].

\bibitem{Crowder:2005nr}
J.~Crowder and N.~J. Cornish, \emph{Beyond {{LISA}}: {{Exploring Future
  Gravitational Wave Missions}}},
  \href{https://doi.org/10.1103/PhysRevD.72.083005}{\emph{Physical Review D}
  {\bfseries 72} (2005) 083005}
  [\href{https://arxiv.org/abs/gr-qc/0506015}{{\ttfamily gr-qc/0506015}}].

\bibitem{Harry:2006fi}
G.~M. Harry, P.~Fritschel, D.~A. Shaddock, W.~Folkner and E.~S. Phinney,
  \emph{Laser interferometry for the {{Big Bang Observer}}},
  \href{https://doi.org/10.1088/0264-9381/23/15/008}{\emph{Classical and
  Quantum Gravity} {\bfseries 23} (2006) 4887}.

\bibitem{Grojean:2006bp}
C.~Grojean and G.~Servant, \emph{Gravitational {{Waves}} from {{Phase
  Transitions}} at the {{Electroweak Scale}} and {{Beyond}}},
  \href{https://doi.org/10.1103/PhysRevD.75.043507}{\emph{Physical Review D}
  {\bfseries 75} (2007) 043507}
  [\href{https://arxiv.org/abs/hep-ph/0607107}{{\ttfamily hep-ph/0607107}}].

\bibitem{Huber:2008hg}
S.~J. Huber and T.~Konstandin, \emph{Gravitational {{Wave Production}} by
  {{Collisions}}: {{More Bubbles}}},
  \href{https://doi.org/10.1088/1475-7516/2008/09/022}{\emph{Journal of
  Cosmology and Astroparticle Physics} {\bfseries 09} (2008) 022}
  [\href{https://arxiv.org/abs/0806.1828}{{\ttfamily 0806.1828}}].

\bibitem{Espinosa:2008kw}
J.~R. Espinosa, T.~Konstandin, J.~M. No and M.~Quiros, \emph{Some
  {{Cosmological Implications}} of {{Hidden Sectors}}},
  \href{https://doi.org/10.1103/PhysRevD.78.123528}{\emph{Physical Review D}
  {\bfseries 78} (2008) 123528}
  [\href{https://arxiv.org/abs/0809.3215}{{\ttfamily 0809.3215}}].

\bibitem{Caprini:2015zlo}
C.~Caprini, M.~Hindmarsh, S.~Huber, T.~Konstandin, J.~Kozaczuk, G.~Nardini
  et~al., \emph{Science with the space-based interferometer {{eLISA}}. {{II}}:
  {{Gravitational}} waves from cosmological phase transitions},
  \href{https://doi.org/10.1088/1475-7516/2016/04/001}{\emph{Journal of
  Cosmology and Astroparticle Physics} {\bfseries 04} (2016) 001}
  [\href{https://arxiv.org/abs/1512.06239}{{\ttfamily 1512.06239}}].

\bibitem{Artymowski:2016tme}
M.~Artymowski, M.~Lewicki and J.~D. Wells, \emph{Gravitational wave and
  collider implications of electroweak baryogenesis aided by non-standard
  cosmology}, \href{https://doi.org/10.1007/JHEP03(2017)066}{\emph{Journal of
  High Energy Physics} {\bfseries 03} (2017) 066}
  [\href{https://arxiv.org/abs/1609.07143}{{\ttfamily 1609.07143}}].

\bibitem{Baldes:2017rcu}
I.~Baldes, \emph{Gravitational waves from the asymmetric-dark-matter generating
  phase transition},
  \href{https://doi.org/10.1088/1475-7516/2017/05/028}{\emph{Journal of
  Cosmology and Astroparticle Physics} {\bfseries 05} (2017) 028}
  [\href{https://arxiv.org/abs/1702.02117}{{\ttfamily 1702.02117}}].

\bibitem{Beniwal:2018hyi}
A.~Beniwal, M.~Lewicki, M.~White and A.~G. Williams, \emph{Gravitational waves
  and electroweak baryogenesis in a global study of the extended scalar singlet
  model}, \href{https://doi.org/10.1007/JHEP02(2019)183}{\emph{Journal of High
  Energy Physics} {\bfseries 02} (2019) 183}
  [\href{https://arxiv.org/abs/1810.02380}{{\ttfamily 1810.02380}}].

\bibitem{Hashino:2018zsi}
K.~Hashino, M.~Kakizaki, S.~Kanemura, P.~Ko and T.~Matsui, \emph{Gravitational
  waves from first order electroweak phase transition in models with the
  {{U}}(1){{{\textsubscript{X}}}} gauge symmetry},
  \href{https://doi.org/10.1007/JHEP06(2018)088}{\emph{JHEP} {\bfseries 06}
  (2018) 088} [\href{https://arxiv.org/abs/1802.02947}{{\ttfamily
  1802.02947}}].

\bibitem{Caprini:2018mtu}
C.~Caprini and D.~G. Figueroa, \emph{Cosmological {{Backgrounds}} of
  {{Gravitational Waves}}},
  \href{https://doi.org/10.1088/1361-6382/aac608}{\emph{Classical and Quantum
  Gravity} {\bfseries 35} (2018) 163001}
  [\href{https://arxiv.org/abs/1801.04268}{{\ttfamily 1801.04268}}].

\bibitem{Bian:2018mkl}
L.~Bian and Y.-L. Tang, \emph{Thermally modified sterile neutrino portal dark
  matter and gravitational waves from phase transition: {{The Freeze-in}}
  case}, \href{https://doi.org/10.1007/JHEP12(2018)006}{\emph{Journal of High
  Energy Physics} {\bfseries 12} (2018) 006}
  [\href{https://arxiv.org/abs/1810.03172}{{\ttfamily 1810.03172}}].

\bibitem{Bian:2018bxr}
L.~Bian and X.~Liu, \emph{Two-step strongly first-order electroweak phase
  transition modified {{FIMP}} dark matter, gravitational wave signals, and the
  neutrino mass},
  \href{https://doi.org/10.1103/PhysRevD.99.055003}{\emph{Physical Review D}
  {\bfseries 99} (2019) 055003}
  [\href{https://arxiv.org/abs/1811.03279}{{\ttfamily 1811.03279}}].

\bibitem{Bian:2019szo}
L.~Bian, W.~Cheng, H.-K. Guo and Y.~Zhang, \emph{Cosmological implications of a
  {{B}} - {{L}} charged hidden scalar: Leptogenesis and gravitational waves},
  \href{https://doi.org/10.1088/1674-1137/ac1e09}{\emph{Chinese Physics C}
  {\bfseries 45} (2021) 113104}
  [\href{https://arxiv.org/abs/1907.13589}{{\ttfamily 1907.13589}}].

\bibitem{Bian:2019kmg}
L.~Bian, Y.~Wu and K.-P. Xie, \emph{Electroweak phase transition with composite
  {{Higgs}} models: Calculability, gravitational waves and collider searches},
  \href{https://doi.org/10.1007/JHEP12(2019)028}{\emph{Journal of High Energy
  Physics} {\bfseries 12} (2019) 028}
  [\href{https://arxiv.org/abs/1909.02014}{{\ttfamily 1909.02014}}].

\bibitem{Caprini:2019egz}
C.~Caprini, M.~Chala, G.~C. Dorsch, M.~Hindmarsh, S.~J. Huber, T.~Konstandin
  et~al., \emph{Detecting gravitational waves from cosmological phase
  transitions with {{LISA}}: An update},
  \href{https://doi.org/10.1088/1475-7516/2020/03/024}{\emph{Journal of
  Cosmology and Astroparticle Physics} {\bfseries 03} (2020) 024}
  [\href{https://arxiv.org/abs/1910.13125}{{\ttfamily 1910.13125}}].

\bibitem{Di:2020ivg}
Y.~Di, J.~Wang, R.~Zhou, L.~Bian, R.-G. Cai and J.~Liu, \emph{Magnetic field
  and gravitational waves from the first-order {{Phase Transition}}},
  \href{https://doi.org/10.1103/PhysRevLett.126.251102}{\emph{Physical Review
  Letters} {\bfseries 126} (2021) 251102}
  [\href{https://arxiv.org/abs/2012.15625}{{\ttfamily 2012.15625}}].

\bibitem{Zhou:2021cfu}
R.~Zhou, L.~Bian and J.~Shu, \emph{Probing new physics for
  {{(g-2){\textsubscript{$\mu$}}}} and gravitational waves},
  \href{https://arxiv.org/abs/2104.03519}{{\ttfamily 2104.03519}}.

\bibitem{Bian:2021dmp}
L.~Bian, Y.-L. Tang and R.~Zhou, \emph{{{FIMP}} dark matter mediated by massive
  gauge boson around the phase transition period and the gravitational waves
  production},  \href{https://arxiv.org/abs/2111.10608}{{\ttfamily
  2111.10608}}.

\bibitem{Altmannshofer:2019zhy}
W.~Altmannshofer, S.~Gori, J.~{Mart{\'i}n-Albo}, A.~Sousa and M.~Wallbank,
  \emph{Neutrino tridents at {{DUNE}}},
  \href{https://doi.org/10.1103/PhysRevD.100.115029}{\emph{Physical Review D:
  Particles and Fields} {\bfseries 100} (2019) 115029}
  [\href{https://arxiv.org/abs/1902.06765}{{\ttfamily 1902.06765}}].

\bibitem{Biswas:2021dan}
A.~Biswas and S.~Khan, \emph{(g - 2){{{\textsubscript{e, $\mu$}}}} and strongly
  interacting dark matter with collider implications},
  \href{https://doi.org/10.1007/JHEP07(2022)037}{\emph{JHEP} {\bfseries 07}
  (2022) 037} [\href{https://arxiv.org/abs/2112.08393}{{\ttfamily
  2112.08393}}].

\bibitem{Bauer:2018onh}
M.~Bauer, P.~Foldenauer and J.~Jaeckel, \emph{Hunting all the hidden photons},
  \href{https://doi.org/10.1007/JHEP07(2018)094}{\emph{JHEP} {\bfseries 07}
  (2018) 094} [\href{https://arxiv.org/abs/1803.05466}{{\ttfamily
  1803.05466}}].

\bibitem{Fermi-LAT:2015kyq}
{\scshape Fermi-LAT} collaboration, \emph{Updated search for spectral lines
  from galactic dark matter interactions with pass 8 data from the fermi large
  area telescope},
  \href{https://doi.org/10.1103/PhysRevD.91.122002}{\emph{Phys. Rev. D}
  {\bfseries 91} (2015) 122002}
  [\href{https://arxiv.org/abs/1506.00013}{{\ttfamily 1506.00013}}].

\bibitem{Pontecorvo:1957qd}
B.~Pontecorvo, \emph{Inverse beta processes and nonconservation of lepton
  charge}, {\emph{Zhurnal Eksperimental'noi i Teoreticheskoi Fiziki} {\bfseries
  7} (1958) 172}.

\bibitem{Maki:1962mu}
Z.~Maki, M.~Nakagawa and S.~Sakata, \emph{Remarks on the unified model of
  elementary particles},
  \href{https://doi.org/10.1143/PTP.28.870}{\emph{Progress of Theoretical
  Physics} {\bfseries 28} (1962) 870}.

\bibitem{MEG:2016leq}
{\scshape MEG} collaboration, \emph{Search for the lepton flavour violating
  decay {{$\mu^+$ \textrightarrow e$^+$ $\gamma$}} with the full dataset of the
  {{MEG}} experiment},
  \href{https://doi.org/10.1140/epjc/s10052-016-4271-x}{\emph{The European
  Physical Journal C: Particles and Fields} {\bfseries 76} (2016) 434}
  [\href{https://arxiv.org/abs/1605.05081}{{\ttfamily 1605.05081}}].

\bibitem{SINDRUM:1987nra}
{\scshape SINDRUM} collaboration, \emph{Search for the decay {$\mu$}+
  \textrightarrow{} e+e+e-},
  \href{https://doi.org/10.1016/0550-3213(88)90462-2}{\emph{Nuclear Physics B}
  {\bfseries 299} (1988) 1}.

\bibitem{Wintz:1998rp}
P.~Wintz, \emph{Results of the {{SINDRUM-II}} experiment}, {\emph{Conf. Proc.
  C} {\bfseries 980420} (1998) 534}.

\bibitem{Ilakovac:1994kj}
A.~Ilakovac and A.~Pilaftsis, \emph{Flavour-{{Violating Charged Lepton Decays}}
  in {{Seesaw-Type Models}}},
  \href{https://arxiv.org/abs/hep-ph/9403398}{{\ttfamily hep-ph/9403398}}.

\bibitem{Lindner:2016bgg}
M.~Lindner, M.~Platscher and F.~S. Queiroz, \emph{A {{Call}} for {{New
  Physics}} : {{The Muon Anomalous Magnetic Moment}} and {{Lepton Flavor
  Violation}}},  \href{https://arxiv.org/abs/1610.06587}{{\ttfamily
  1610.06587}}.

\bibitem{CHARM:1985nku}
{\scshape CHARM} collaboration, \emph{A search for decays of heavy neutrinos in
  the mass range 0.5\textendash 2.8 {{GeV}}},
  \href{https://doi.org/10.1016/0370-2693(86)91601-1}{\emph{Physics Letters B}
  {\bfseries 166} (1986) 473}.

\bibitem{CHARMII:1994jjr}
{\scshape CHARM II} collaboration, \emph{Search for heavy isosinglet
  neutrinos}, \href{https://doi.org/10.1016/0370-2693(94)01422-9}{\emph{Physics
  Letters B} {\bfseries 343} (1995) 453}.

\bibitem{Belle:2013ytx}
{\scshape Belle} collaboration, \emph{Search for heavy neutrinos at {{Belle}}},
   \href{https://arxiv.org/abs/1301.1105}{{\ttfamily 1301.1105}}.

\bibitem{DELPHI:1996qcc}
{\scshape DELPHI} collaboration, \emph{Search for neutral heavy leptons
  produced in {{Z}} decays},
  \href{https://doi.org/10.1007/s002880050370}{\emph{Zeitschrift f\"ur Physik}
  {\bfseries 74} (1997) 57}.

\bibitem{Krasnov:2019kdc}
I.~Krasnov, \emph{On {{DUNE}} prospects in the search for sterile neutrinos},
  \href{https://doi.org/10.1103/PhysRevD.100.075023}{\emph{Physical Review D}
  {\bfseries 100} (2019) 075023}
  [\href{https://arxiv.org/abs/1902.06099}{{\ttfamily 1902.06099}}].

\bibitem{Ballett:2019bgd}
P.~Ballett, T.~Boschi and S.~Pascoli, \emph{Heavy {{Neutral Leptons}} from
  low-scale seesaws at the {{DUNE Near Detector}}},
  \href{https://doi.org/10.1007/JHEP03(2020)111}{\emph{Journal of High Energy
  Physics} {\bfseries 03} (2020) 111}
  [\href{https://arxiv.org/abs/1905.00284}{{\ttfamily 1905.00284}}].

\bibitem{SHiP:2018xqw}
{\scshape SHiP} collaboration, \emph{Sensitivity of the {{SHiP}} experiment to
  {{Heavy Neutral Leptons}}},
  \href{https://arxiv.org/abs/1811.00930}{{\ttfamily 1811.00930}}.

\bibitem{Chou:2016lxi}
J.~P. Chou, D.~Curtin and H.~J. Lubatti, \emph{New {{Detectors}} to {{Explore}}
  the {{Lifetime Frontier}}},
  \href{https://doi.org/10.1016/j.physletb.2017.01.043}{\emph{Physics Letters
  B} {\bfseries 767} (2017) 29}
  [\href{https://arxiv.org/abs/1606.06298}{{\ttfamily 1606.06298}}].

\bibitem{Chun:2019nwi}
E.~J. Chun, A.~Das, S.~Mandal, M.~Mitra and N.~Sinha, \emph{Sensitivity of
  {{Lepton Number Violating Meson Decays}} in {{Different Experiments}}},
  \href{https://doi.org/10.1103/PhysRevD.100.095022}{\emph{Physical Review D}
  {\bfseries 100} (2019) 095022}
  [\href{https://arxiv.org/abs/1908.09562}{{\ttfamily 1908.09562}}].

\bibitem{Blondel:2014bra}
{\scshape FCC-ee study Team} collaboration, \emph{Search for {{Heavy Right
  Handed Neutrinos}} at the {{FCC-ee}}},
  \href{https://arxiv.org/abs/1411.5230}{{\ttfamily 1411.5230}}.

\bibitem{Alimena:2022hfr}
J.~Alimena, P.~Azzi, M.~Bauer, A.~Blondel, M.~Drewes, R.~G. Suarez et~al.,
  \emph{Searches for {{Long-Lived Particles}} at the {{Future FCC-ee}}},
  \href{https://arxiv.org/abs/2203.05502}{{\ttfamily 2203.05502}}.

\bibitem{Drewes:2019fou}
M.~Drewes and J.~Hajer, \emph{Heavy {{Neutrinos}} in displaced vertex searches
  at the {{LHC}} and {{HL-LHC}}},
  \href{https://doi.org/10.1007/JHEP02(2020)070}{\emph{Journal of High Energy
  Physics} {\bfseries 02} (2020) 070}
  [\href{https://arxiv.org/abs/1903.06100}{{\ttfamily 1903.06100}}].

\bibitem{Antusch:2017hhu}
S.~Antusch, E.~Cazzato and O.~Fischer, \emph{Sterile neutrino searches via
  displaced vertices at {{LHCb}}},
  \href{https://doi.org/10.1016/j.physletb.2017.09.057}{\emph{Physics Letters
  B} {\bfseries 774} (2017) 114}
  [\href{https://arxiv.org/abs/1706.05990}{{\ttfamily 1706.05990}}].

\bibitem{NuTeV:1999kej}
{\scshape NuTeV, E815} collaboration, \emph{Search for {{Neutral Heavy
  Leptons}} in a {{High-Energy Neutrino Beam}}},
  \href{https://doi.org/10.1103/PhysRevLett.83.4943}{\emph{Physical Review
  Letters} {\bfseries 83} (1999) 4943}
  [\href{https://arxiv.org/abs/hep-ex/9908011}{{\ttfamily hep-ex/9908011}}].

\bibitem{FMMF:1994yvb}
{\scshape FMMF} collaboration, \emph{Search for neutral weakly interacting
  massive particles in the {{Fermilab Tevatron}} wideband neutrino beam},
  \href{https://doi.org/10.1103/PhysRevD.52.6}{\emph{Physical Review D}
  {\bfseries 52} (1995) 6}.

\bibitem{Feng:2017uoz}
J.~L. Feng, I.~Galon, F.~Kling and S.~Trojanowski, \emph{{{FASER}}: {{ForwArd
  Search ExpeRiment}} at the {{LHC}}},
  \href{https://doi.org/10.1103/PhysRevD.97.035001}{\emph{Physical Review D}
  {\bfseries 97} (2018) 035001}
  [\href{https://arxiv.org/abs/1708.09389}{{\ttfamily 1708.09389}}].

\bibitem{Dercks:2018wum}
D.~Dercks, H.~K. Dreiner, M.~Hirsch and Z.~S. Wang, \emph{Long-{{Lived
  Fermions}} at {{AL3X}}},
  \href{https://doi.org/10.1103/PhysRevD.99.055020}{\emph{Physical Review D}
  {\bfseries 99} (2019) 055020}
  [\href{https://arxiv.org/abs/1811.01995}{{\ttfamily 1811.01995}}].

\bibitem{Alloul:2013bka}
A.~Alloul, N.~D. Christensen, C.~Degrande, C.~Duhr and B.~Fuks,
  \emph{{{FeynRules}} 2.0 - {{A}} complete toolbox for tree-level
  phenomenology},
  \href{https://doi.org/10.1016/j.cpc.2014.04.012}{\emph{Computer Physics
  Communications} {\bfseries 185} (2014) 2250}
  [\href{https://arxiv.org/abs/1310.1921}{{\ttfamily 1310.1921}}].

\bibitem{Belyaev:2012qa}
A.~Belyaev, N.~D. Christensen and A.~Pukhov, \emph{{{CalcHEP}} 3.4 for collider
  physics within and beyond the {{Standard Model}}},
  \href{https://doi.org/10.1016/j.cpc.2013.01.014}{\emph{Computer Physics
  Communications} {\bfseries 184} (2013) 1729}
  [\href{https://arxiv.org/abs/1207.6082}{{\ttfamily 1207.6082}}].

\bibitem{Belanger:2006is}
G.~Belanger, F.~Boudjema, A.~Pukhov and A.~Semenov, \emph{{{micrOMEGAs2}}.0: A
  program to calculate the relic density of dark matter in a generic model},
  \href{https://doi.org/10.1016/j.cpc.2006.11.008}{\emph{Computer Physics
  Communications} {\bfseries 176} (2007) 367}
  [\href{https://arxiv.org/abs/hep-ph/0607059}{{\ttfamily hep-ph/0607059}}].

\bibitem{Carena:2019une}
M.~Carena, Z.~Liu and Y.~Wang, \emph{Electroweak phase transition with
  spontaneous {{Z}}{$_2$}-breaking},
  \href{https://doi.org/10.1007/JHEP08(2020)107}{\emph{JHEP} {\bfseries 08}
  (2020) 107} [\href{https://arxiv.org/abs/1911.10206}{{\ttfamily
  1911.10206}}].

\bibitem{CMS:2018uag}
{\scshape CMS} collaboration, \emph{Combined measurements of {{Higgs}} boson
  couplings in proton\textendash proton collisions at {{s=13\,TeV }}},
  \href{https://doi.org/10.1140/epjc/s10052-019-6909-y}{\emph{The European
  Physical Journal C: Particles and Fields} {\bfseries 79} (2019) 421}
  [\href{https://arxiv.org/abs/1809.10733}{{\ttfamily 1809.10733}}].

\bibitem{ATLAS:2016neq}
{\scshape ATLAS, CMS} collaboration, \emph{Measurements of the {{Higgs}} boson
  production and decay rates and constraints on its couplings from a combined
  {{ATLAS}} and {{CMS}} analysis of the {{LHC}} pp collision data at {{ s=7 }}
  and 8 {{TeV}}}, \href{https://doi.org/10.1007/JHEP08(2016)045}{\emph{JHEP}
  {\bfseries 08} (2016) 045}
  [\href{https://arxiv.org/abs/1606.02266}{{\ttfamily 1606.02266}}].

\bibitem{Berlin:2014tja}
A.~Berlin, D.~Hooper and S.~D. McDermott, \emph{Simplified {{Dark Matter
  Models}} for the {{Galactic Center Gamma-Ray Excess}}},
  \href{https://doi.org/10.1103/PhysRevD.89.115022}{\emph{Physical Review D}
  {\bfseries 89} (2014) 115022}
  [\href{https://arxiv.org/abs/1404.0022}{{\ttfamily 1404.0022}}].

\bibitem{Junnarkar:2013ac}
P.~Junnarkar and A.~{Walker-Loud}, \emph{The {{Scalar Strange Content}} of the
  {{Nucleon}} from {{Lattice QCD}}},
  \href{https://doi.org/10.1103/PhysRevD.87.114510}{\emph{Physical Review D}
  {\bfseries 87} (2013) 114510}
  [\href{https://arxiv.org/abs/1301.1114}{{\ttfamily 1301.1114}}].

\bibitem{LZ:2022ufs}
{\scshape LZ} collaboration, \emph{First {{Dark Matter Search Results}} from
  the {{LUX-ZEPLIN}} ({{LZ}}) {{Experiment}}},
  \href{https://arxiv.org/abs/2207.03764}{{\ttfamily 2207.03764}}.

\bibitem{PandaX-II:2017hlx}
{\scshape PandaX-II} collaboration, \emph{Dark {{Matter Results From}}
  54-{{Ton-Day Exposure}} of {{PandaX-II Experiment}}},
  \href{https://doi.org/10.1103/PhysRevLett.119.181302}{\emph{Physical Review
  Letters} {\bfseries 119} (2017) 181302}
  [\href{https://arxiv.org/abs/1708.06917}{{\ttfamily 1708.06917}}].

\bibitem{Liu:2022zgu}
{\scshape PandaX} collaboration, \emph{The first results of {{PandaX-4T}}},
  \href{https://doi.org/10.1142/S0218271822300075}{\emph{International Journal
  of Modern Physics D} {\bfseries 31} (2022) 2230007}.

\bibitem{DarkSide:2018kuk}
{\scshape DarkSide} collaboration, \emph{{{DarkSide-50}} 532-day {{Dark Matter
  Search}} with {{Low-Radioactivity Argon}}},
  \href{https://doi.org/10.1103/PhysRevD.98.102006}{\emph{Physical Review D}
  {\bfseries 98} (2018) 102006}
  [\href{https://arxiv.org/abs/1802.07198}{{\ttfamily 1802.07198}}].

\bibitem{DarkSide:2018bpj}
{\scshape DarkSide} collaboration, \emph{Low-{{Mass Dark Matter Search}} with
  the {{DarkSide-50 Experiment}}},
  \href{https://doi.org/10.1103/PhysRevLett.121.081307}{\emph{Physical Review
  Letters} {\bfseries 121} (2018) 081307}
  [\href{https://arxiv.org/abs/1802.06994}{{\ttfamily 1802.06994}}].

\bibitem{Ibe:2017yqa}
M.~Ibe, W.~Nakano, Y.~Shoji and K.~Suzuki, \emph{Migdal {{Effect}} in {{Dark
  Matter Direct Detection Experiments}}},
  \href{https://doi.org/10.1007/JHEP03(2018)194}{\emph{Journal of High Energy
  Physics} {\bfseries 03} (2018) 194}
  [\href{https://arxiv.org/abs/1707.07258}{{\ttfamily 1707.07258}}].

\bibitem{SuperCDMS:2018gro}
{\scshape SuperCDMS} collaboration, \emph{Search for {{Low-Mass Dark Matter}}
  with {{CDMSlite Using}} a {{Profile Likelihood Fit}}},
  \href{https://doi.org/10.1103/PhysRevD.99.062001}{\emph{Physical Review D}
  {\bfseries 99} (2019) 062001}
  [\href{https://arxiv.org/abs/1808.09098}{{\ttfamily 1808.09098}}].

\bibitem{CRESST:2017ues}
{\scshape CRESST} collaboration, \emph{Results on {{MeV-scale}} dark matter
  from a gram-scale cryogenic calorimeter operated above ground},
  \href{https://doi.org/10.1140/epjc/s10052-017-5223-9}{\emph{The European
  Physical Journal C} {\bfseries 77} (2017) 637}
  [\href{https://arxiv.org/abs/1707.06749}{{\ttfamily 1707.06749}}].

\bibitem{CRESST:2019jnq}
{\scshape CRESST} collaboration, \emph{First results from the {{CRESST-III}}
  low-mass dark matter program},
  \href{https://doi.org/10.1103/PhysRevD.100.102002}{\emph{Physical Review D}
  {\bfseries 100} (2019) 102002}
  [\href{https://arxiv.org/abs/1904.00498}{{\ttfamily 1904.00498}}].

\bibitem{Fermi-LAT:2015att}
{\scshape Fermi-LAT} collaboration, \emph{Searching for {{Dark Matter
  Annihilation}} from {{Milky Way Dwarf Spheroidal Galaxies}} with {{Six
  Years}} of {{Fermi-LAT Data}}},
  \href{https://doi.org/10.1103/PhysRevLett.115.231301}{\emph{Physical Review
  Letters} {\bfseries 115} (2015) 231301}
  [\href{https://arxiv.org/abs/1503.02641}{{\ttfamily 1503.02641}}].

\bibitem{Leane:2018kjk}
R.~K. Leane, T.~R. Slatyer, J.~F. Beacom and K.~C.~Y. Ng, \emph{{{GeV-Scale
  Thermal WIMPs}}: {{Not Even Slightly Dead}}},
  \href{https://doi.org/10.1103/PhysRevD.98.023016}{\emph{Physical Review D}
  {\bfseries 98} (2018) 023016}
  [\href{https://arxiv.org/abs/1805.10305}{{\ttfamily 1805.10305}}].

\bibitem{Bergstrom:2013jra}
L.~Bergstrom, T.~Bringmann, I.~Cholis, D.~Hooper and C.~Weniger, \emph{New
  limits on dark matter annihilation from {{AMS}} cosmic ray positron data},
  \href{https://doi.org/10.1103/PhysRevLett.111.171101}{\emph{Physical Review
  Letters} {\bfseries 111} (2013) 171101}
  [\href{https://arxiv.org/abs/1306.3983}{{\ttfamily 1306.3983}}].

\bibitem{ATLAS:2019cid}
{\scshape ATLAS} collaboration, \emph{Combination of searches for invisible
  {{Higgs}} boson decays with the {{ATLAS}} experiment},
  \href{https://doi.org/10.1103/PhysRevLett.122.231801}{\emph{Physical Review
  Letters} {\bfseries 122} (2019) 231801}
  [\href{https://arxiv.org/abs/1904.05105}{{\ttfamily 1904.05105}}].

\bibitem{Covi:1999ty}
L.~Covi, J.~E. Kim and L.~Roszkowski, \emph{Axinos as cold dark matter},
  \href{https://doi.org/10.1103/PhysRevLett.82.4180}{\emph{Phys. Rev. Lett.}
  {\bfseries 82} (1999) 4180}
  [\href{https://arxiv.org/abs/hep-ph/9905212}{{\ttfamily hep-ph/9905212}}].

\bibitem{Chiang:2018gsn}
C.-W. Chiang, Y.-T. Li and E.~Senaha, \emph{Revisiting electroweak phase
  transition in the standard model with a real singlet scalar},
  \href{https://doi.org/10.1016/j.physletb.2018.12.017}{\emph{Physics Letters
  B} {\bfseries 789} (2019) 154}
  [\href{https://arxiv.org/abs/1808.01098}{{\ttfamily 1808.01098}}].

\bibitem{Croon:2020cgk}
D.~Croon, O.~Gould, P.~Schicho, T.~V.~I. Tenkanen and G.~White,
  \emph{Theoretical uncertainties for cosmological first-order phase
  transitions}, \href{https://doi.org/10.1007/JHEP04(2021)055}{\emph{JHEP}
  {\bfseries 04} (2021) 055}
  [\href{https://arxiv.org/abs/2009.10080}{{\ttfamily 2009.10080}}].

\bibitem{Nielsen:1975fs}
N.~K. Nielsen, \emph{On the gauge dependence of spontaneous symmetry breaking
  in gauge theories},
  \href{https://doi.org/10.1016/0550-3213(75)90301-6}{\emph{Nuclear Physics B}
  {\bfseries 101} (1975) 173}.

\bibitem{Fukuda:1975di}
R.~Fukuda and T.~Kugo, \emph{Gauge invariance in the effective action and
  potential}, \href{https://doi.org/10.1103/PhysRevD.13.3469}{\emph{Physical
  Review D: Particles and Fields} {\bfseries 13} (1976) 3469}.

\bibitem{Patel:2011th}
H.~H. Patel and M.~J. {Ramsey-Musolf}, \emph{Baryon {{Washout}}, {{Electroweak
  Phase Transition}}, and {{Perturbation Theory}}},
  \href{https://doi.org/10.1007/JHEP07(2011)029}{\emph{Journal of High Energy
  Physics} {\bfseries 07} (2011) 029}
  [\href{https://arxiv.org/abs/1101.4665}{{\ttfamily 1101.4665}}].

\bibitem{Chiang:2017zbz}
C.-W. Chiang and E.~Senaha, \emph{On gauge dependence of gravitational waves
  from a first-order phase transition in classical scale-invariant {{U}}(1)'
  models}, \href{https://doi.org/10.1016/j.physletb.2017.09.064}{\emph{Physics
  Letters B} {\bfseries 774} (2017) 489}
  [\href{https://arxiv.org/abs/1707.06765}{{\ttfamily 1707.06765}}].

\bibitem{Schicho:2022wty}
P.~Schicho, T.~V.~I. Tenkanen and G.~White, \emph{Combining thermal resummation
  and gauge invariance for electroweak phase transition},
  \href{https://arxiv.org/abs/2203.04284}{{\ttfamily 2203.04284}}.

\bibitem{Dolan:1973qd}
L.~Dolan and R.~Jackiw, \emph{Symmetry behavior at finite temperature},
  \href{https://doi.org/10.1103/PhysRevD.9.3320}{\emph{Physical Review D}
  {\bfseries 9} (1974) 3320}.

\bibitem{Parwani:1991gq}
R.~R. Parwani, \emph{Resummation in a hot scalar field theory},
  \href{https://doi.org/10.1103/PhysRevD.45.4695}{\emph{Physical Review D}
  {\bfseries 45} (1992) 4695}
  [\href{https://arxiv.org/abs/hep-ph/9204216}{{\ttfamily hep-ph/9204216}}].

\bibitem{Carrington:1991hz}
M.~E. Carrington, \emph{Effective potential at finite temperature in the
  standard model},
  \href{https://doi.org/10.1103/PhysRevD.45.2933}{\emph{Physical Review D}
  {\bfseries 45} (1992) 2933}.

\bibitem{Chao:2014ina}
W.~Chao, \emph{First order electroweak phase transition triggered by the
  {{Higgs}} portal vector dark matter},
  \href{https://doi.org/10.1103/PhysRevD.92.015025}{\emph{Physical Review D:
  Particles and Fields} {\bfseries 92} (2015) 015025}
  [\href{https://arxiv.org/abs/1412.3823}{{\ttfamily 1412.3823}}].

\bibitem{Breitbach:2018ddu}
M.~Breitbach, J.~Kopp, E.~Madge, T.~Opferkuch and P.~Schwaller, \emph{Dark,
  cold, and noisy: {{Constraining}} secluded hidden sectors with gravitational
  waves}, \href{https://doi.org/10.1088/1475-7516/2019/07/007}{\emph{JCAP}
  {\bfseries 07} (2019) 007}
  [\href{https://arxiv.org/abs/1811.11175}{{\ttfamily 1811.11175}}].

\bibitem{Borah:2021ocu}
D.~Borah, A.~Dasgupta and S.~K. Kang, \emph{Gravitational waves from a dark
  {{U}}(1){{D}} phase transition in light of {{NANOGrav}} 12.5 yr data},
  \href{https://doi.org/10.1103/PhysRevD.104.063501}{\emph{Physical Review D:
  Particles and Fields} {\bfseries 104} (2021) 063501}
  [\href{https://arxiv.org/abs/2105.01007}{{\ttfamily 2105.01007}}].

\bibitem{Cline:2012hg}
J.~M. Cline and K.~Kainulainen, \emph{Electroweak baryogenesis and dark matter
  from a singlet {{Higgs}}},
  \href{https://doi.org/10.1088/1475-7516/2013/01/012}{\emph{JCAP} {\bfseries
  01} (2013) 012} [\href{https://arxiv.org/abs/1210.4196}{{\ttfamily
  1210.4196}}].

\bibitem{Vaskonen:2016yiu}
V.~Vaskonen, \emph{Electroweak baryogenesis and gravitational waves from a real
  scalar singlet},
  \href{https://doi.org/10.1103/PhysRevD.95.123515}{\emph{Physical Review D:
  Particles and Fields} {\bfseries 95} (2017) 123515}
  [\href{https://arxiv.org/abs/1611.02073}{{\ttfamily 1611.02073}}].

\bibitem{Kurup:2017dzf}
G.~Kurup and M.~Perelstein, \emph{Dynamics of electroweak phase transition in
  singlet-scalar extension of the standard model},
  \href{https://doi.org/10.1103/PhysRevD.96.015036}{\emph{Physical Review D:
  Particles and Fields} {\bfseries 96} (2017) 015036}
  [\href{https://arxiv.org/abs/1704.03381}{{\ttfamily 1704.03381}}].

\bibitem{Ellis:2018mja}
J.~Ellis, M.~Lewicki and J.~M. No, \emph{On the maximal strength of a
  first-order electroweak phase transition and its gravitational wave signal},
  \href{https://doi.org/10.1088/1475-7516/2019/04/003}{\emph{Journal of
  Cosmology and Astroparticle Physics} {\bfseries 2019} (2019) 003}
  [\href{https://arxiv.org/abs/1809.08242}{{\ttfamily 1809.08242}}].

\bibitem{Biondini:2022ggt}
S.~Biondini, P.~Schicho and T.~V.~I. Tenkanen, \emph{Strong electroweak phase
  transition in {{t}}-channel simplified dark matter models},
  \href{https://arxiv.org/abs/2207.12207}{{\ttfamily 2207.12207}}.

\bibitem{Sakharov:1967dj}
A.~D. Sakharov, \emph{Violation of {{{\emph{CP}}}} in variance, {{{\emph{C}}}}
  asymmetry, and baryon asymmetry of the universe},
  \href{https://doi.org/10.1070/PU1991v034n05ABEH002497}{\emph{Soviet Physics
  Uspekhi} {\bfseries 5} (1967) 32}.

\bibitem{Wainwright:2011kj}
C.~L. Wainwright, \emph{{{CosmoTransitions}}: {{Computing}} cosmological phase
  transition temperatures and bubble profiles with multiple fields},
  \href{https://doi.org/10.1016/j.cpc.2012.04.004}{\emph{Computer Physics
  Communications} {\bfseries 183} (2012) 2006}
  [\href{https://arxiv.org/abs/1109.4189}{{\ttfamily 1109.4189}}].

\bibitem{Ellis:2019oqb}
J.~Ellis, M.~Lewicki, J.~M. No and V.~Vaskonen, \emph{Gravitational wave energy
  budget in strongly supercooled phase transitions},
  \href{https://doi.org/10.1088/1475-7516/2019/06/024}{\emph{Journal of
  Cosmology and Astroparticle Physics} {\bfseries 06} (2019) 024}
  [\href{https://arxiv.org/abs/1903.09642}{{\ttfamily 1903.09642}}].

\bibitem{Ellis:2020awk}
J.~Ellis, M.~Lewicki and J.~M. No, \emph{Gravitational waves from first-order
  cosmological phase transitions: Lifetime of the sound wave source},
  \href{https://doi.org/10.1088/1475-7516/2020/07/050}{\emph{Journal of
  Cosmology and Astroparticle Physics} {\bfseries 07} (2020) 050}
  [\href{https://arxiv.org/abs/2003.07360}{{\ttfamily 2003.07360}}].

\bibitem{Guo:2020grp}
H.-K. Guo, K.~Sinha, D.~Vagie and G.~White, \emph{Phase transitions in an
  expanding universe: Stochastic gravitational waves in standard and
  non-standard histories},
  \href{https://doi.org/10.1088/1475-7516/2021/01/001}{\emph{Journal of
  Cosmology and Astroparticle Physics} {\bfseries 01} (2021) 001}
  [\href{https://arxiv.org/abs/2007.08537}{{\ttfamily 2007.08537}}].

\bibitem{Schmitz:2020syl}
K.~Schmitz, \emph{New sensitivity curves for gravitational-wave signals from
  cosmological phase transitions},
  \href{https://doi.org/10.1007/JHEP01(2021)097}{\emph{JHEP} {\bfseries 01}
  (2021) 097} [\href{https://arxiv.org/abs/2002.04615}{{\ttfamily
  2002.04615}}].

\bibitem{Ringwald:2020vei}
A.~Ringwald, K.~Saikawa and C.~Tamarit, \emph{Primordial gravitational waves in
  a minimal model of particle physics and cosmology},
  \href{https://doi.org/10.1088/1475-7516/2021/02/046}{\emph{JCAP} {\bfseries
  02} (2021) 046} [\href{https://arxiv.org/abs/2009.02050}{{\ttfamily
  2009.02050}}].

\bibitem{Banerjee:2015gca}
S.~Banerjee, P.~S.~B. Dev, A.~Ibarra, T.~Mandal and M.~Mitra, \emph{Prospects
  of heavy neutrino searches at future lepton colliders},
  \href{https://doi.org/10.1103/PhysRevD.92.075002}{\emph{Phys. Rev. D}
  {\bfseries 92} (2015) 075002}
  [\href{https://arxiv.org/abs/1503.05491}{{\ttfamily 1503.05491}}].

\bibitem{Steinhardt:1981ct}
P.~J. Steinhardt, \emph{Relativistic detonation waves and bubble growth in
  false vacuum decay},
  \href{https://doi.org/10.1103/PhysRevD.25.2074}{\emph{Physical Review D}
  {\bfseries 25} (1982) 2074}.

\end{thebibliography}
